\documentclass[11pt,a4paper]{article}
\pdfoutput=1

\usepackage{jcappub}

\usepackage{mathrsfs}
\usepackage{bbm}
\usepackage{bm}
\usepackage{dsfont}
\usepackage{tensor}
\usepackage{xspace}
\usepackage{lmodern}
\usepackage{wasysym}
\usepackage{microtype}
\usepackage{empheq}
\usepackage{ifthen}
\usepackage{amsmath}
\usepackage{wrapfig}
\usepackage{cleveref}
\crefname{section}{sec.}{secs.}
\Crefname{section}{Section}{Sections}
\crefname{subsection}{subsec.}{subsecs.}
\Crefname{subsection}{Subsection}{Subsections}
\crefname{subsubsection}{subsubsec.}{subsubsecs.}
\Crefname{subsubsection}{Subsubsection}{Subsubsections}
\usepackage{xcolor}
\usepackage{import}
\usepackage{siunitx}
\usepackage{natbib}
\DeclareSIUnit{\parsec}{pc}
\DeclareSIUnit{\astronomicalunit}{AU}

\newcommand\ph{\ensuremath{\varphi}}

\newcommand\define{\equiv}

\newcommand\vect[1]{\boldsymbol{#1}}
\newcommand\mat[1]{\boldsymbol{#1}}
\newcommand\cplx[1]{\underline{#1}}
\newcommand\ex[1]{\mathrm{e}^{#1}}
\newcommand\ii{\mathrm{i}}

\renewcommand\Re{\mathrm{Re}}
\renewcommand\Im{\mathrm{Im}}

\newcommand\e[1]{_{\mathrm{#1}}}

\newcommand{\dd}{\mathrm{d}}

\newcommand{\pd}[3][]{\frac{\partial^{#1} #2}{\partial {#3}^{#1}}}
\newcommand{\ddf}[3][]{\frac{\dd^{#1} #2}{\dd {#3}^{#1}}}

\newcommand{\delimiters}[4][]{
\ifthenelse{ \equal{#1}{1} }{  #2 #3 #4  }
					{ \ifthenelse{\equal{#1}{2}}{ \big#2 #3 \big#4 }
						{ \ifthenelse{\equal{#1}{3}}{ \Big#2 #3 \Big#4 }
							{ \ifthenelse{\equal{#1}{4}}{ \bigg#2 #3 \bigg#4 }
								{ \ifthenelse{\equal{#1}{5}}{ \Bigg#2 #3 \Bigg#4 }
									{ \left#2 #3 \right#4 }
								}
							}
						}
					}
													}
													
\newcommand{\pa}[2][]{\delimiters[#1]{(}{#2}{)}}
\newcommand{\pac}[2][]{\delimiters[#1]{[}{#2}{]}}
\newcommand{\paac}[2][]{\delimiters[#1]{\{}{#2}{\}}}
\newcommand{\abs}[2][]{\delimiters[#1]{|}{#2}{|}}
\newcommand{\ev}[2][]{\delimiters[#1]{\langle}{#2}{\rangle}}

\definecolor{blue4}{RGB}{0,0,143}
\definecolor{red4}{RGB}{143,0,0}
\definecolor{orange}{RGB}{255,128,0}
\definecolor{darkcyan}{RGB}{0,128,128}
\definecolor{olive}{RGB}{0,128,0}
\definecolor{purple}{RGB}{128,0,128}
\definecolor{cyan2}{RGB}{0,255,255}
\definecolor{fushia}{RGB}{255,0,255}
\definecolor{mygray}{gray}{0.5}
\definecolor{lightgray}{gray}{0.85}

\newcommand{\amp}{\mathcal{A}}
\newcommand{\Prob}{\mathrm{Prob}}

\newcommand{\changed}[1]{#1}

\title{Accurate modelling of extragalactic microlensing by compact objects}

\author[a]{Víctor Boscá,}
\author[a,b]{Pierre Fleury,}
\author[a]{Juan García-Bellido}

\affiliation[a]{Instituto de Física Teórica UAM-CSIC,
Universidad Autónoma de Madrid,\\
Cantoblanco, 28049 Madrid, Spain}
\affiliation[b]{Université Paris-Saclay, CNRS, CEA,\\
Institut de physique théorique,
91191, Gif-sur-Yvette, France}

\emailAdd{victor.bosca@uam.es}
\emailAdd{pierre.fleury@ipht.fr}
\emailAdd{juan.garciabellido@uam.es}

\abstract{
Microlensing of extragalactic sources, in particular the probability of significant amplifications, is a potentially powerful probe of the abundance of compact objects outside the halo of the Milky Way. Accurate experimental constraints require an equally accurate theoretical model for the amplification statistics produced by such a population.
In this article, we argue that the simplest (strongest-lens) model does not meet this demanding requirement. We thus propose an elaborate practical modelling scheme for extragalactic microlensing.
We derive from first principles an expression for the amplification probability that consistently allows for: (i) the coupling between microlenses; (ii) realistic perturbations from the cosmic large-scale structure; (iii) extended-source corrections.
An important conclusion is that the external shear applied on the dominant microlens, both by the other lenses and by the large-scale structure, is practically negligible. Yet, the predictions of our approach can still differ by a factor of a few with respect to existing models of the literature. Updated constraints on the abundance of compact objects accounting for such discrepancies may be required.
}


\preprint{IFT-UAM/CSIC-22-51}

\date{\today}

\begin{document}

\maketitle
\flushbottom

\section{Introduction}
\label{sec:introduction}

Needless to say the nature of the dark matter (DM) is one of the most open questions in modern science. Since the first detection of a gravitational-wave signal reported by the LIGO/Virgo Collaboration~\cite{PhysRevLett.116.241103}, an old DM candidate got back under spotlight, namely primordial black holes (PBHs)~\cite{Bird:2016dcv,Clesse:2016vqa,Sasaki:2016jop}. Originally considered by Zel'dovich \& Novikov~\cite{1967SvA....10..602Z} and Hawking~\cite{10.1093/mnras/152.1.75}, PBHs are to date the only known example of massive compact halo object (MACHO)~\cite{1991ApJ...366..412G} that could make up a significant fraction of the DM (see e.g. ref.~\cite{Carr:2016drx} and references therein). Other historical MACHO candidates, such as neutron stars, planets or brown dwarfs, are excluded by a variety of cosmological constraints implying that DM must be non-baryonic; those notably include the precision measurement of the CMB acoustic peaks
and the abundance of light elements predicted in the context of the hot big-bang nucleosynthesis -- see e.g. chapters~6--8 of ref.~\cite{1804633} for a detailed historical review.

A suitable probe of the clumpiness of the DM is gravitational lensing, which has the virtue of being directly sensitive to the distribution of mass, as opposed to the astronomical observations which rely on luminous matter. In particular, microlensing is found to be a useful tool to detect compact objects. The idea is that, as a compact object crosses the line of sight of a distance source, it temporarily magnifies its apparent brightness~\cite{1986ApJ...301..503P}. This idea was put into practice by the MACHO experiment~\cite{MACHO:1995udp}, which looked for microlensing events in the Magellanic Clouds for several years in the 90s by monitoring thousands of stars hoping that some compact objects in the mass range $3 \times 10^{-4}$ to $0.06\,M_{\odot}$ would cross the line of sight and amplify their brightness. Among the 9.5 million light curves that were analysed, only 3 events consistent with microlensing were found. After similar analyses were conducted by the EROS and OGLE collaborations~\cite{https://doi.org/10.48550/arxiv.astro-ph/0212176,10.1111/j.1365-2966.2011.19243.x}, the possibility that the dark halo of our galaxy is made of compact objects in that mass range was excluded. Those constraints keep being discussed in the literature; for instance ref.~\cite{2015A&A...575A.107H} argues that detection method presented inconsistencies; ref.~\cite{Calcino_2018} suggests that the constraints may be alleviated if the compact objects are clustered. Importantly, galactic microlensing is inefficient at detecting high-mass compact objects~\cite{Sato_Polito_2019}.

Press and Gunn~\cite{1973ApJ...185..397P} proposed to use quasar microlensing for detecting a cosmologically significant density of compact bodies. This idea was further investigated by Hawkins~\cite{1993Natur.366..242H, https://doi.org/10.48550/arxiv.astro-ph/9811019} and Hawkins and Verón~\cite{10.1093/mnras/260.1.202} who studied the structure of DM using quasar light curves, and Schneider~\cite{1993A&A...279....1S}, who constrained the population of compact objects that could make up the total matter density. The difficulty of constraining the abundance of high-mass compact objects in our galactic halo motivated the observation of microlensing effects in images of multiple imaged quasars~\cite{Mediavilla:2009um, 1991AJ....102.1939W, 10.1093/mnras/283.1.225}. Microlensing can also manifest as flux ratio anomalies between multiple quasar images \cite{1995ApJ...443...18W, Schechter_2002, 10.1111/j.1365-2966.2011.19982.x}, which may be interpreted as a direct proof for the presence of CDM substructure around lensing galaxies.   

Another method to constrain compact DM, motivated by the recent discoveries of the highly magnified stars MACS J1149 Lensed Star 1 (``Icarus'') \cite{https://doi.org/10.48550/arxiv.1706.10279} and WHL0137-LS (``Earendel'') \cite{article}, both visible at cosmological distances, uses caustic-crossing events in giant arcs. This mechanism allows the detection of compact objects in the subsolar-mass regime. 

One possible alternative to detect extragalactic microlensing due to compact objects beyond the solar mass is via supernova (SN) lensing. Refdsal~\cite{1964MNRAS.128..295R} was the first to study the possibility of observing this phenomenon and use SNe as a cosmological probe. Since then there has been many research programmes looking for strong gravitational lensing of SNe~\cite{Suyu_2020}, where microlensing is a potentially worrisome source of noise. The idea of using SN microlensing as a signal, to constrain the abundance of compact objects was first proposed by Linder, Schneider and Wagoner~\cite{1988ApJ...324..786L}. Rauch~\cite{1991ApJ...374...83R} and Metcalf and Silk~\cite{1999ApJ...519L...1M} only considered two scenarios: all and none of the DM in the form of compact objects. Seljak and Holz~\cite{Seljak:1999tm} already contemplated that a fraction of DM in compact objects can be measured with any given SN survey~\cite{Bergstrom:1999xh}. The constraints were updated 7 years later by Metcalf and Silk~\cite{2007PhRvL..98g1302M}, and most recently by Zumalacárregui and Seljak~\cite{Zumalacarregui:2017qqd}; see also refs.~\cite{Jonsson:2005qv,Dhawan:2017kft}.

From the theoretical perspective, all the aforementioned methods share the need of an accurate modelling of the microlensing statistics, and in particular of the probability density function of the lensing amplification, $p(A)$. This theoretical effort started in the 80s, and include the works of Peacock~\cite{10.1093/mnras/199.4.987}; Turner, Ostriker, and Gott~\cite{1984ApJ...284....1T}; and Dyer~\cite{1984ApJ...287...26D}. Years later, Schneider~\cite{1987A&A...171...49S} and Seitz and Schneider~\cite{1994A&A...288....1S} determined that the probability distribution of a point source exhibits a $A^{-3}$ behaviour for large amplification. Still more elaborate analyses helped understanding the non-linear interaction between lenses~\cite{1987A&A...171...49S, 1984JApA....5..235N, 1986ApJ...310..568B, 1990ApJ...357...23L, 1992ApJ...389...63M, 1997ApJ...489..508K, 1997ApJ...489..522L}. Besides, a number of ray-shooting numerical simulations were performed to assess the accuracy of those theoretical works~\cite{1986A&A...166...36K, 1992ApJ...386...19W, 1991ApJ...374...83R, 1995MNRAS.276..103L}.

However, those past analysis generally focused on a fraction of the effects potentially affecting the amplification statistics. In this article, we thus extend and improve upon them, by proposing an accurate modelling of the statistics of extragalactic microlensing from first principles. Our model accounts for line-of-sight effects and lens-lens coupling in the mild-optical-depth regime, and extended-source corrections. The end product is a semi-analytical expression for the amplification probability, in a realistic universe whose DM is made of a certain fraction of compact objects.

The remainder of the article is organised as follows. In \cref{sec:optical_depth}, we discuss on the notion of microlensing optical depth, its role in amplification statistics, and we evaluate its relevant values in practice. In \cref{sec:point_lens_with_perturbations} we account for the environment and line-of-sight corrections of a single point lens; we turn this result into a probability density function for the amplification in \cref{sec:amplification_probabilities}, where we also compare the predictions of the most recent analysis to our approach. We consider extended-source corrections in \cref{sec:extended_sources} and conclude in \cref{sec:conclusion}.

\paragraph{Conventions and notation.} We adopt units in which the speed of light is unity, $c=1$. The background cosmological model is set to have a spatially flat ($K=0$) Friedmann-Lemaître-Robertson-Walker geometry with the 2018 \emph{Planck} best-fit parameters~\cite{Planck:2018vyg}. Bold symbols denote two-dimensional or three-dimensional vectors. The probability density function (PDF) of a random variable $x$ is denoted with with a lower-case $p(x)$. We use and upper-case $P$ to denote the complementary cumulative distribution function (CDF), defined as
\begin{equation}
P(x) \define \Prob(>x) = \int_x^\infty \dd x' \; p(x') \ .
\end{equation}
In conditional probabilities, variables are separated from conditions with a semicolon; for instance, $p(x;y)$ denotes the PDF of $x$ under the condition that $y$ is fixed.

\section{Optical depth and extragalactic microlensing}
\label{sec:optical_depth}

This section is a preliminary discussion on the notion of microlensing optical depth, denoted~$\tau$, which will be central in the discussions of this paper. After providing definitions of $\tau$, and illustrating its role in amplification statistics, we demonstrate that microlensing by extragalactic compact objects is characterised by a low, although not very low, optical depth.

\subsection{Intuition and definitions}

Consider a population of compact objects distributed in space. In this article, compactness will be defined in the sense of lensing rather than in the sense of gravitation: a compact object will loosely refer to a celestial body capable of producing multiple images and strong amplifications of point sources. Let us model such objects as point lenses. Each lens is then fully characterised by its \emph{Einstein radius}
\begin{equation}
\label{eq:einsteinradius}
\theta_{\mathrm{E}}
\equiv
\sqrt{\frac{4GmD_{\mathrm{ds}}}{D_\mathrm{os}D_\mathrm{od}}} \ ,
\end{equation}
where $m$ is the mass of the lens, $D_{\mathrm{od}}$ is angular-diameter distance to the lens (or deflector), $D_{\mathrm{os}}$ is the angular-diameter distance to the light source (in the absence of the lens) and $D_{\mathrm{ds}}$ is the angular-diameter distance to the source as seen from the lens.

The Einstein radius technically represents the size of the ring that would be observed if a point source were exactly aligned with the lens. But it also gives an idea of the lensing \emph{cross section} of the lens. If the angle~$\beta$ separating the source and the lens on the celestial sphere is comparable to $\theta\e{E}$, then the image's total flux is appreciably amplified compared to the source; if on the contrary $\beta\gg\theta\e{E}$, then the amplification is close to unity. More precisely, the amplification factor $A\define F\e{o}/F\e{s}$ between the observed flux~$F\e{o}$ and the unlensed source's flux~$F\e{s}$ reads, for a point lens~\cite{1992grle.book.....S},
\begin{equation}
\label{eq:A_u}
A = \frac{u^2 + 2}{u\sqrt{u^2+4}} \ ,
\qquad
u \define \frac{\beta}{\theta\e{E}} \ .
\end{equation}
For $u=1$, i.e. when the source is at the verge of the Einstein disk, we have $A_1=3/\sqrt{5}\approx 1.34$.

Assuming a fixed distance to a light source, we may now picture the Einstein disks of our population of lenses covering part of the celestial sphere. The probability that a certain light source gets significantly amplified is then naturally quantified by the fraction of the sky that is covered by Einstein disks. This fraction is called the \emph{microlensing optical depth},
\begin{empheq}[box=\fbox]{equation}
\label{eq:optical_depth_definition}
\tau
\define \frac{1}{4\pi} \sum_\ell \pi\theta\e{E,\ell}^2
= \Sigma \ev[2]{\pi \theta\e{E}^2} ,
\end{empheq}
where $\Sigma$ denotes the angular density of lenses, i.e. the number of lenses per unit solid angle in the sky, and $\ev{\ldots}$ denotes a statistical expectation value. If the distribution of lenses is inhomogeneous, then $\tau$ may be considered a field on the sphere.

It is instructive to express the optical depth in terms of more standard cosmological quantities. Suppose that the compact lenses are placed in a homogeneous-isotropic FLRW universe with zero spatial curvature. Denote with $\rho\e{c}(t, \vect{x})$ their contribution to the physical cosmic matter density at time $t$ and position $\vect{x}$. Then \cref{eq:optical_depth_definition} reads (see e.g. ref.~\cite{Fleury:2019xzr})
\begin{equation}
\label{eq:optical_depth_rho}
\tau
= 4\pi G \int_0^{\chi\e{s}} \dd\chi \
    \frac{\chi(\chi\e{s}-\chi)}{\chi\e{s}} \,
    a^2(\chi) \, \rho\e{c}(\chi) \ ,
\end{equation}
where $\chi$ and $\chi\e{s}$ respectively denote the comoving distances of the lenses and of the source from the observer; $a$ denotes the cosmic scale factor, and the notation $a(\chi), \rho\e{c}(\chi)$ mean that those quantities are evaluated at $\chi$ down the FLRW light cone, that is at conformal time $\eta(\chi)=\eta_0-\chi$ if $\eta_0$ means today. It is also implicit that $\rho\e{c}$ is spatially evaluated along a straight line, thereby giving $\tau$ an angular dependency from the inhomogeneity of $\rho\e{c}$.

\subsection{Amplification probability at very low optical depth}
\label{subsec:strongest_lens_approach}

The problem of determining the statistics of microlensing amplifications turns out to be quite simple in the low-optical depth regime, $\tau\ll 1$. In this case, which corresponds to lenses being rare and well separated from each other, the total amplification produced on a given light source is well approximated by the amplification of the strongest lens of the population, i.e. with the smallest reduced impact parameter~$u$ to the source. This shall be referred to as the \emph{strongest-lens prescription}. It is then quite straightforward to derive the (complementary) cumulative distribution function (CDF) and probability density function (PDF) of the amplification~\cite{1986MNRAS.223..113P, Fleury:2019xzr}
\begin{align}
\label{eq:CDF_A_low_optical_depth}
P(A; \tau)
&=
1 - \exp\pac{ -2\tau\pa{\frac{A}{\sqrt{A^2-1}}-1} } ,
\\
\label{eq:PDF_A_low_optical_depth}
p(A; \tau)
&=
- \pd{P}{A}
= \frac{2\tau}{(A^2 - 1)^{3/2}} \exp\pac{ -2\tau\pa{\frac{A}{\sqrt{A^2-1}}-1} } .
\end{align}
The PDF displays the well-known asymptotic behaviour $p(A\gg 1;\tau)\propto 2\tau/A^3$ for high amplifications. It is also easy to check from \cref{eq:PDF_A_low_optical_depth} that the average amplification reads
$
\ev{A} = 1 + 2\tau
$
at lowest order in $\tau\ll 1$, which means that the average magnification due to a sparse population of lenses equals the amplification that would be produced by the same matter density if it were smoothly distributed in space.\footnote{This result was first obtained in 1976 by Weinberg~\cite{Weinberg:1976jq}, who thereby showed the important result that, at linear order, the average luminosity distance measured in a clumpy Universe is the same as in the underlying homogeneous model. This was later generalised at any order by refs.~\cite{kibble2005average, 2008MNRAS.386..230W}; see also ref.~\cite{2021A&A...655A..54B} for details.} Another remarkable property of the amplification PDF/CDF at very low optical depths is that \emph{it does not depend on the mass of the deflectors}, but only on the optical depth. In that sense a sparse population of high-mass lenses is statistically indistinguishable from an abundant population of low-mass lenses.

\subsection{Distribution of optical depth in a realistic universe}
\label{subsec:distribution_optical_depth}

Given the simplicity of the amplification statistics in the low-optical depth regime, the first question that we need to address is whether or not this regime is a good description of extragalactic microlensing. A first way to address this question consists in estimating the optical depth in a realistic inhomogeneous Universe containing a population of compact objects. Let $\alpha$ be the fraction of the total matter that consists of compact objects. For simplicity, the distribution of compact objects is assumed to closely follow the total matter density field: in a small region with density~$\rho(t, \vect{x})$, there is a population of compact objects with mean density
\begin{equation}
\rho\e{c}(t, \vect{x}) = \alpha \, \rho(t, \vect{x}) \ .
\end{equation}
We assume that $\alpha$ is constant in space and time. A concrete example of this scenario would be if a fraction $f\e{PBH}=\alpha/0.83$ of the DM were made of PBHs.

\begin{figure}[t]
\centering
\begin{minipage}{0.5\columnwidth}
\includegraphics[width=\columnwidth]{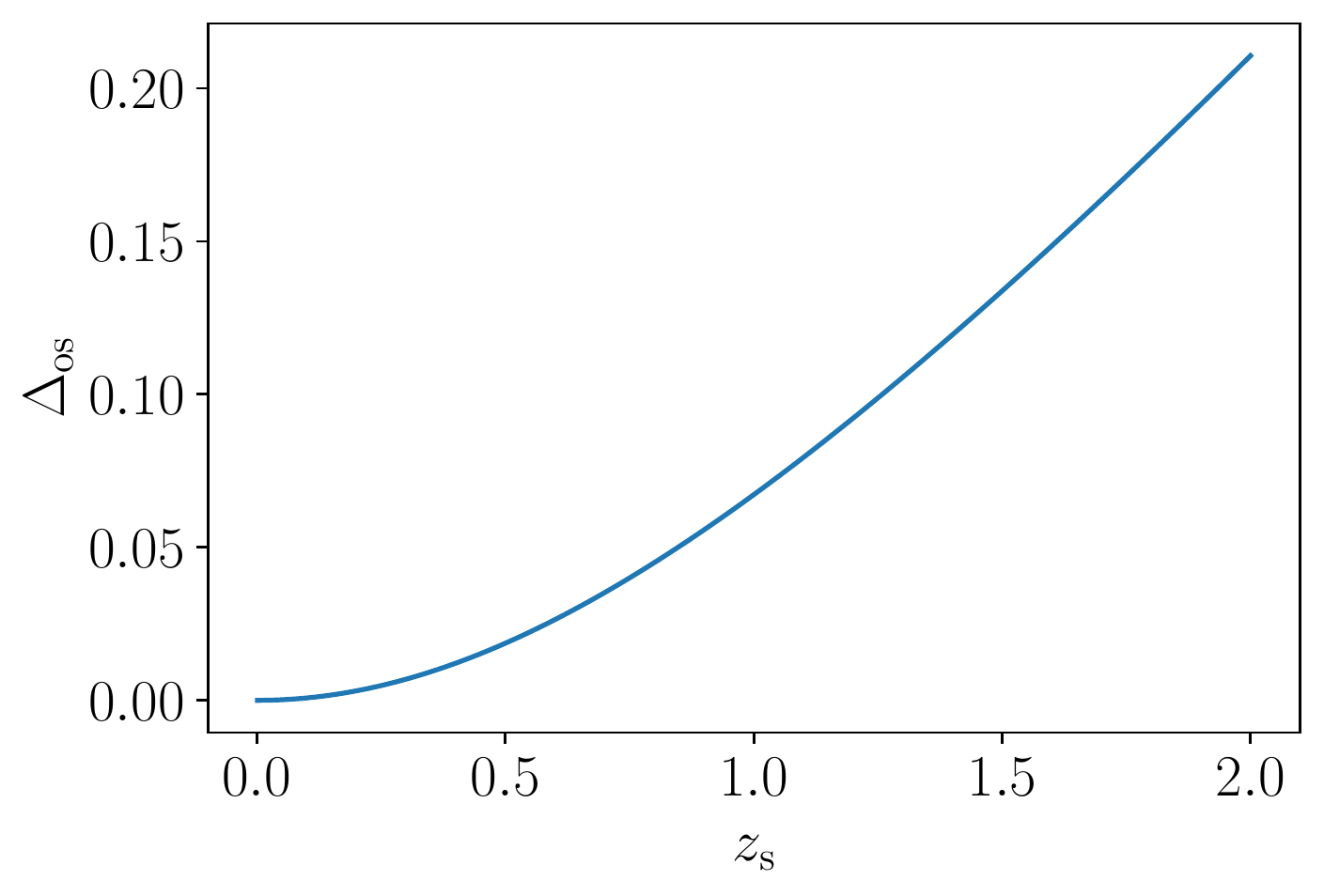}
\end{minipage}
\hfill
\begin{minipage}{0.4\columnwidth}
\caption{Evolution of the mean projected density term~$\Delta\e{os}(z\e{s})$ defined in \cref{eq:Delta_os} as a function of the redshift~$z\e{s}$ of the source. We note that this quantity is non-negligible at high redshift, reaching about \SI{10}{\percent} between $z=1$ and $1.5$.}
\label{fig:Delta_os}
\end{minipage}
\end{figure}

In such conditions, if we split the total matter density into cosmic mean~$\bar{\rho}(t)$ and large-scale perturbations as $\rho(t, \vect{x}) = \bar{\rho}(t) [1+\delta(t, \vect{x})]$, where $\delta(t, \vect{x})$ denotes the density contrast, then the optical depth~\eqref{eq:optical_depth_rho} takes the form
\begin{equation}
\label{eq:optical_depth_alpha}
\tau = \alpha (\Delta\e{os} + \bar{\kappa}\e{os}) \ ,
\end{equation}
with\footnote{The reason why we specified the subscript ``os'' in $\Delta\e{os}, \bar{\kappa}\e{os}$ will be clearer in \cref{sec:point_lens_with_perturbations}.}
\begin{align}
\label{eq:Delta_os}
\Delta\e{os}
&=
4\pi G \bar{\rho}_0
\int_0^{\chi\e{s}} \dd\chi \;
\frac{\chi(\chi\e{s}-\chi)}{\chi\e{s}} \,
\frac{1}{a(\chi)}\ ,
\\
\label{eq:kappa_bar_os}
\bar{\kappa}\e{os}
&=
4\pi G \bar{\rho}_0
\int_0^{\chi\e{s}} \dd\chi \;
\frac{\chi(\chi\e{s}-\chi)}{\chi\e{s}} \,
\frac{\delta(\chi)}{a(\chi)} \ ,
\end{align}
where $\bar{\rho}_0=\bar{\rho}(t_0)$ denotes today's cosmic mean density. If the Universe were homogeneous on astronomical scales ($\delta=0$), then we would have $\tau=\alpha\Delta\e{os}$; this quantity thus represents the contribution of the mean cosmic density to the optical depth.\footnote{Note also that $-\Delta\e{os}$ represents the convergence of Zel'dovich's ``empty-beam''~\cite{1964SvA.....8...13Z}, i.e., the negative convergence that would apply if light were propagating through an empty Universe compared to FLRW.} The evolution of $\Delta\e{os}$ with the redshift~$z\e{s}$ of the source is depicted in \cref{fig:Delta_os}.

The second quantity in \cref{eq:optical_depth_alpha}, $\bar{\kappa}\e{os}$, is a projection of the total density perturbation along the line of sight; it coincides with the weak-lensing convergence that would occur if matter were entirely diffuse, i.e., if the compact objects were smoothed out. For an overdense line of sight, $\bar{\kappa}\e{os}>0$, there are more compact objects and hence $\tau$ increases. We estimate the distribution of $\bar{\kappa}\e{os}$ from a combination of (i) publicly available numerical results from ray tracing in an $N$-body simulation~\cite{Breton:2018wzk} and (ii) standard cosmological calculations; see \cref{app:fit_simulation} for details on the simulation and our fitting functions.

\begin{figure}
\centering
\begin{minipage}{0.3\columnwidth}
\caption{Simulated map of the microlensing optical depth~$\tau$ (in logarithmic scale) in a universe with a fraction~$\alpha=0.5$ of compact objects, and a source at $z\e{s}=0.45$. The sky is dominated by very low values of $\tau$, with rare occurrences of mild values.}
\label{fig:map_tau}
\end{minipage}
\hfill
\begin{minipage}{0.65\columnwidth}
\includegraphics[width=\columnwidth]{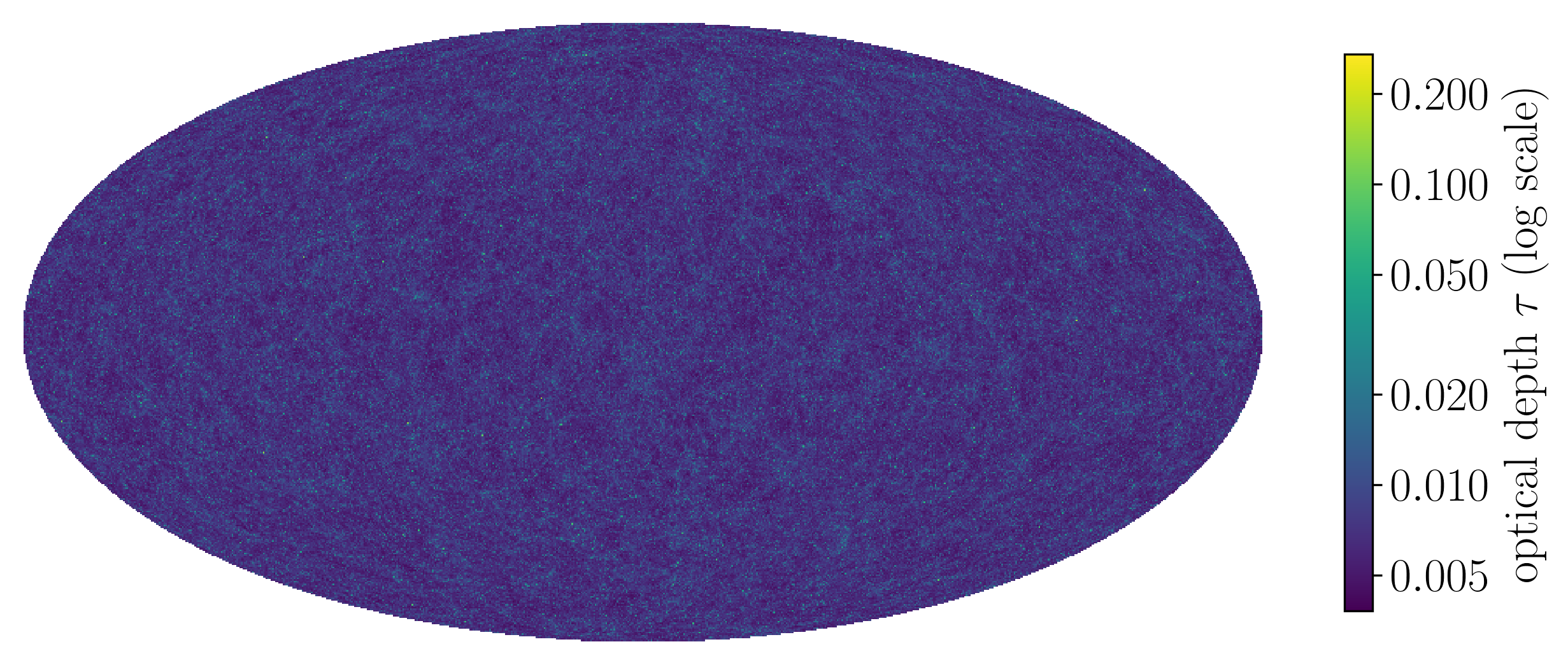}
\end{minipage}
\end{figure}

\begin{figure}
\centering
\includegraphics[width=0.49\columnwidth]{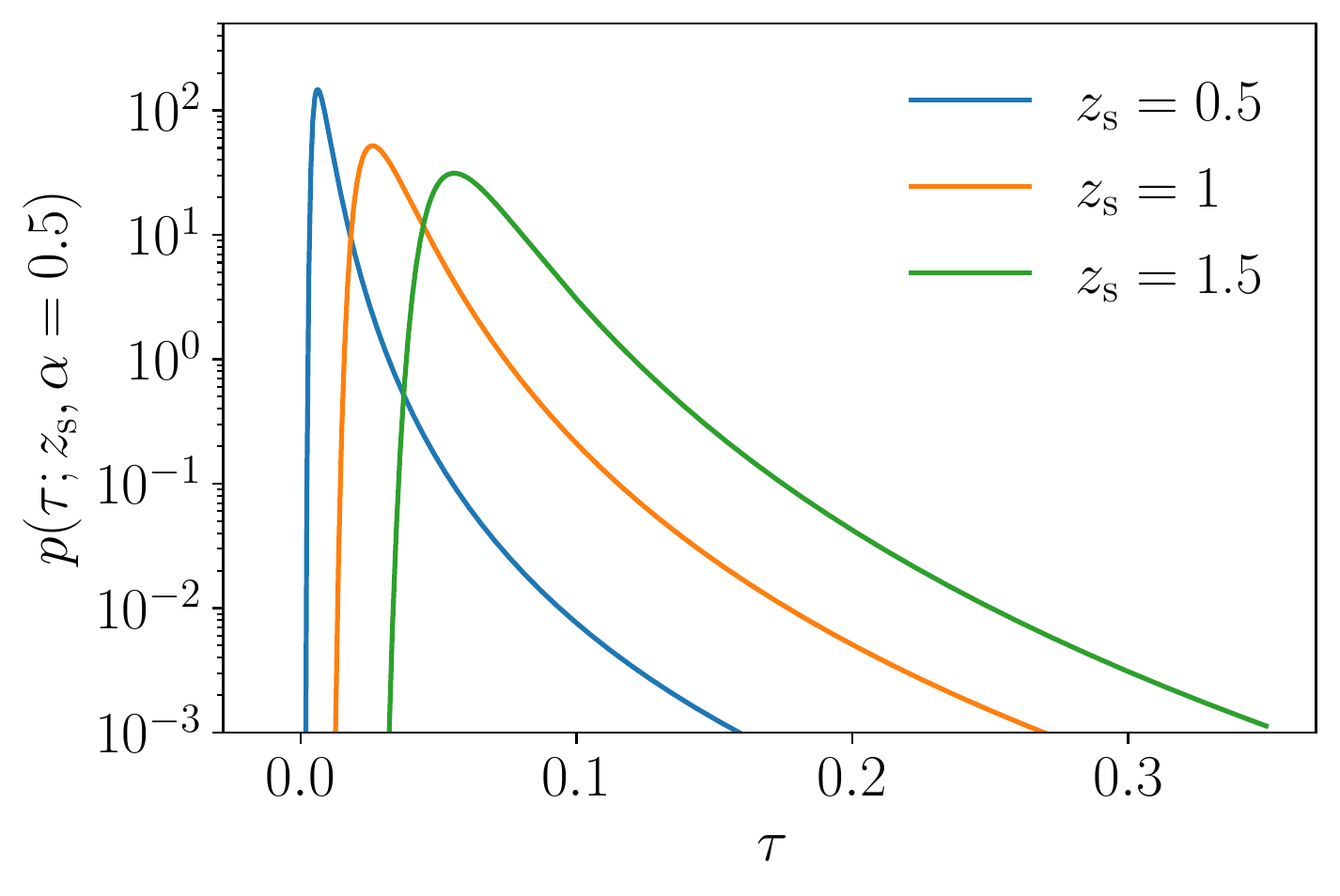}
\hfill
\includegraphics[width=0.49\columnwidth]{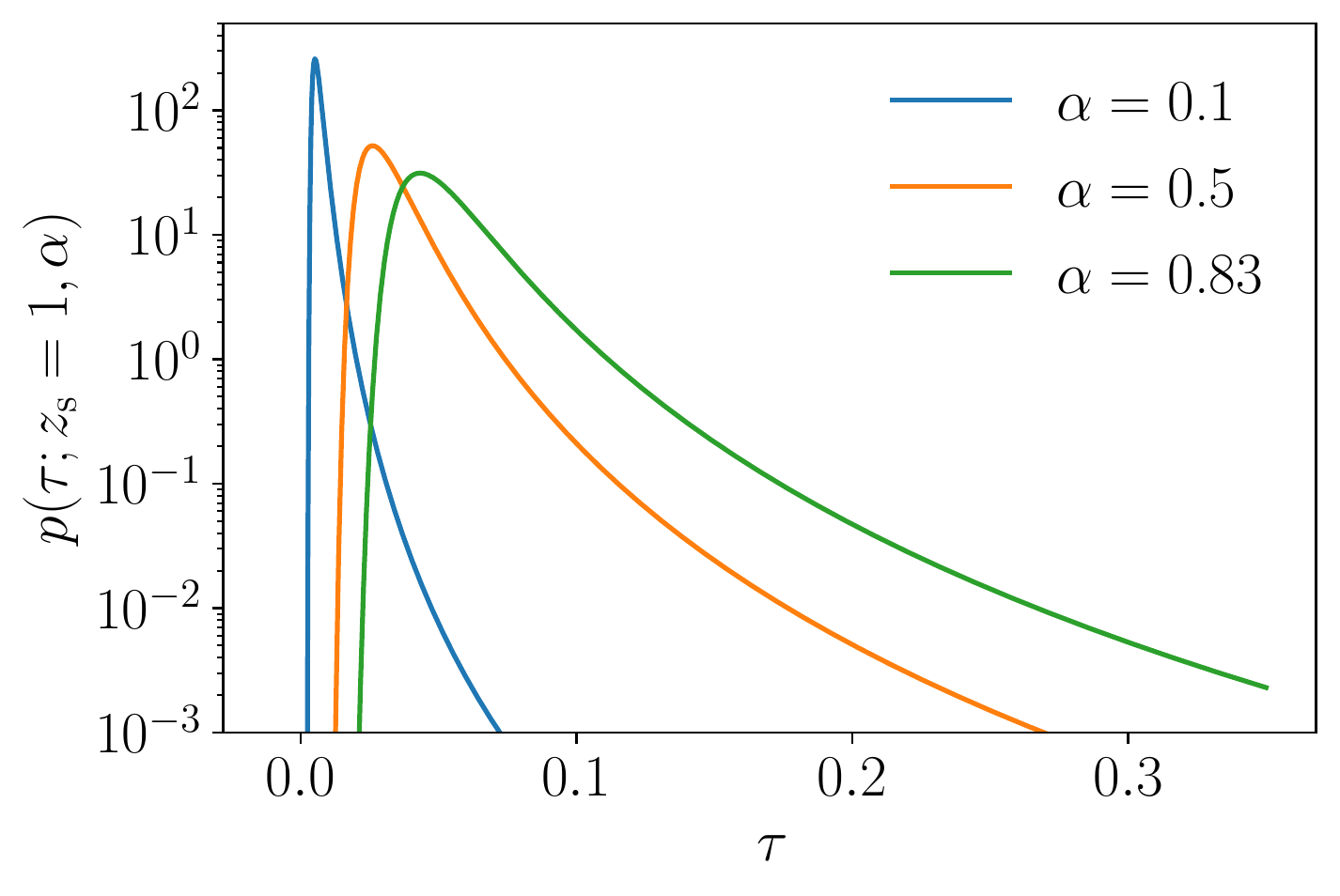}
\caption{Probability density~$p(\tau; z\e{s},\alpha)$ of the optical depth $\tau$ for a source at redshift $z\e{s}$ in a realistic universe made of a fraction $\alpha$ of compact objects. \emph{Left}: effect the source redshift for $\alpha=0.5$. \emph{Right}: effect of the fraction of compact objects for $z\e{s}=1$.}
\label{fig:PDF_tau}
\end{figure}

For the sake of illustration, \cref{fig:map_tau} shows a sky map of the optical depth~$\tau$ for $z\e{s}=0.45$ and a fraction~$\alpha=0.5$ of compact objects. More quantitatively, our prediction for the PDF of the optical depth~$\tau$ is shown in \cref{fig:PDF_tau} for various values of the fraction of compact objects~$\alpha$ and of the source redshift~$z\e{s}$. We can see that, except for sources located at high redshift and for a fraction of compact objects approaching unity, the optical depth remains at most on the percent order for most of the lines of sight.

\subsection{Relevant optical depths are not that low}

The distributions shown in \cref{fig:PDF_tau} indicate that, except in rather extreme cases, most of the celestial sphere is characterised by a very low optical depth, thereby  suggesting that the model of \cref{eq:CDF_A_low_optical_depth} may be a good description of the amplification probabilities. However, this conclusion must be nuanced as we wish to focus on mild to high amplifications. Suppose for instance that we seek a microlensing signal in the Hubble diagram of type-Ia SNe, such as in refs.~\cite{1999ApJ...519L...1M, Seljak:1999tm, 2007PhRvL..98g1302M, Zumalacarregui:2017qqd}. To be detectable, the effect of microlensing should be larger than the intrinsic dispersion of SN magnitudes, $\sigma\e{int}\sim 0.1~\mathrm{mag}$~\cite{2011ApJS..192....1C}. The decrease of an SN magnitude by $3\sigma\e{int}$ would be equivalent to an amplification factor $A=10^{6\sigma\e{int}/5}\approx 1.3$, which is considerable.

Although regions with large optical depth~$\tau$ are rare, they are also expected to produce more detectable amplifications than the low-$\tau$ regions. The relevant question then becomes: are detectable amplifications mostly lying in low-$\tau$ regions, which cover most of the sky, or in the rarer but more efficient high-$\tau$ regions?

To answer this question, we adopt the following protocol. Let us focus on events with amplification $A>A_1=\sqrt{3}/5\approx 1.34$ corresponding to sources falling within the Einstein disk of a lens, and which coincidentally produce a $3\sigma$ effect on type-Ia SNe. In a region with optical depth~$\tau$, this has a probability $P(A_1;\tau)=1-\ex{-\tau}$ in the strongest-lens approach~\eqref{eq:CDF_A_low_optical_depth}. So for the entire sky, the probability of such a high-amplification event would be
\begin{equation}
P(A_1; z\e{s},\alpha)
= 
\int_0^\infty \dd\tau \; p(\tau;z\e{s},\alpha) \pa{1-\ex{-\tau}} ,
\end{equation}
for a fraction~$\alpha$ of compact objects and a source at $z\e{s}$. Now suppose that we mask all the regions of the sky with an optical depth larger than $\tau\e{m}$; the probability would become
\begin{equation}
P_{\tau\e{m}}(A_1; z\e{s},\alpha)
=
\frac
{\int_0^{\tau\e{m}} \dd\tau \; p(\tau;z\e{s},\alpha) \pa{1-\ex{-\tau}}}
{\int_0^{\tau\e{m}} \dd\tau \; p(\tau;z\e{s},\alpha)} \ .
\end{equation}
The ratio of those probabilities,
\begin{equation}
R(\tau\e{m}, z\e{s}, \alpha)
\define
\frac{P_{\tau\e{m}}(A_1; z\e{s},\alpha)}
{P(A_1; z\e{s},\alpha)} \ ,
\end{equation}
then defines the fraction of high-amplification events that survive the masking operation; in other words, $R(\tau\e{m})$ is the proportion of high-amplification events happening in regions whose optical depth is lower than $\tau\e{m}$.

\begin{figure}
\centering
\includegraphics[width=0.49\columnwidth]{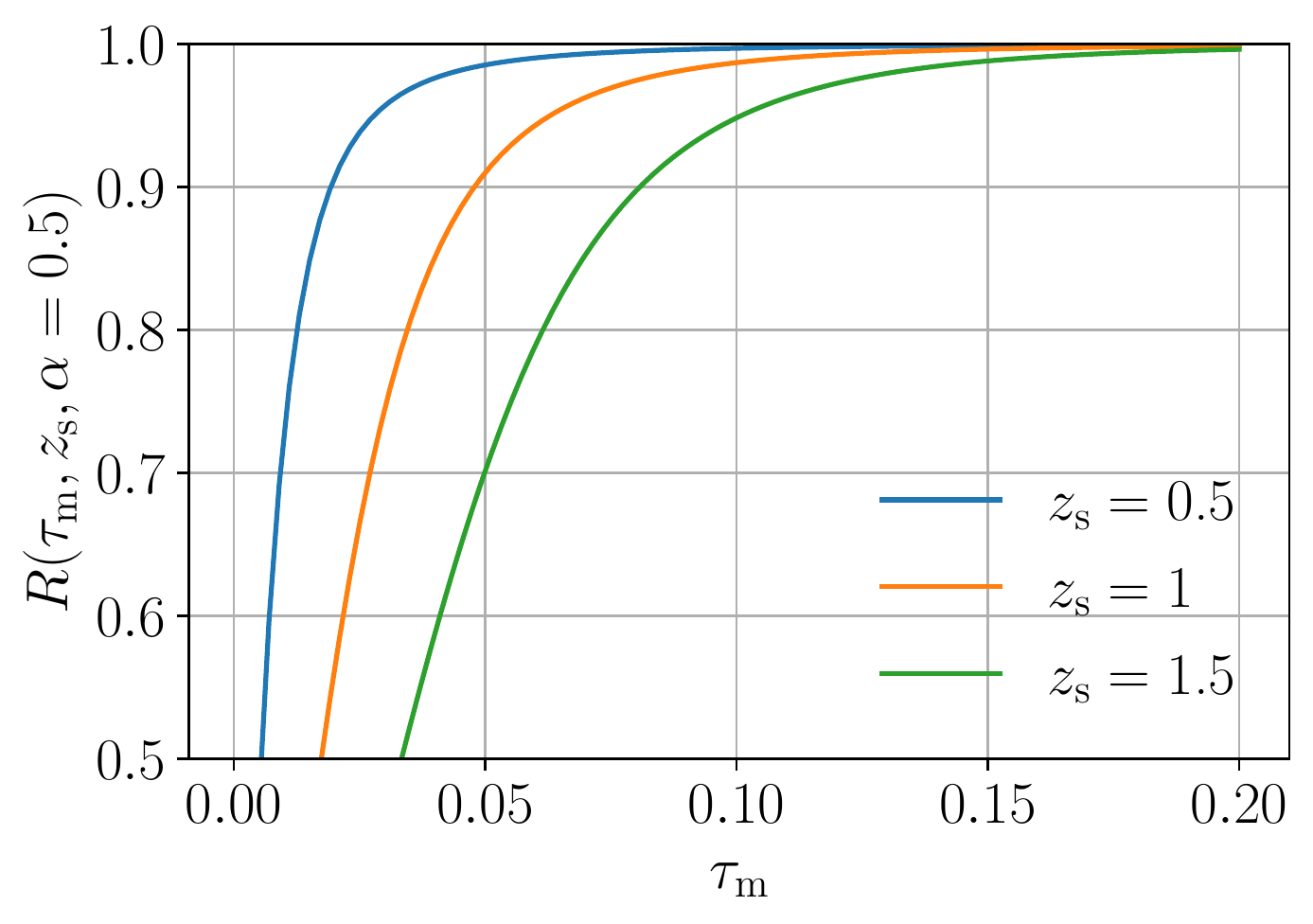}
\hfill
\includegraphics[width=0.49\columnwidth]{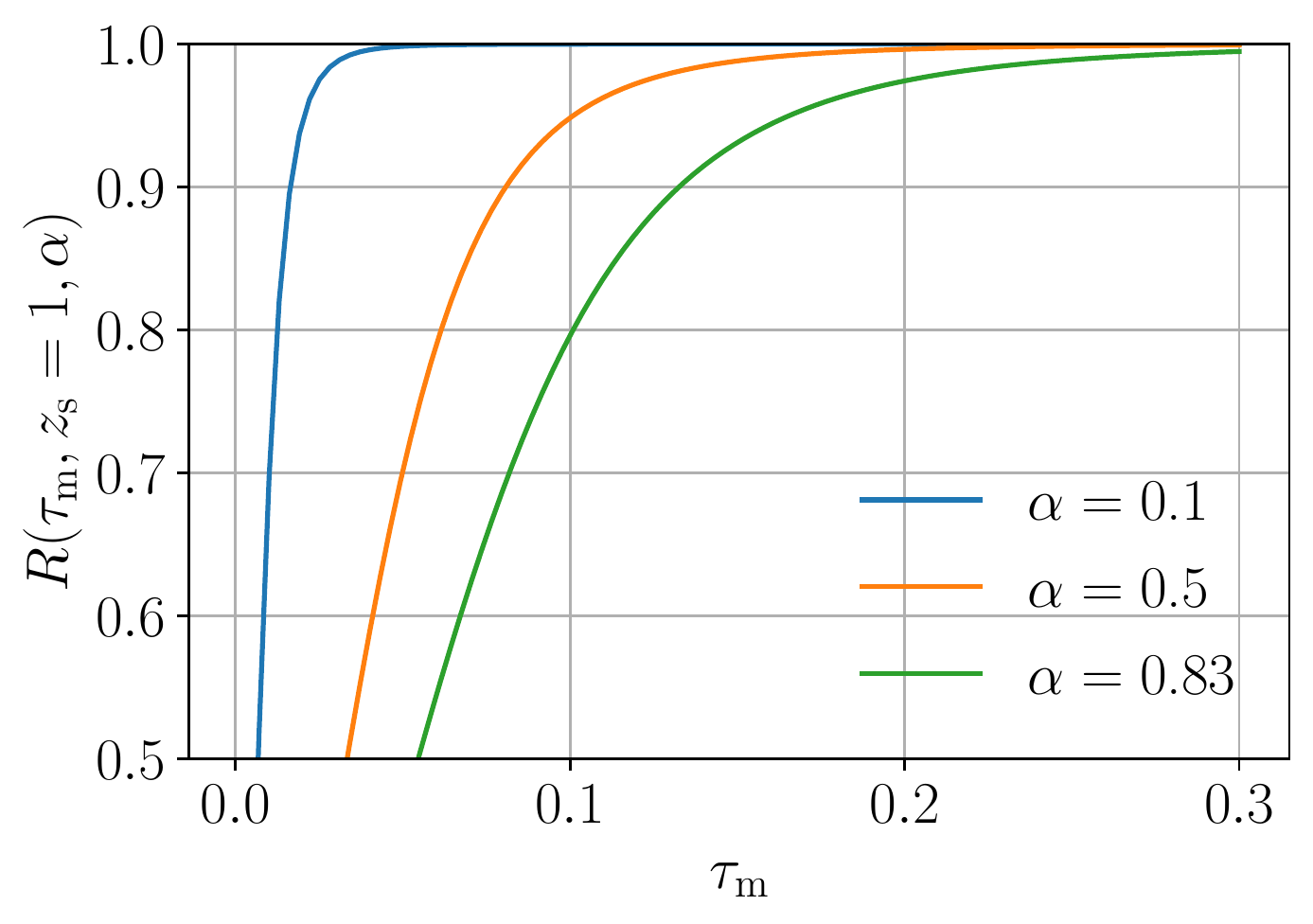}
\caption{Proportion~$R(\tau\e{m}, z\e{s}, \alpha)$ of the microlensing events with amplification $A>A_1\approx 1.34$ occurring within regions of the sky with optical depth lower than $\tau\e{m}$, for a source at $z\e{s}$ in a Universe with a fraction $\alpha$ of compact objects. \emph{Left}: effect of the redshift of the source for $\alpha=0.5$. \emph{Right}: effect of the fraction of compact objects for $z\e{s}=1$.}
\label{fig:masking}
\end{figure}

The evolution of the ratio~$R(\tau\e{m}; z\e{s}, \alpha)$ as a function of $\tau\e{m}$ is depicted in \cref{fig:masking} for various values of $z\e{s}, \alpha$. For sources at high redshift, and for a non-negligible fraction of compact objects, we see that in order to properly account for, say, \SI{99}{\percent} of the high-amplification events, we must allow the optical depth to reach values larger than $0.1$. Hence, as far as high amplifications ($A>1.34$) are concerned, the relevant regions of the sky do not necessarily have very low optical depths.

The general conclusion of the analysis conducted in this section is that we must a priori go beyond the simple model given by \cref{eq:CDF_A_low_optical_depth,eq:PDF_A_low_optical_depth} in order to accurately model the statistics of extragalactic microlensing. This will be the purpose of the next two sections, where we propose a complete set of corrections to the strongest-lens approach.

\section{Point lens with environment and line-of-sight perturbations}
\label{sec:point_lens_with_perturbations}

The discussion of \cref{sec:optical_depth} suggests that the simplest modelling of extragalactic microlensing statistics -- the strongest-lens approach sketched in \cref{subsec:strongest_lens_approach} -- may not be sufficient, because the relevant optical depths are low but not extremely low. In this context, we shall thus add perturbative corrections to this simple approach. We assume that when a source's light is significantly amplified, lensing is still mostly due to a single lens, which we may call the \emph{dominant lens}. However, we now allow for corrections due to the rest of the Universe -- large-scale matter inhomogeneities, their substructure and the other compact lenses altogether -- which we shall treat as tidal perturbations to the dominant lens.

In this section, we consider the problem of a single point lens that is perturbed by the presence of matter lumps in its environment and along the line of sight. We demonstrate that this problem can be suitably reformulated as a point lens with an external shear, and we derive the expression of the angular differential cross section~$\Omega(A)$ of the amplification.

\subsection{Description of the set-up}
\label{subsec:description_set-up}

\begin{figure}[t]
    \centering
\begingroup%
  \makeatletter%
  \providecommand\color[2][]{%
    \errmessage{(Inkscape) Color is used for the text in Inkscape, but the package 'color.sty' is not loaded}%
    \renewcommand\color[2][]{}%
  }%
  \providecommand\transparent[1]{%
    \errmessage{(Inkscape) Transparency is used (non-zero) for the text in Inkscape, but the package 'transparent.sty' is not loaded}%
    \renewcommand\transparent[1]{}%
  }%
  \providecommand\rotatebox[2]{#2}%
  \newcommand*\fsize{\dimexpr\f@size pt\relax}%
  \newcommand*\lineheight[1]{\fontsize{\fsize}{#1\fsize}\selectfont}%
  \ifx\svgwidth\undefined%
    \setlength{\unitlength}{436.44735416bp}%
    \ifx\svgscale\undefined%
      \relax%
    \else%
      \setlength{\unitlength}{\unitlength * \real{\svgscale}}%
    \fi%
  \else%
    \setlength{\unitlength}{\svgwidth}%
  \fi%
  \global\let\svgwidth\undefined%
  \global\let\svgscale\undefined%
  \makeatother%
  \begin{picture}(1,0.32320221)%
    \lineheight{1}%
    \setlength\tabcolsep{0pt}%
    \put(0,0){\includegraphics[width=\unitlength,page=1]{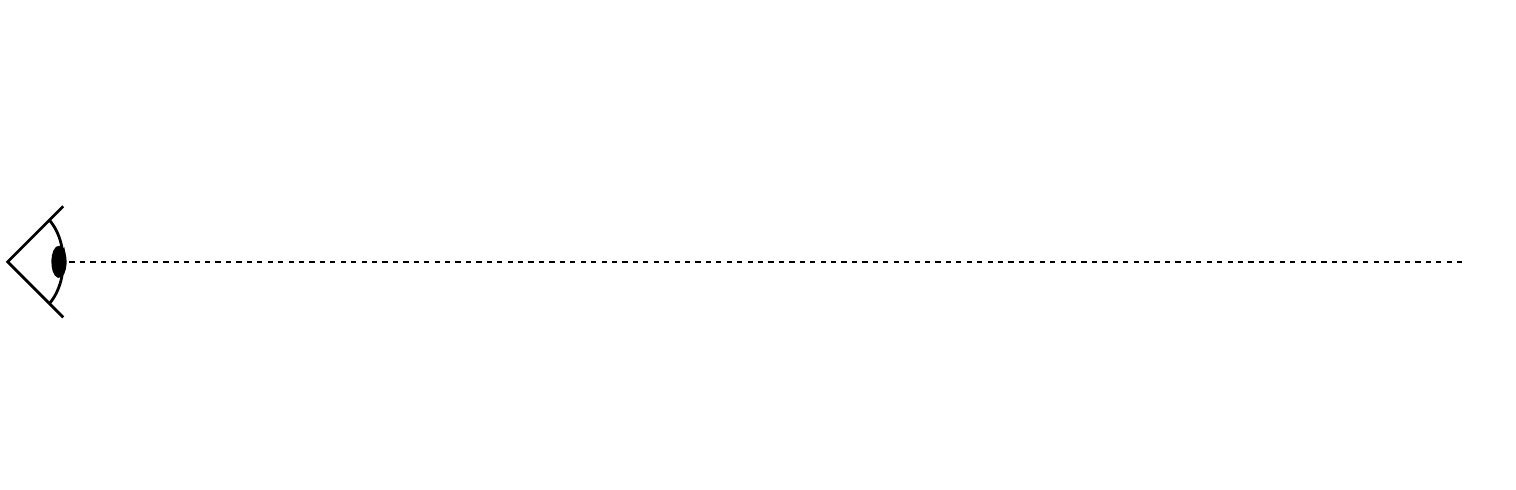}}%
    \put(-0.00142307,0.21237379){\makebox(0,0)[lt]{\lineheight{1.25}\smash{\begin{tabular}[t]{l}observer\end{tabular}}}}%
    \put(0.91042742,0.18116401){\color[rgb]{0,0,0}\makebox(0,0)[lt]{\lineheight{1.25}\smash{\begin{tabular}[t]{l}source\end{tabular}}}}%
    \put(0,0){\includegraphics[width=\unitlength,page=2]{set-up.pdf}}%
    \put(0.36260433,0.2330145){\color[rgb]{0.66666667,0,0}\makebox(0,0)[lt]{\lineheight{1.25}\smash{\begin{tabular}[t]{l}dominant lens\end{tabular}}}}%
    \put(0,0){\includegraphics[width=\unitlength,page=3]{set-up.pdf}}%
    \put(0.63730355,0.2991721){\color[rgb]{0,0,0.50196078}\makebox(0,0)[lt]{\lineheight{1.25}\smash{\begin{tabular}[t]{l}diffuse matter\end{tabular}}}}%
    \put(0,0){\includegraphics[width=\unitlength,page=4]{set-up.pdf}}%
    \put(0.62629984,0.03015255){\makebox(0,0)[lt]{\lineheight{1.25}\smash{\begin{tabular}[t]{l}compact objects\end{tabular}}}}%
    \put(0,0){\includegraphics[width=\unitlength,page=5]{set-up.pdf}}%
  \end{picture}%
\endgroup%

    \caption{An extragalactic point-like source of light is observed through the inhomogeneous Universe. A fraction $\alpha$ of the matter density is made of compact objects, $\rho\e{c}(t, \vect{x})=\alpha\rho(t, \vect{x})$, while the rest is treated as diffuse matter. Since the microlensing optical depth associated with compact objects is small, any significant amplification is dominated by a single lens: the dominant lens.}
    \label{fig:set-up}
\end{figure}

The concrete situation that we consider is depicted in \cref{fig:set-up}. An extragalactic point-like source (supernova or quasar) at $z\e{s}$ is observed through an inhomogeneous universe. On large scales, we assume that the inhomogeneity of the matter distribution is well described by the $\Lambda$CDM cosmological model. On small scales, we assume that a fraction~$\alpha$ of the total matter density is made of compact objects, which we model as point masses.\footnote{\changed{Although binary systems are generally common in the Universe, the separation between members of a binary is generally much smaller than their Einstein radius; hence binaries practically behave as point lenses.}} Just like in \cref{subsec:distribution_optical_depth}, we assume that the distribution of compact objects closely follows the total matter density field: in a region with density~$\rho(t, \vect{x})$, there is a Poisson-distributed population of compact objects with mean density $\rho\e{c}(t, \vect
x) \define \alpha\rho(t, \vect{x})$. We assume that $\alpha$ is constant in space and time.

\paragraph{Dominant lens} We define the dominant lens as the one that would produce the strongest amplification~$A$ if it were alone in the Universe. Equivalently, it is the lens with the smallest reduced impact parameter~$\beta/\theta\e{E}$, where $\beta$ is the angle between the unlensed source position and the lens position, and $\theta\e{E}$ its Einstein radius. We shall denote with a ``d'' subscript the quantities associated with the dominant lens, e.g., its redshift~$z\e{d}$.

\paragraph{Tidal perturbations} All the other inhomogeneities of the Universe, which includes both astronomical structures and the non-dominant compact objects, are treated in the \emph{tidal regime}. In the terminology of ref.~\cite{Fleury:2020cal}, this means that apart from the immediate vicinity of the dominant lens, light propagates through a smooth space-time geometry. This is equivalent to stating that the angle between multiple images produced by the dominant lens, which is on the order of its Einstein radius~$\theta\e{E}$, is much smaller than the typical scale over which the gravitational field produced by the other inhomogeneities changes appreciably. This notably requires all the non-dominant compact objects to lie far from the line of sight. In practice, the tidal approximation means that non-dominant inhomogeneities only produce weak-lensing convergence and shear which perturb the behaviour of the dominant lens.

\subsection{Lens equation and equivalent lens}

We now discuss the lens equation associated with the set-up described in \cref{subsec:description_set-up}. We then show that, with a suitable change of variables, it may be turned into the lens equation of a point lens with external shear.

\subsubsection{Lens equation with tidal perturbations}

\begin{figure}
    \centering
    \begin{minipage}{0.65\columnwidth}
    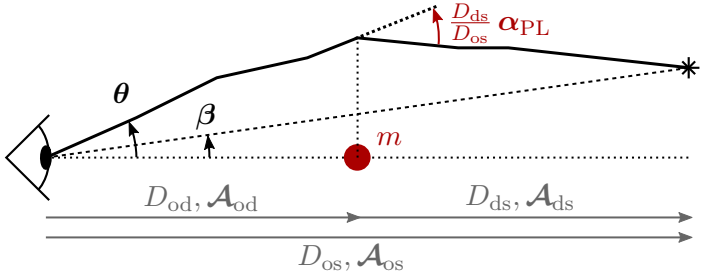
    \end{minipage}
    \hfill
    \begin{minipage}{0.34\columnwidth}
    \caption{Schematic representation of the quantities involved in the lens equation~\eqref{eq:lens_equation}. The total displacement $\vect{\theta}-\vect{\beta}$ is caused by the dominant point-like lens and the tidal distortions~$\mat{\amp}\e{od}, \mat{\amp}\e{ds}, \mat{\amp}\e{os}$.}
    \label{fig:lens_equation}
    \end{minipage}
\end{figure}

The relevant quantities defined below are depicted in \cref{fig:lens_equation}. The line of sight is conventionally set as the direction in which the main lens is observed. With respect to that origin, we call $\vect{\beta}$ the unlensed position of the source. Throughout this article, ``unlensed'' will refer to the case where light would propagate in the reference FLRW model. We denote with $\vect{\theta}$ the observed position of an image of the source.

The lens equation is the relation between $\vect{\theta}$ and $\vect{\beta}$. For a dominant point lens with tidal perturbation along the line of sight, it takes the form~\cite{1987ApJ...316...52K, 1994A&A...287..349S, 1996ApJ...468...17B, Schneider:1997bq, McCully:2013fga, Birrer:2016xku, Fleury:2020cal, Fleury:2021tke}
\begin{empheq}[box=\fbox]{equation}
\label{eq:lens_equation}
\vect{\beta}
= \mat{\amp}\e{os} \, \vect{\theta}
    - \mat{\amp}\e{ds} \,
        \vect{\alpha}\e{PL}(\mat{\amp}\e{od}\vect{\theta}) \ ,
\end{empheq}
where we have introduced some notation. The heart of \cref{eq:lens_equation} is the displacement angle~$\vect{\alpha}\e{PL}(\vect{\theta})$ of the dominant (point-like) lens only. In the absence of other inhomogeneities, we would simply have $\vect{\beta}=\vect{\theta}-\vect{\alpha}\e{PL}(\vect{\theta})$. Its explicit expression is
\begin{equation}
\vect{\alpha}\e{PL}(\vect{\theta}) = \frac{\theta\e{E}^2 \vect{\theta}}{|\vect{\theta}|^2} \define \frac{\theta\e{E}^2}{\vect{\theta}} \ ,
\end{equation}
where $\theta\e{E}$ is the unperturbed angular Einstein radius,
\begin{equation}
\theta\e{E}^2
\define
\frac{4 G m D\e{ds}}{D\e{od} D\e{os}}
= \frac{4 G m (\chi\e{s}-\chi\e{d})}{a(\chi\e{d})\chi\e{d}\chi\e{s}} \ ,
\end{equation}
and $m$ the mass of the dominant lens.

The three quantities~$\mat{\amp}\e{od}, \mat{\amp}\e{os}, \mat{\amp}\e{ds}$ are $2\times 2$ distortion matrices which encode the tidal perturbations along the line of sight. They are defined as follows: \emph{in the absence of the main lens}, for an observer at (a) and source at (b), the unlensed position~$\vect{\beta}\e{ab}$ and lensed position~$\vect{\theta}\e{ab}$ of the source are related by $\vect{\beta}\e{ab}=\mat{\amp}\e{ab}\vect{\theta}\e{ab}$. Thus, in the absence of the dominant lens [$\vect{\alpha}\e{PL}(\vect{\theta})=\vect{0}$] the lens equation would reduce to $\vect{\beta}=\mat{\amp}\e{os}\vect{\theta}$, which corresponds to standard weak lensing~\cite{Bartelmann:1999yn}. The distortion matrices may be decomposed as
\begin{equation}
\mat{\amp}\e{ab}
=
\mat{1} -
\begin{bmatrix}
\kappa\e{ab}+\Re(\gamma\e{ab}) & \Im(\gamma\e{ab})-\omega\e{ab} \\
\Im(\gamma\e{ab})+\omega\e{ab} & \kappa\e{ab}-\Re(\gamma\e{ab})
\end{bmatrix} ,
\qquad
\mathrm{a}, \mathrm{b} \in \{\mathrm{o}, \mathrm{d}, \mathrm{s}\} \ .
\end{equation}
In this decomposition, $\kappa\e{ab}\in\mathbb{R}$ represents the convergence that is be produced by the diffuse matter from (a) to (b); the symmetric trace-free part, encoded in $\gamma\e{ab}\in\mathbb{C}$, represents the shear produced in the same interval; the anti-symmetric part~$\omega\e{ab}\in\mathbb{R}$ represents the solid rotation of images from (a) to (b).

In the following, we shall work \emph{at first order in the shear}, $\gamma\e{ab}\ll 1$. In that regime, it can be shown that the rotation is a second-order quantity~$\omega\sim |\gamma|^2$ (see ref.~\cite{Fleury:2015hgz}, sec.~2.3.2); we shall thus neglect~$\omega\e{ab}$. However, as will be clearer in the very next paragraph, the convergence may reach values exceeding \SI{10}{\percent}, hence we shall work non-perturbatively in $\kappa\e{ab}$.

\subsubsection{Physical origin of the convergence and shear}
\label{subsubsec:physical_origin_convergence_shear}

Let us now elaborate on the convergences~$\kappa\e{ab}$ and shears~$\gamma\e{ab}$ appearing in the distortion matrices that enter in the lens equation~\eqref{eq:lens_equation}.

\paragraph{Convergence is due to the diffuse matter} that is intercepted by the line of sight. More precisely, $\kappa\e{ab}$ represents the excess (or deficit) of focusing from diffuse matter, with respect to the homogeneous FLRW reference, for a source located at (b) and observed from (a). Its explicit expression is
\begin{align}
\kappa\e{ab}
&\define
4\pi G
\int_{\chi\e{a}}^{\chi\e{b}}
\dd\chi \; \frac{(\chi-\chi\e{a})(\chi\e{b}-\chi)}{\chi\e{b} - \chi\e{a}} \, a^2(\chi)
\pac{(1-\alpha)\rho - \bar{\rho}}
\\
&= (1-\alpha)\bar{\kappa}\e{ab} - \alpha\Delta\e{ab} \ .
\label{eq:kappa_ab}
\end{align}
where $\Delta\e{ab}$ and $\bar{\kappa}\e{ab}$ are generalisations of the $\Delta\e{os}$ and $\bar{\kappa}\e{os}$ defined in \cref{eq:Delta_os,eq:kappa_bar_os},
\begin{align}
\label{eq:Delta_ab}
\Delta\e{ab}
&\define
4\pi G \bar{\rho}_0
\int_{\chi\e{a}}^{\chi\e{b}}
\dd\chi \; \frac{(\chi-\chi\e{a})(\chi\e{b}-\chi)}{\chi\e{b} - \chi\e{a}} \,
\frac{1}{a(\chi)} \ ,
\\
\label{eq:kappa_bar_ab}
\bar{\kappa}\e{ab}
&\define
4\pi G \bar{\rho}_0
\int_{\chi\e{a}}^{\chi\e{b}}
\dd\chi \; \frac{(\chi-\chi\e{a})(\chi\e{b}-\chi)}{\chi\e{b} - \chi\e{a}} \, \frac{\delta(\chi)}{a(\chi)} \ .
\end{align}
The first term in \cref{eq:kappa_ab} is quite intuitive; since the fraction of diffuse matter is $1-\alpha$, any excess~$\bar{\kappa}\e{ab}$ in total projected density translates into $(1-\alpha)\bar{\kappa}\e{ab}$ from its diffuse component. The second term is more subtle; it encodes the deficit of diffuse matter, relative to FLRW, that occurs as one turns a fraction $\alpha$ of it into compact matter. In the extreme case $\alpha=1$, there is no diffuse matter at all, which implies a significant focusing deficit, $\kappa\e{ab}=-\Delta\e{ab}$, with respect to FLRW -- this is Zel'dovich's empty-beam case~\cite{1964SvA.....8...13Z}. The presence of $\Delta\e{ab}$ in $\kappa\e{ab}$ is the reason why the convergence can reach relatively large values and should not be treated at linear order. Note finally that \cref{eq:optical_depth_alpha,eq:kappa_ab} imply the following relation between convergences and optical depth: $\kappa\e{os} = \bar{\kappa}\e{os}-\tau$.

\paragraph{Shear is due to both diffuse and compact matter} unlike convergence. This is because shear is associated with long-range tidal forces generated by any matter lump. For a source located at (b) and observed from (a), we may decompose the total shear as
\begin{equation}
\label{eq:shear_macro_micro}
\gamma\e{ab} = \bar{\gamma}\e{ab} + s\e{ab} \ .
\end{equation}
In \cref{eq:shear_macro_micro}, $\bar{\gamma}\e{ab}$ is the \emph{macroshear} associated with the smooth density contrast, that is, the shear that would be produced on a beam of light in the absence of any compact object near the line of sight. Its formal expression is~\cite{Fleury:2018cro}
\begin{equation}
\label{eq:macroshear}
\bar{\gamma}\e{ab}
= -4\pi G\bar{\rho}_0
	\int_{\chi\e{a}}^{\chi\e{b}} \dd\chi \; \frac{(\chi-\chi\e{a})(\chi\e{b}-\chi)}{\chi\e{b} - \chi\e{a}}
	\int_{\mathbb{R}^2} \frac{\dd^2\vect{r}}{\pi r^2} \;
	    \frac{\delta(\chi, \vect{r})}{a(\chi)} \, \ex{2\ii\ph} \ ,
\end{equation}
where $\vect{r}=r(\cos\ph, \sin\ph)$ denotes the physical transverse position of a point, orthogonally to the line of sight, $\dd^2\vect{r}=r \dd r \dd\ph$. Specifically, $r$ is the distance between a point and the line of sight and $\ph$ is its polar angle about that axis.

The second term of \cref{eq:shear_macro_micro}, $s\e{ab}$, is the \emph{microshear} produced by compact objects in the vicinity of the line of sight, except the dominant lens. If the region between (a) and (b) contains $N$ point-like lenses labelled with $\ell$, then the microshear reads~\cite{Fleury:2021tke}
\begin{equation}
\label{eq:microshear}
s\e{ab}
=
-4\pi G \sum_{\ell=1}^{N} \frac{(\chi_\ell - \chi\e{a})(\chi\e{b}-\chi_\ell)}{\chi\e{b}-\chi\e{a}} \,
                        \frac{a(\chi_\ell) m_\ell}{\pi r_\ell^2} \, \ex{2\ii\ph_\ell} \ ,
\end{equation}
where $m_\ell$ is the mass of lens~$\ell$, $\chi_\ell$ its comoving distance from the observer, $r_\ell$ its physical distance from the optical axis and $\ph_\ell$ its polar angle about it. Note that \cref{eq:macroshear} is nothing but the continuous limit of \cref{eq:microshear}.

The careful reader may have noticed that the macroshear~$\bar{\gamma}\e{ab}$ does not come with any prefactor~$(1-\alpha)$. Such a prefactor could be expected indeed, to avoid double-counting the shear of compact matter, which should be encoded in $s\e{ab}$ already. The reason is that, in the following, we shall compute $s\e{ab}$ as if the compact objects were randomly distributed transversely to the line of sight. Hence, $s\e{ab}$ will not account for the large-scale clustering of those objects. Because they follow the total matter density contrast~$\delta(t, \vect{x})$ on large scales, their contribution to cosmic shear is essentially the same as if they were replaced by diffuse matter. Therefore, $\bar{\gamma}\e{ab}$ is unchanged under changes of the compact matter fraction~$\alpha$. Increasing $\alpha$ only produces more shear via $s\e{ab}$.

\subsubsection{Equivalent lens}
\label{subsubsec:equivalent_lens}

The lens equation~\eqref{eq:lens_equation} contains a priori nine real parameters besides the dominant lens's Einstein radius: the three convergences and the six shear components. But only a few specific combinations of those parameters turn out to be relevant to the problem of amplification probabilities. Multiplying \cref{eq:lens_equation} to the left with $(1-\kappa\e{od})(1-\kappa\e{os})^{-1}(1-\kappa\e{ds})\mat{\amp}\e{ds}^{-1}$ and working at first order in the shears, we find the \emph{equivalent lens equation}
\begin{empheq}[box=\fbox]{equation}
\label{eq:equivalent_lens}
\tilde{\vect{\beta}}
=
(1-\vect{\Gamma}) \, \tilde{\vect{\theta}}
- \frac{\tilde{\theta}\e{E}^2}{\tilde{\vect{\theta}}} \ ,
\end{empheq}
whose new variables are
\begin{align}
\label{eq:beta_tilde}
\tilde{\vect{\beta}}
&\equiv
\frac{(1-\kappa\e{od})(1-\kappa\e{ds})}{1-\kappa\e{os}} \,
\mat{\amp}\e{ds}^{-1} \vect{\beta} \ ,
\\
\label{eq:theta_tilde}
\tilde{\vect{\theta}}
&\equiv \mat{\amp}\e{od} \vect{\theta} \ ,
\end{align}
and the new parameters read
\begin{align}
\label{eq:theta_E_tilde}
\tilde{\theta}\e{E}
&\equiv
\sqrt{\frac{(1-\kappa\e{od})(1-\kappa\e{ds})}{1-\kappa\e{os}}} \,
\theta\e{E} \ ,
\\
\vect{\Gamma}
&\equiv \begin{bmatrix}
\Re(\gamma) & \Im(\gamma) \\
\Im(\gamma) & -\Re(\gamma)
\end{bmatrix} ,
\qquad
\gamma
\equiv
\frac{\gamma\e{os}}{1-\kappa\e{os}}
- \frac{\gamma\e{od}}{1-\kappa\e{od}}
- \frac{\gamma\e{ds}}{1-\kappa\e{ds}} \ .
\end{align}

In other words, under the linear change of variables~$(\vect{\beta}, \vect{\theta})\mapsto (\tilde{\vect{\beta}}, \tilde{\vect{\theta}})$, our initial problem of a point lens with generic tidal perturbations has turned into the much simpler \cref{eq:equivalent_lens}, which describes a point mass with an external shear\footnote{This particular shear combination~$\gamma$ is different from the line-of-sight shear combination~$\gamma\e{LOS}=\gamma\e{os}+\gamma\e{od}-\gamma\e{ds}$ that was isolated in ref.~\cite{Fleury:2021tke}, even in the absence of the convergences.}~$\gamma$ in the same plane. This equivalent problem has been well studied since the 1980s~\cite{1984JApA....5..235N, 1992ApJ...389...63M, Kofman:1996es}.

\subsection{Amplification cross section}
\label{subsec:amplification_cross_section}

Let us use the equivalent lens model~\eqref{eq:equivalent_lens} to derive the amplification cross section. We define the \emph{differential amplification cross section}~$\Omega(A)$ so that $\Omega(A)\dd A$ is the angular area (solid angle) of the region of the sky where the amplification is between $A$ and $A+\dd A$.

\subsubsection{For the equivalent lens}

\begin{figure}
    \centering
    \begin{minipage}{0.54\columnwidth}
    \includegraphics[width=\columnwidth]{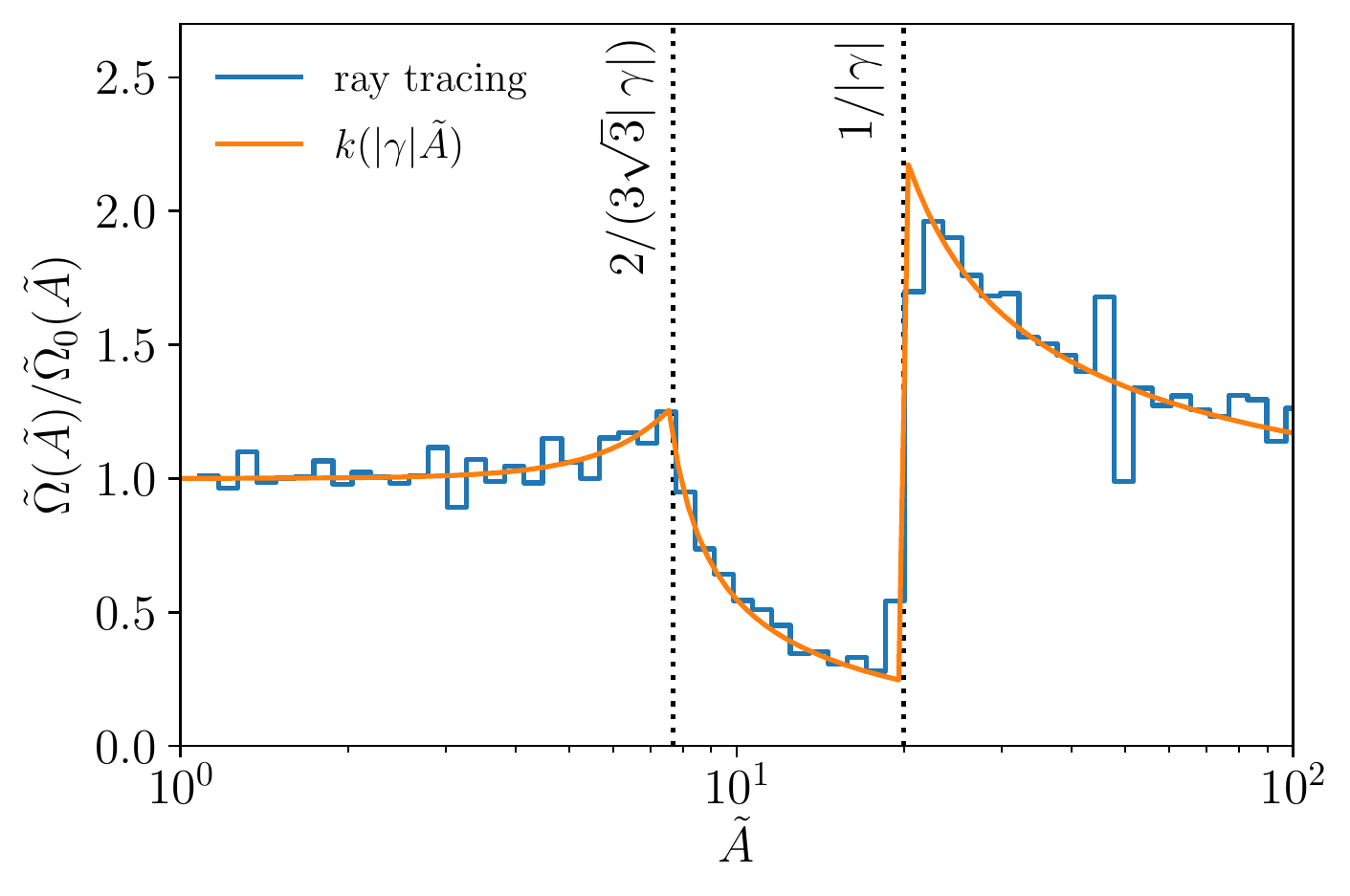}
    \end{minipage}
    \hfill
    \begin{minipage}{0.45\columnwidth}
    \vspace*{-0.8cm}
    \begin{flushright}
    \includegraphics[width=0.93\columnwidth]{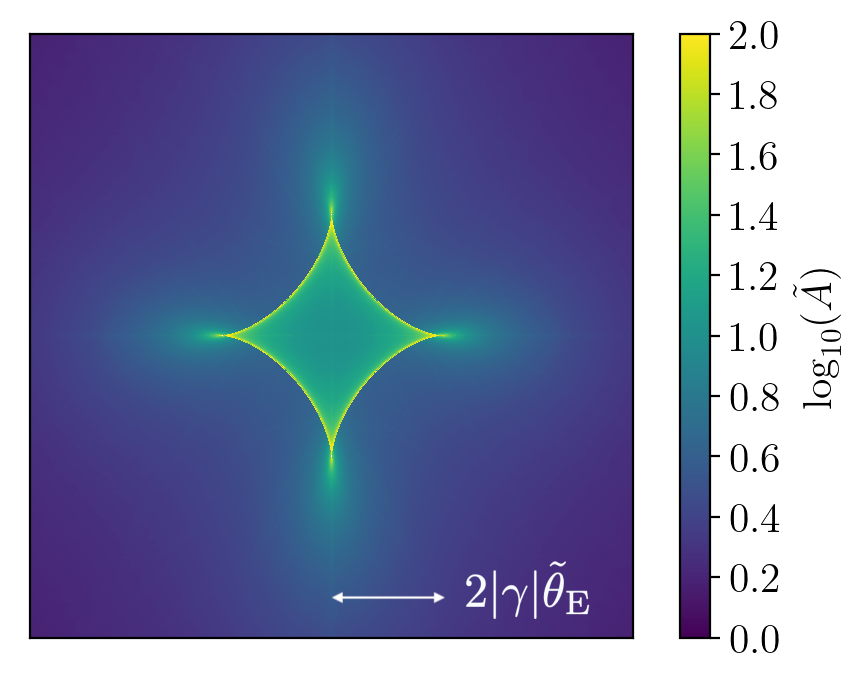}
    \end{flushright}
    \end{minipage}
    \caption{Properties of the equivalent point lens with external shear. \textit{Left panel}: Ratio of the amplification cross section~$\tilde{\Omega}(\tilde{A})$ in the presence of shear (here $\gamma=0.05$) with the no-shear case~$\tilde{\Omega}_0(\tilde{A})$~\eqref{eq:cross_section_no_shear}. The solid-line histogram shows results from ray tracing, while the dashed line shows the analytical fit~$k(|\gamma|\tilde{A})$ of \cref{eq:function_Kofman+1997} as proposed in ref.~\cite{Kofman:1996es}. \textit{Right panel}: Amplification map~$\tilde{A}(\tilde{\vect{\beta}})$. The astroid-shaped caustic has a size $2|\gamma|\tilde{\theta}\e{E}$; inside the astroid a source has four images, outside it has two images, and exactly on it it has three images among which one is infinitely amplified.}
    \label{fig:point_lens_shear}
\end{figure}

We first work in the twiddled world described by \cref{eq:equivalent_lens}. If the effective shear~$\gamma$ were zero, then the problem would reduce to a single point lens whose constant-amplification contours are circles, with radius~$\tilde{\beta}_0(\tilde{A})=\tilde{\theta}\e{E}\,[2\tilde{A}/(\tilde{A}^2-1)^{1/2}-2]^{1/2}$. The cross section would thus read, in this simple case,
\begin{equation}
\label{eq:cross_section_no_shear}
\tilde{\Omega}_0(\tilde{A})
= 2\pi \tilde{\beta}_0(\tilde{A}) \left| \ddf{\tilde{\beta}_0}{\tilde{A}} \right|
= \frac{2\pi\tilde{\theta}\e{E}^2}{(\tilde{A}^2-1)^{3/2}} \ .
\end{equation}

The problem is more involved in the presence of shear. In that case the source plane displays two distinct regions separated by an astroid-shaped caustic (see right panel of \cref{fig:point_lens_shear}). Outside the caustic a source has two images while inside it has four images. The shear~$\gamma$ fixes the size and orientation of the astroid. To the best of our knowledge there is no analytic expression for the amplification~$\tilde{A}(\tilde{\vect{\beta}})$ or its contours in that case. Nevertheless, Nityanda \& Ostriker~\cite{1984JApA....5..235N} noticed the remarkable fact that for low values of the shear~$|\gamma|$, corrections the the amplification cross section should only depend on the product $|\gamma|A$. More than a decade later, Kofman et al.~\cite{Kofman:1996es} further pushed this idea by writing
\begin{equation}
\label{eq:cross_section_Kofman}
\tilde{\Omega}(\tilde{A})
= \tilde{\Omega}_0(\tilde{A}) \, k(|\gamma|\tilde{A}) \ ,
\end{equation}
where $\tilde{\Omega}_0(\tilde{A})$ is the no-shear cross section of \cref{eq:cross_section_no_shear}, while $k$ is a function fitted from numerical simulations,\footnote{In ref.~\cite{Kofman:1996es}, that function was denoted with $\ph$. We changed the notation so as to avoid confusion with the polar angle of \cref{subsubsec:physical_origin_convergence_shear}, and adopted ``$k$'' instead, in honour of Lev Kofman.}
\begin{equation}
\label{eq:function_Kofman+1997}
k(x)
=
\begin{cases}
1 + 7.7 \, x^{3.5}
& \displaystyle x\leq \frac{2}{3\sqrt{3}} \ ,
\\[4mm]
\displaystyle
\frac{0.17}{(x-0.33)^{1/2}} + \frac{0.023}{x-0.33} 
& \displaystyle \frac{2}{3\sqrt{3}} \leq x \leq 1 \ ,
\\[4mm]
\displaystyle
1 + \frac{0.85}{x} + \frac{0.37}{x^5}
& \displaystyle x\geq 1 \ .
\end{cases}
\end{equation}
For the sake of completeness, we have reproduced in \cref{fig:point_lens_shear} the comparison between numerical ray tracing and Kofman et al.'s result~\eqref{eq:cross_section_Kofman} for $\gamma=0.05$. The simulation uses inverse ray tracing with a simple adaptive mesh refinement, see ref.~\cite{Fleury:2019xzr} for details. Agreement is excellent.

\subsubsection{Back to the original problem}
\label{subsubsec:back_original_problem}

We now translate the results of the twiddled world in terms of~$\Omega(A)$. The first step is to express $\tilde{A}$ in terms of $A$. At first order in the shear, we find
\begin{equation}
\label{eq:amplification_tilde}
\tilde{A}
\define
\frac{\dd^2\tilde{\vect{\theta}}}{\dd^2\tilde{\vect{\beta}}}
=
\frac{(1-\kappa\e{os})^2}{(1-\kappa\e{od})^2(1-\kappa\e{ds})^2} \,
\frac{\det\mat{\amp}\e{od}}{\det\mat{\amp}\e{ds}^{-1}} \,
\frac{\dd^2\vect{\theta}}{\dd^2\vect{\beta}}
= (1-\kappa\e{os})^2 A \ .
\end{equation}
The conversion of $\tilde{\Omega}$ into $\Omega$ must take two aspects into account. On the one hand, since those are differential cross sections, their relation involves the Jacobian~$|\dd\tilde{A}/\dd A|$, just like when one changes variables in a probability density function, for example. Second, since $\tilde{\Omega}$ is a cross section in the twiddled source plane, it is expressed in the twiddled units~$[\tilde{\beta}]^2$, which differ from the units~$[\beta]^2$ of the original source plane. Taking both aspects into account yields
\begin{equation}
\label{eq:relation_Omega_Omega_tilde}
\Omega(A) \, \dd A
=
\frac{\dd^2\vect{\beta}}{\dd^2\tilde{\vect{\beta}}} \times \tilde{\Omega}(\tilde{A}) \, \dd\tilde{A} \ .
\end{equation}

Substituting \cref{eq:beta_tilde,eq:amplification_tilde,eq:cross_section_Kofman} into \cref{eq:relation_Omega_Omega_tilde}, and still working at first order in the shear, we find the following elegant expression for the cross section,
\begin{empheq}[box=\fbox]{equation}
\label{eq:cross_section_final}
\Omega(A)
=
2\pi\vartheta\e{E}^2 \, k\pa{\frac{|\gamma| A}{A\e{min}}} \,
\frac{A\e{min}^2}{(A^2 - A\e{min}^2)^{3/2}} \ ,
\end{empheq}
where $A\e{min} \define (1-\kappa\e{os})^{-2}$ is the minimal amplification in this setup, i.e. the amplification that would be observed if the dominant lens were infinitely far from the line of sight, so that only the weak-lensing convergence is at play. It is implicit that $\Omega(A<A\e{min})=0$. Besides, we have introduced $\vartheta\e{E}$ such that
\begin{equation}
\vartheta\e{E}^2
\define
\frac{(1-\kappa\e{ds}) \, \theta\e{E}^2}{(1-\kappa\e{od})(1-\kappa\e{os})} 
=
\frac{4 G m (1-\kappa\e{ds})D\e{ds}}{(1-\kappa\e{od})D\e{od} (1-\kappa\e{os})D\e{os}} \ .
\end{equation}
Physically, $\vartheta\e{E}$ represents the the lensed Einstein radius, i.e. the size of the Einstein ring that would be observed if the source were perfectly aligned with the dominant lens in the presence of the external convergences; this can be checked by setting $\vect{\beta}=\vect{0}$ in \cref{eq:lens_equation}.

Summarising, the convergence due to diffuse matter has two distinct effect on $\Omega(A)$: (i) they rescale amplifications according to $A\to A/A\e{min} = (1-\kappa\e{os})^2 A$, thereby fixing the minimum amplification accessible to the system; (ii) they rescale the dominant lens's Einstein radius as $\theta\e{E}\to\vartheta\e{E}$, thereby changing the cross section directly. The shear, due to both diffuse and compact matter, affects $\Omega(A)$ via Kofman et al.'s function~$k(|\gamma|A/A\e{min})$ only.

\section{Amplification probabilities}
\label{sec:amplification_probabilities}

In the previous section, we have derived the amplification cross section~$\Omega(A)$ for a single point lens with perturbations~\eqref{eq:cross_section_final}. We shall now turn this result into a PDF for the amplification, $p(A)$, using the statistical properties of the dominant lens and its perturbations.

\subsection{Amplification probability for a single lens}
\label{sec:amplificationprobabilityonelens}

\begin{figure}[h]
    \centering
    \begin{minipage}{0.65\columnwidth}
\begingroup%
  \makeatletter%
  \providecommand\color[2][]{%
    \errmessage{(Inkscape) Color is used for the text in Inkscape, but the package 'color.sty' is not loaded}%
    \renewcommand\color[2][]{}%
  }%
  \providecommand\transparent[1]{%
    \errmessage{(Inkscape) Transparency is used (non-zero) for the text in Inkscape, but the package 'transparent.sty' is not loaded}%
    \renewcommand\transparent[1]{}%
  }%
  \providecommand\rotatebox[2]{#2}%
  \newcommand*\fsize{\dimexpr\f@size pt\relax}%
  \newcommand*\lineheight[1]{\fontsize{\fsize}{#1\fsize}\selectfont}%
  \ifx\svgwidth\undefined%
    \setlength{\unitlength}{295.08797956bp}%
    \ifx\svgscale\undefined%
      \relax%
    \else%
      \setlength{\unitlength}{\unitlength * \real{\svgscale}}%
    \fi%
  \else%
    \setlength{\unitlength}{\svgwidth}%
  \fi%
  \global\let\svgwidth\undefined%
  \global\let\svgscale\undefined%
  \makeatother%
  \begin{picture}(1,0.24758099)%
    \lineheight{1}%
    \setlength\tabcolsep{0pt}%
    \put(0,0){\includegraphics[width=\unitlength,page=1]{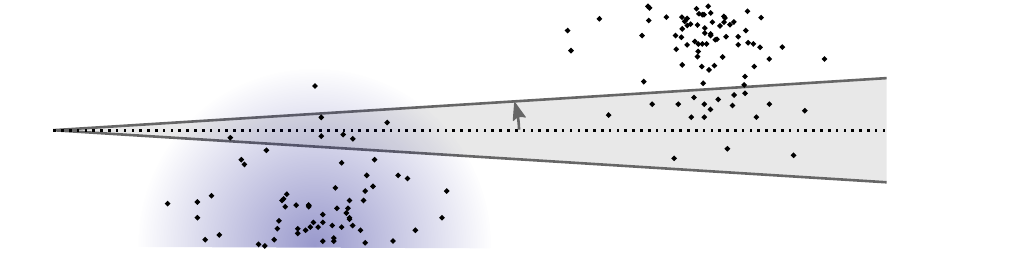}}%
    \put(0.48112446,0.16901564){\color[rgb]{0.4,0.4,0.4}\makebox(0,0)[lt]{\lineheight{1.25}\smash{\begin{tabular}[t]{l}$\Theta$\end{tabular}}}}%
    \put(0,0){\includegraphics[width=\unitlength,page=2]{mesoscopic_cone.pdf}}%
    \put(0.84512068,0.0064864){\color[rgb]{0.4,0.4,0.4}\makebox(0,0)[lt]{\lineheight{1.25}\smash{\begin{tabular}[t]{l}$z\e{s}$\end{tabular}}}}%
    \put(0,0){\includegraphics[width=\unitlength,page=3]{mesoscopic_cone.pdf}}%
  \end{picture}%
\endgroup%

    \end{minipage}
    \hfill
    \begin{minipage}{0.34\columnwidth}
    \caption{Mesoscopic cone with half angle~$\Theta$ at the observer, containing a large number of compact objects, but across which the macroscopic quantities $\tau, \bar{\kappa}, \bar{\gamma}$ can be considered constant.}
    \label{fig:mesosopic_cone}
    \end{minipage}
\end{figure}

In realistic scenarios where the mass of the compact objects is small (e.g. comparable to a solar mass), then the typical angle separating two such objects is much smaller than the angular scales over which $\bar{\kappa}, \bar{\gamma}, \tau$ are changing appreciably. Thus, we may consider a ``mesoscopic'' cone with half angle $\Theta$ at the observer which contains a large number of compact objects, but across which the macroscopic quantities~$\bar{\kappa}, \bar{\gamma}, \tau$ are constant (see \cref{fig:mesosopic_cone}). Since those empirically show significant changes on the arcmin scale, we have $\Theta\ll 1$.

The first step of our calculation consists in expressing the PDF~$p_1(A)$ of the amplification due to one dominant lens in the mesoscopic cone. If all the parameters entering the amplification cross section~\eqref{eq:cross_section_final} were fixed, then we would have by definition $p_1(A)=\Omega(A)/(\pi\Theta^2)$. But since the properties of the main lens -- namely its mass~$m$ and comoving distance~$\chi$ from the observer -- and the microshear~$s=\gamma-\bar{\gamma}$ vary a lot across the mesoscopic cone, we must marginalise over their statistical distribution,
\begin{equation}
\label{eq:p_1_definition}
p_1(A)
=
\frac{1}{\pi\Theta^2}
\int \dd m \, \dd\chi \, \dd^2 s \; p(m, \chi, s) \, \Omega(A; m, \chi, |\bar{\gamma} + s|) \ ,
\end{equation}
where $\Omega$ depends on $m$ via the Einstein radius of the dominant lens, $\theta\e{E}^2\propto m$, and on $\chi$ via $\theta\e{E}$ and the (od), (ds) convergences and shears. We did not explicitly include the fixed macroscopic parameters~$\tau, \bar{\kappa}\e{os}, \bar{\gamma}\e{os}$ to alleviate notation.

\subsubsection{Approximations}
\label{subsubsec:approximations}

In order to model the joint distribution $p(m, \chi, s)$, we make the following assumptions:
\begin{enumerate}
    \item The mass~$m$ of the dominant lens is uncorrelated with the other parameters. Since $\Omega\propto m$, this implies that we may simply replace $m$ by its average value $\ev{m}$ in the remainder of this calculation.
    \item Compact objects are randomly distributed in space and their {\em comoving} number density~$n\e{c}$ is constant within the mesoscopic cone. This implies, in particular, that $p(\chi)=3\chi^2/\chi\e{s}^3$.
\end{enumerate}

Besides, in order to simplify the evaluation of the various convergences and shears involved in $\Omega(A)$, we shall adopt the following \emph{mean-field approximation}:
\begin{equation}
\label{eq:mean_field_approximation}
\bar{\kappa}\e{ab}
\approx \pa{\frac{\chi\e{b}-\chi\e{a}}{\chi\e{s}}}^2 \bar{\kappa}\e{os} \ ,
\qquad
\bar{\gamma}\e{ab}
\approx \pa{\frac{\chi\e{b}-\chi\e{a}}{\chi\e{s}}}^2 \bar{\gamma}\e{os} \ .
\end{equation}
The intuition behind this approximation appears by examining the integrals~\eqref{eq:kappa_bar_ab} and \eqref{eq:macroshear} defining $\bar{\kappa}\e{ab}$ and $\bar{\gamma}\e{ab}$. Consider all the possible lines of sight with the same fixed $\bar{\kappa}\e{os}$, $\bar{\gamma}\e{os}$. They are in principle quite diverse, because the matter density contrast~$\delta$ may display significant variations along them, and hence they may have a variety of $\bar{\kappa}\e{od}, \bar{\kappa}\e{ds}, \bar{\gamma}\e{od}, \bar{\gamma}\e{ds}$. However, on average all the elements $\dd\chi$ along the line of sight should conspire so as to produce the required $\bar{\kappa}\e{os}, \bar{\gamma}\e{os}$. If we neglect the effect of dark energy on structure formation, we know that $\delta \propto a$, which motivates us to consider that the mean-field contribution of $\delta(\chi)/a(\chi)$ to $\bar{\kappa}\e{ab}, \bar{\gamma}\e{ab}$ is independent of $\chi$. As the latter is taken off the integrals over $\chi$, \cref{eq:kappa_bar_ab,eq:macroshear} imply
\begin{equation}
\bar{\kappa}\e{ab}, \bar{\gamma}\e{ab}
\propto
\int_{\chi\e{a}}^{\chi\e{b}}\dd\chi \;
\frac{(\chi - \chi\e{a})(\chi\e{b} - \chi)}{\chi\e{b} - \chi\e{a}}
\propto (\chi\e{b} - \chi\e{a})^2 \ ,
\end{equation}
whence \cref{eq:mean_field_approximation}. We shall also apply a similar rule to the full convergence $\kappa\e{ab}$.

The difficult step then consists in determining the distribution for the microshear, and evaluating its consequences on $p_1(A)$.

\subsubsection{Distribution of the microshear}

The statistics of the shear caused by a random distribution of point masses has been, in fact, a well-know problem for a long time. It was first considered for masses placed in the same plane by Nityananda \& Ostriker~\cite{1984JApA....5..235N}, using one of the methods exposed in the famous review~\cite{1943RvMP...15....1C} by Chandrasekhar in 1943 on statistical problems in astrophysics. However, the very last step of the calculation was only performed three years later by Schneider~\cite{1987A&A...179...71S}. The result was then generalised to any lens profile by Lee \& Spergel~\cite{1997ApJ...489..508K} and finally to multiple lens planes in Lee et al.~\cite{1997ApJ...489..522L}.

\begin{figure}[t]
    \centering
    \includegraphics[width=0.49\columnwidth]{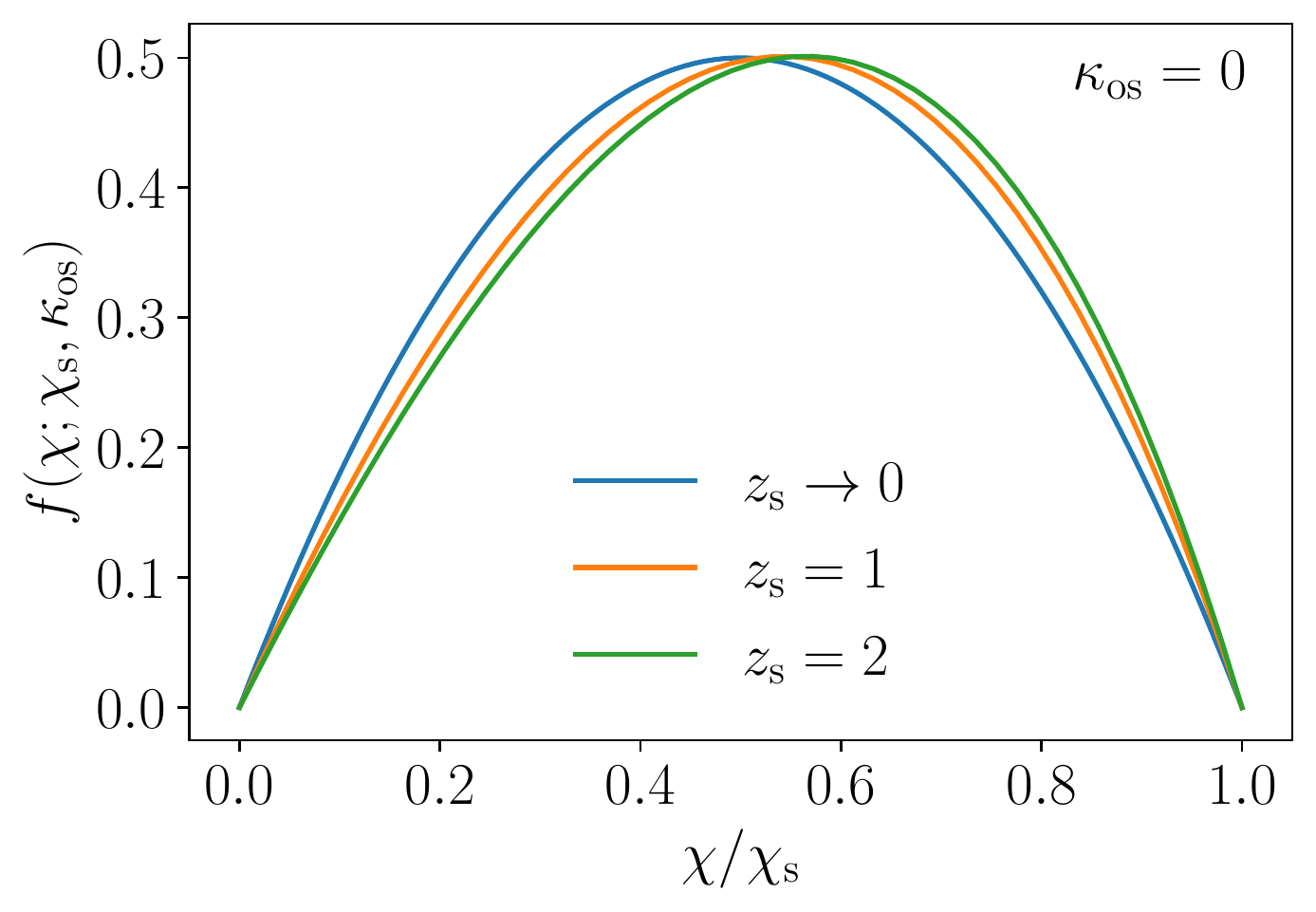}
    \hfill
    \includegraphics[width=0.49\columnwidth]{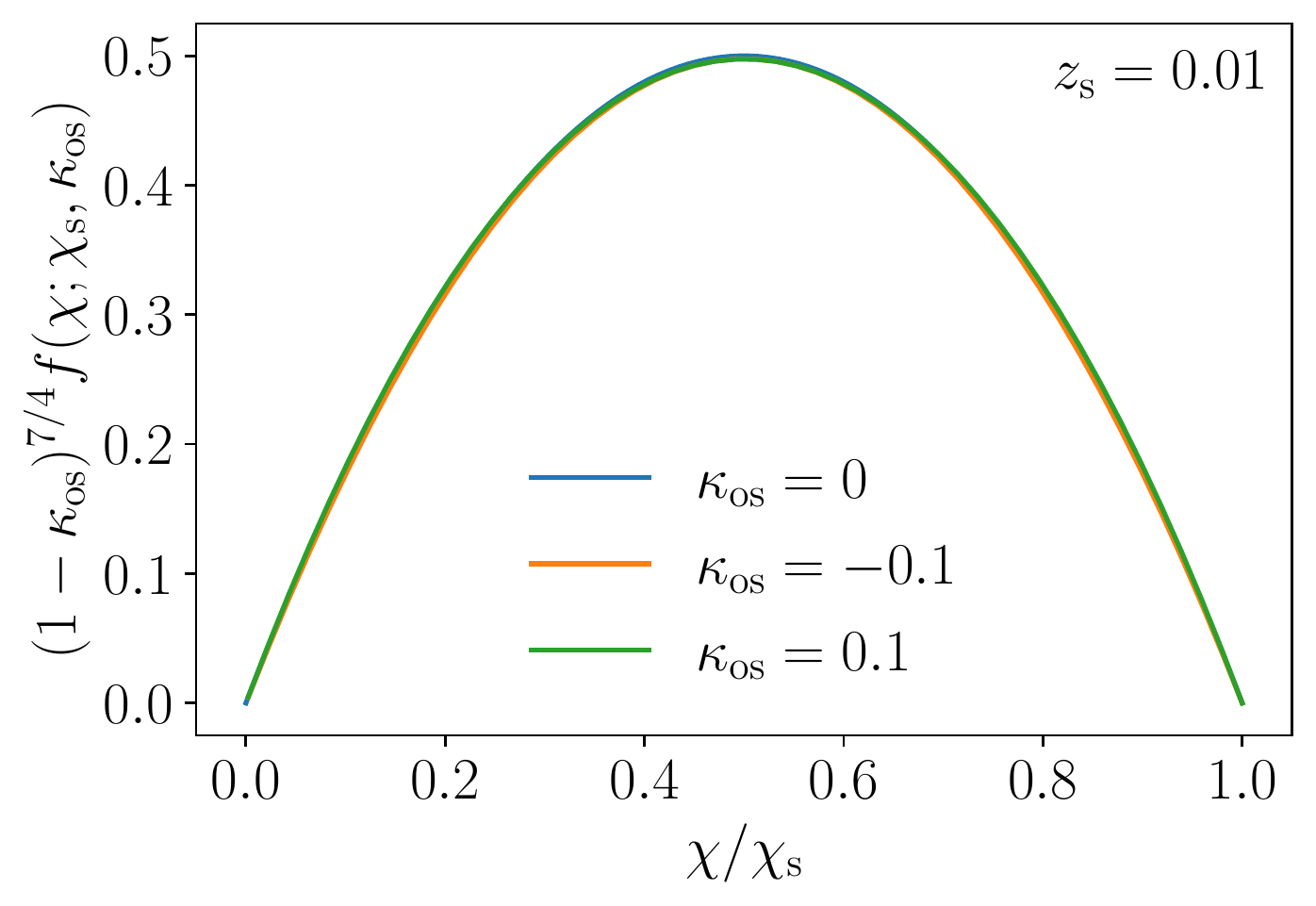}
    \caption{Factor $f(\chi;\chi\e{s}, \kappa\e{s})$ defined in \cref{eq:effective_optical_depth} as a function of the ratio $\chi/\chi\e{s}$. \textit{Left}: showing the dependence on the source redshift, by comparing $z\e{s}\to 0$ to $z\e{s} = 1, 2$ for $\kappa\e{os}=\kappa\e{od}=\kappa\e{ds}=0$. \textit{Right}: showing the dependence in $\kappa\e{os}$ in the mean-field approximation, for $z\e{s}\ll 1$.}
    \label{fig:function_f}
\end{figure}

In the case that we are interested in here, if the dominant lens is fixed at a comoving distance~$\chi$ from the observer, then the reduced microshear $s=s\e{os}/(1-\kappa\e{os})-s\e{od}/(1-\kappa\e{od})-s\e{ds}/(1-\kappa\e{ds})$ caused by the other compact objects has an amplitude~$S\define |s|$ distributed as
\begin{equation}
\label{eq:distribution_microshear}
p(S; \chi) \, \dd S
= \frac{f(\chi) \tau S \, \dd S}{[f^2(\chi)\tau^2+S^2]^{3/2}}
\ ,
\qquad
P(S; \chi) = \pac{1 - \frac{S^2}{f^2(\chi)\tau^2}}^{-1/2} \ .
\end{equation}
\Cref{eq:distribution_microshear} is controlled by an effective optical depth $f \tau$, with\footnote{We shall often omit the fixed variables $\chi\e{s}, \kappa\e{os}$ and just write $f(\chi)$ instead of $f(\chi;\chi\e{s}, \kappa\e{os})$, just like we do not specify the dependence of $\tau$ on those parameters.}
\begin{align}
\label{eq:effective_optical_depth}
f(\chi; \chi\e{s}, \kappa\e{os})
&\define
\frac{
    \int_0^{\chi} \frac{\dd\chi'}{a(\chi')}
        \pac{
            \frac{\chi'(\chi\e{s}-\chi')}{(1-\kappa\e{os})\chi\e{s}}
            - \frac{\chi'(\chi-\chi')}{[1-\kappa\e{od}(\chi)]\chi}
            }
    +
    \int_{\chi}^{\chi\e{s}} \frac{\dd\chi'}{a(\chi')}
        \pac{
            \frac{\chi'(\chi\e{s}-\chi')}{(1-\kappa\e{os})\chi\e{s}}
            - \frac{(\chi'-\chi)(\chi\e{s}-\chi')}{[1-\kappa\e{ds}(\chi)](\chi\e{s}-\chi)} }
    }
    {
    \int_0^{\chi\e{s}} \frac{\dd\chi'}{a(\chi')} \;
        \frac{\chi'(\chi\e{s}-\chi')}{\chi\e{s}}
    }
\\
&\approx
\frac{1}{(1-\kappa\e{os})^{7/4}} \,
\frac{2\chi}{\chi\e{s}} \pa{1-\frac{\chi}{\chi\e{s}}} .
\end{align}
The last approximation holds when the scale factor can be considered constant in the integrals (i.e. for $\chi\e{s}\to 0$) and in the mean-field approximation for $\kappa\e{od}, \kappa\e{ds}$. The $(1-\kappa\e{os})^{-7/4}$ is empirical. The shape of the function $f(\chi)$ for various values of the source redshift and external convergence $\kappa\e{os}$ is depicted in \cref{fig:function_f}. Since the derivation of \cref{eq:distribution_microshear} in ref.~\cite{1997ApJ...489..522L} uses different conventions and notation, we propose a full derivation in \cref{app:microshear_distribution} for completeness. 

Let us finally point out that \cref{eq:distribution_microshear} is actually an approximation where high values of $S$ are overestimated. Indeed, the compact objects responsible for the microshear are, by definition, non-dominant lenses. As such, their individual shear should not exceed the one that would be produced by the dominant lens if it were alone. So in principle $p(S)$ should also depend on, e.g., the impact parameter of the dominant lens~$\beta\e{d}$, which would set an upper bound on $S$. This upper bound would go to infinity as $\beta\e{d}\rightarrow 0$, i.e. for large values of the amplification. Albeit more rigorous, these considerations would significantly complicate the treatment of the problem. We thus choose to ignore them, with the perspective of placing an upper bound on the effect of the microshear on $p(A)$.

\subsubsection{The macroshear is negligible}

The total shear~$\gamma=\bar{\gamma}+s$ is the sum of the microshear~$s$ discussed above with the macroshear~$\bar{\gamma}$ due to the large-scale structure. While the distribution of microshear has a heavy tail, $\Prob(>S)\propto S^{-1}$, it turns out that the macroshear does not share this property, because the structures producing it are more diffuse. As shown in \cref{app:fit_simulation_shear}, the conditional PDF of the macroshear at fixed convergence is surprisingly well fit by a two-dimensional Gaussian distribution, which therefore predicts very few high values for the macroshear.

Note that $s$ must be compared with $\bar{\gamma} \define \bar{\gamma}\e{os}/(1-\kappa\e{os})-\bar{\gamma}\e{od}/(1-\kappa\e{od})-\bar{\gamma}\e{ds}/(1-\kappa\e{ds})$ rather than with $\bar{\gamma}\e{os}$ alone. The difficulty is that ray tracing in numerical simulations is performed for a unique observer at present time; they allow one to compute $\bar{\gamma}\e{os}, \bar{\gamma}\e{od}$, but not $\bar{\gamma}\e{ds}$. To circumvent this issue we apply again the mean-field approximation introduced in \cref{subsubsec:approximations}, which yields
\begin{equation}
\bar{\gamma}
\approx f \, \bar{\gamma}\e{os} \ ,
\label{eq:mean_field_shear}
\end{equation}
with $f$ defined in \cref{eq:effective_optical_depth}. It is not surprising to find here the same correction factor~$f$ as for the effective optical depth of microshear: both have the same origin. Within that approximation, the PDF of $|\bar{\gamma}|$ is obtained from \cref{eq:PDF_macroshear_fit} by simple rescaling.

\begin{figure}[t]
    \centering
    \includegraphics[width=0.49\columnwidth]{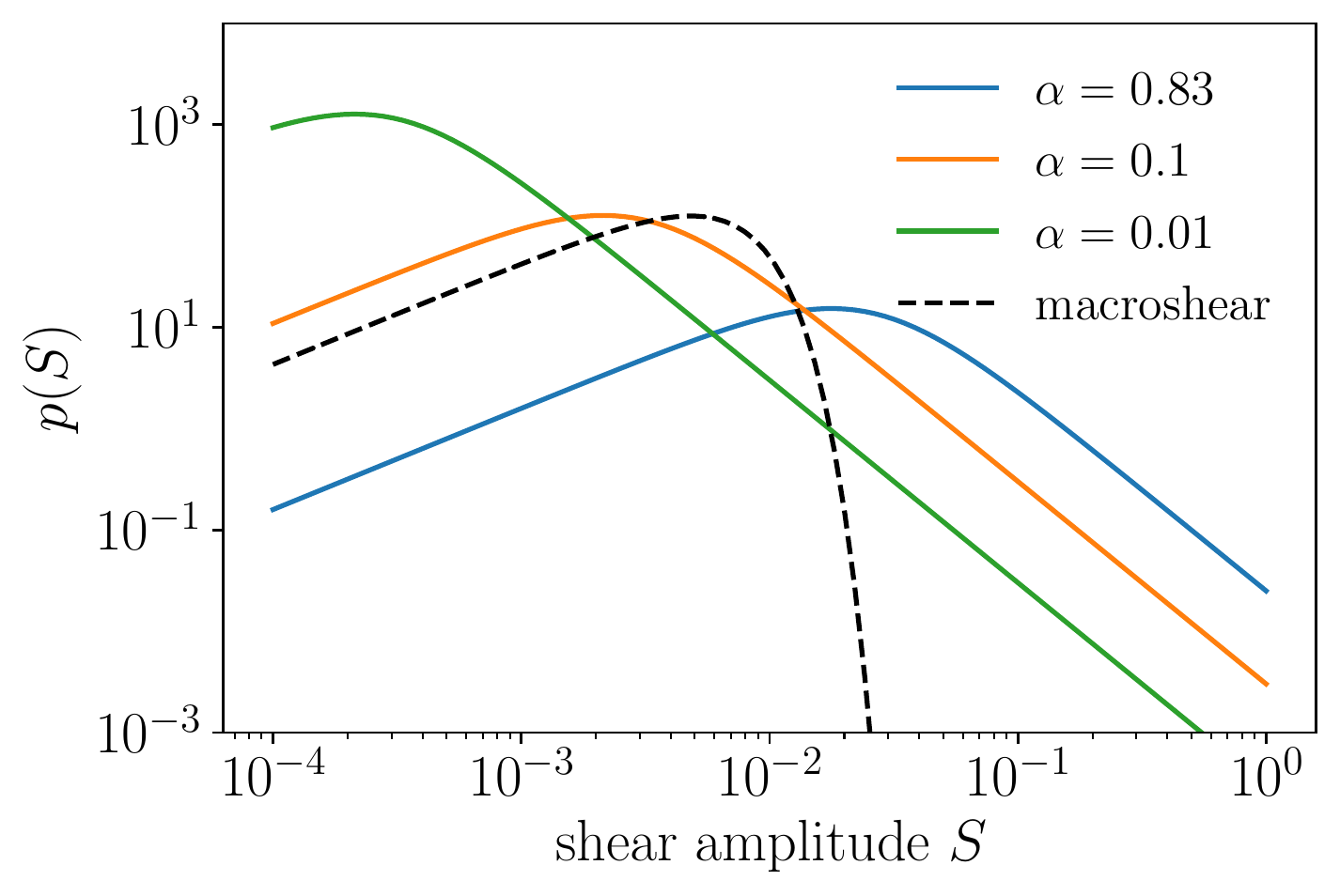}
    \hfill
    \includegraphics[width=0.49\columnwidth]{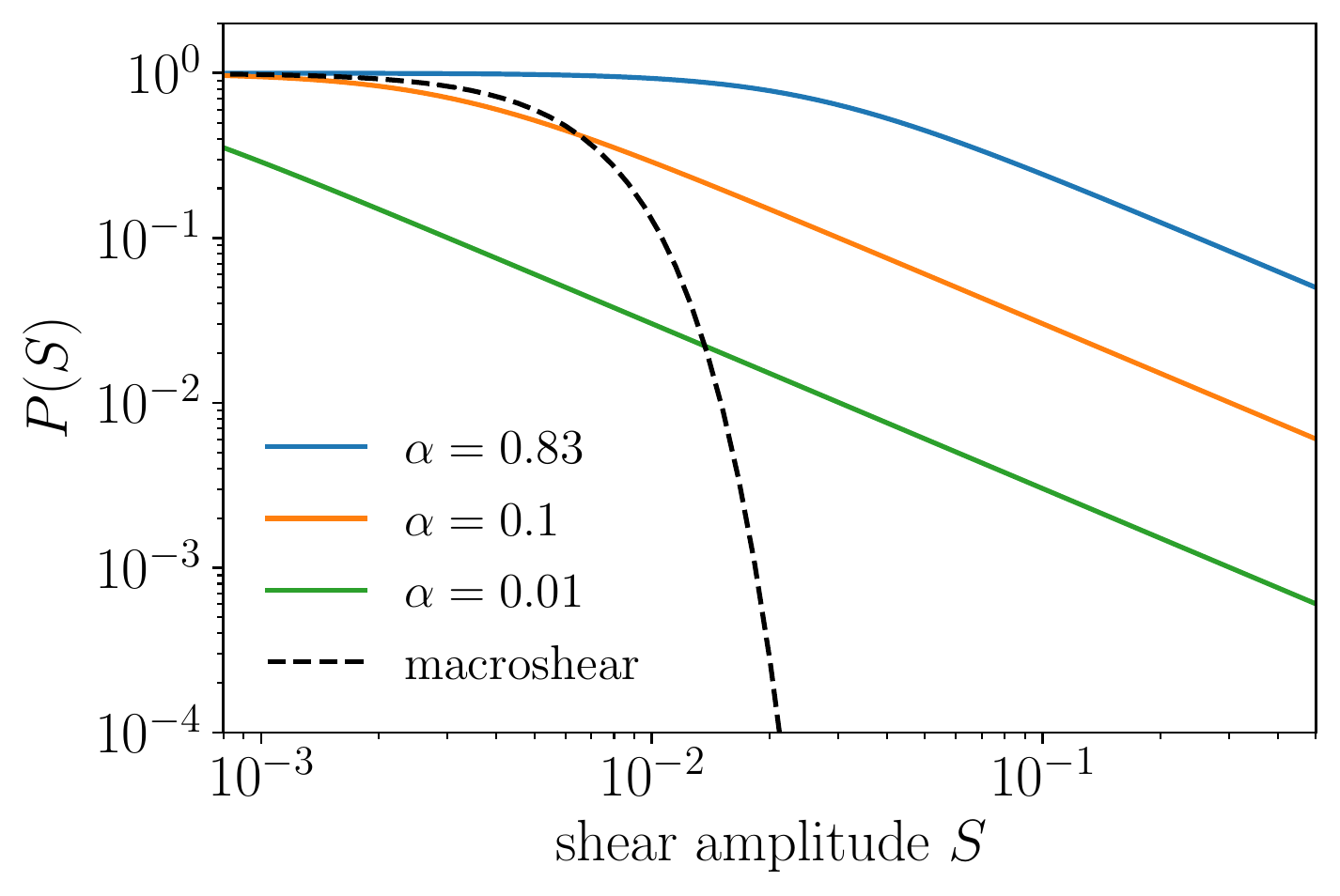}
    \caption{Distributions of the amplitude of microshear (solid lines), if the source located at $z\e{s}=0.95$ and the dominant lens at $z=0.5$. The left panel shows PDFs while the right panel shows CDFs. Three values for the fraction of compact matter are considered, $\alpha=0.83, 0.1, 0.01$. Dashed lines indicate the distributions of the amplitude of the macroshear $|\bar{\gamma}|\approx f|\gamma\e{os}|$, for lines of sight with $\bar{\kappa}\e{os}=0$.}
    \label{fig:comparison_micro_macro}
\end{figure}

\Cref{fig:comparison_micro_macro} compares the distributions of the microshear and macroshear amplitudes, for a source at $z\e{s}=0.95$, a dominant lens at $z=0.5$, and for a line of sight with $\bar{\kappa}\e{os}=0$ for simplicity. Three values for the fraction of compact matter are considered, $\alpha=0.83, 0.1, 0.01$ -- the first case would correspond to the whole DM being made of compact objects. Those values correspond, respectively, to the effective optical depths $f\tau=2.5\times 10^{-2}, 3.0\times 10^{-3}, 3.0\times 10^{-4}$. Although the macroshear is generally not negligible compared to the microshear, especially when $\alpha$ is small, it is unable to produce large amplitudes. But large values of the shear are necessary to produce changes in $\Omega(A)$ at reasonable amplifications, $A<10$ (see \cref{subsec:amplification_cross_section}). In the situation illustrated here, $|\bar{\gamma}|<\SI{3}{\percent}$ which would only affect $\Omega(A\gtrsim 13)$. Summarising, when macroshear is comparable to, or even larger than, microshear, then both have a negligible impact on the amplification statistics anyway. We shall thus neglect macroshear from now on, and replace $\Omega(A;m, \chi, |\bar{\gamma}+s|)$ with $\Omega(A;m, \chi, S)$ in \cref{eq:p_1_definition}.

\subsubsection{Final expression of \texorpdfstring{$p_1(A)$}{p1A}}
\label{subsubsec:p_1_final}

Substituting, in \cref{eq:p_1_definition}, the probability density
$
p(m, s, \chi)
= p(m) \, p(S; \chi) \, p(\chi)
$
-- where $p(S;\chi)$ is given by \cref{eq:distribution_microshear} --  and the expression~\eqref{eq:cross_section_final} of $\Omega(A; m, \chi, S)$, and performing the change of variable $S \mapsto y\define S/f \tau$, we find
\begin{equation}
\label{eq:p_1_intermediate}
p_1(A)
= \frac{2}{\Theta^2} \,
    \frac{A\e{min}^2}{(A^2 - A\e{min}^2)^{3/2}}
    \int_0^{\chi\e{s}} \dd \chi \; p(\chi)
    \ev[2]{\vartheta\e{E}^2(\chi)}_m
    K\pac{\frac{f(\chi)\tau A}{A\e{min}}} \ ,
\end{equation}
where $\ev{\ldots}_m$ denotes an average over the mass~$m$ of the dominant lens, and\footnote{In ref.~\cite{Kofman:1996es}, the function~$K(x)$ is denoted by $f_1(x)$. The approximation in the second equality of \cref{eq:integral_K_approx} was proposed in ref.~\cite{Kofman:1996es} and its comparison with the exact result is shown in fig.~4 therein.}
\begin{equation}
\label{eq:integral_K_approx}
K(x)
\define \int_0^\infty \dd y \;\frac{y\,k(xy)}{(1+y^2)^{3/2}}
\approx 1 - 0.81 \, x^2(1-3x)\pa{1 + \frac{3}{2}\,x^{3/2}}^{-8/3}
\ .
\end{equation}

The last step of the simplification of $p_1(A)$ consists in fully isolating the effect of the (micro)shear. For that purpose, we may multiply and divide \cref{eq:p_1_intermediate} with the average value of the weakly lensed squared Einstein radius,
\begin{equation}
\label{eq:mean_lensed_einstein_radius}
\ev[2]{\vartheta\e{E}^2}
\define
\int_0^{\chi\e{s}} \dd\chi \; p(\chi)
\ev[2]{\vartheta\e{E}^2(\chi)}_m
=
\frac{4G\ev{m}}{(1-\kappa\e{os})\chi\e{s}}
\int_0^{\chi\e{s}}\frac{\dd\chi}{a(\chi)} \; 
\frac{1-\kappa\e{ds}(\chi)}{1-\kappa\e{od}(\chi)} \,
\chi(\chi\e{s}-\chi)
\ ,
\end{equation}
to get
\begin{empheq}[box=\fbox]{equation}
\label{eq:p_1_final}
p_1(A)
=
\frac{2\ev{\vartheta\e{E}^2}}{\Theta^2}\,
\frac{A\e{min}^2}{(A^2 - A\e{min}^2)^{3/2}} \,
\mathcal{K}\pac{\frac{\tau}{(1-\kappa\e{os})^{7/4}}\frac{A}{A\e{min}}}
\ ,
\end{empheq}
with the last function that we shall define in this derivation:
\begin{align}
\mathcal{K}(x)
\define
\frac{
    \int_0^{\chi\e{s}} \frac{\dd\chi}{a(\chi)} \;
    \frac{1-\kappa\e{ds}(\chi)}{1-\kappa\e{od}(\chi)} \,
    \chi(\chi\e{s}-\chi) \,
    K[(1-\kappa\e{os})^{7/4}f(\chi) x]
    }
    {
    \int_0^{\chi\e{s}} \frac{\dd\chi}{a(\chi)} \;
    \frac{1-\kappa\e{ds}(\chi)}{1-\kappa\e{od}(\chi)} \,
    \chi(\chi\e{s}-\chi) \,
    } \ .
\label{eq:integral_cal_K_definition}
\end{align}
The presence of the $(1-\kappa\e{os})^{7/4}$ factor in the argument of $K$ in \cref{eq:integral_cal_K_definition} is designed to absorb the empirical dependence on $\kappa\e{os}$ in $f$, and hence make $\mathcal{K}(x)$ practically insensitive to $\kappa\e{os}$.

\begin{figure}[t]
    \centering
    \includegraphics[width=0.49\columnwidth]{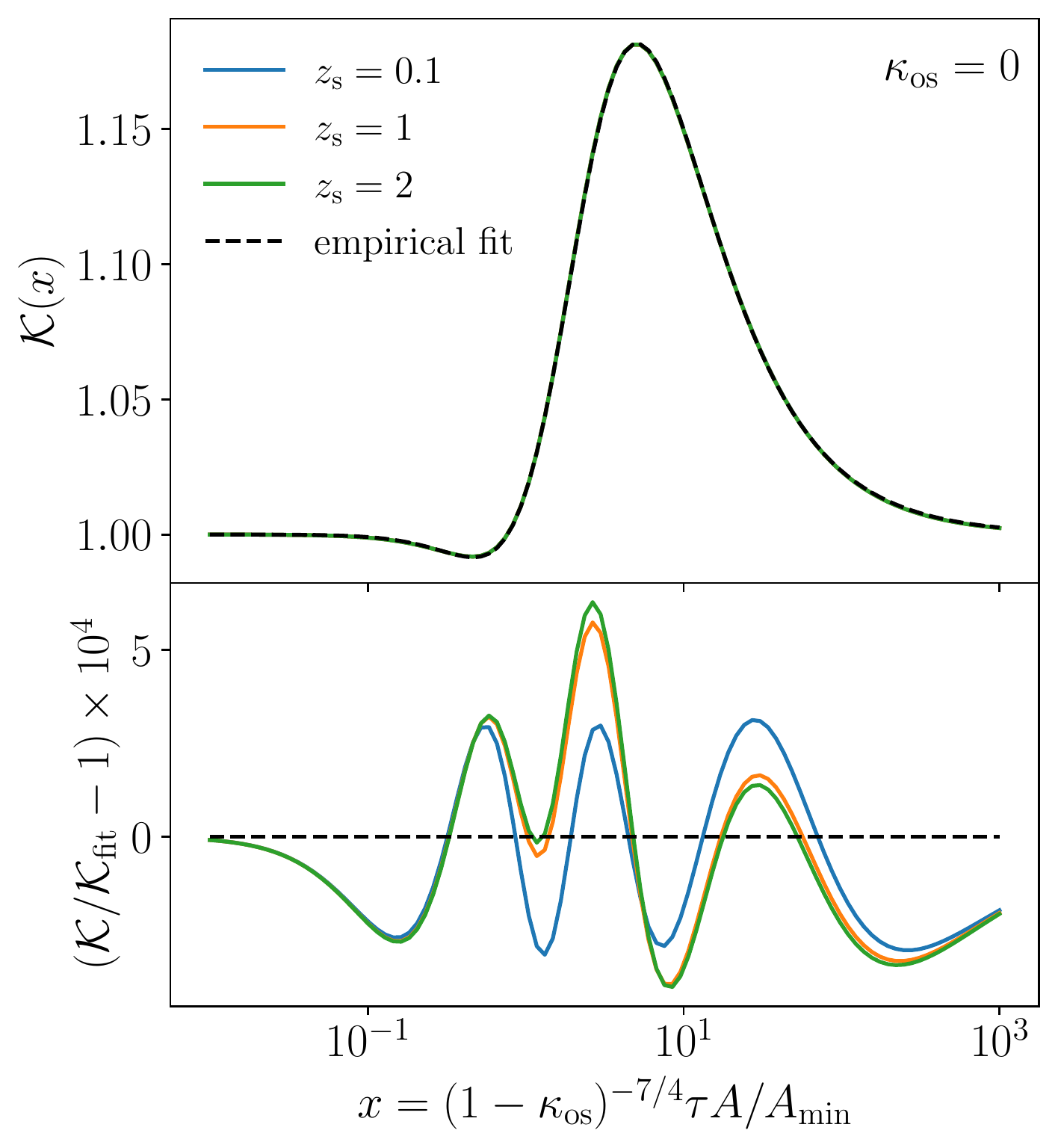}
    \hfill
    \includegraphics[width=0.49\columnwidth]{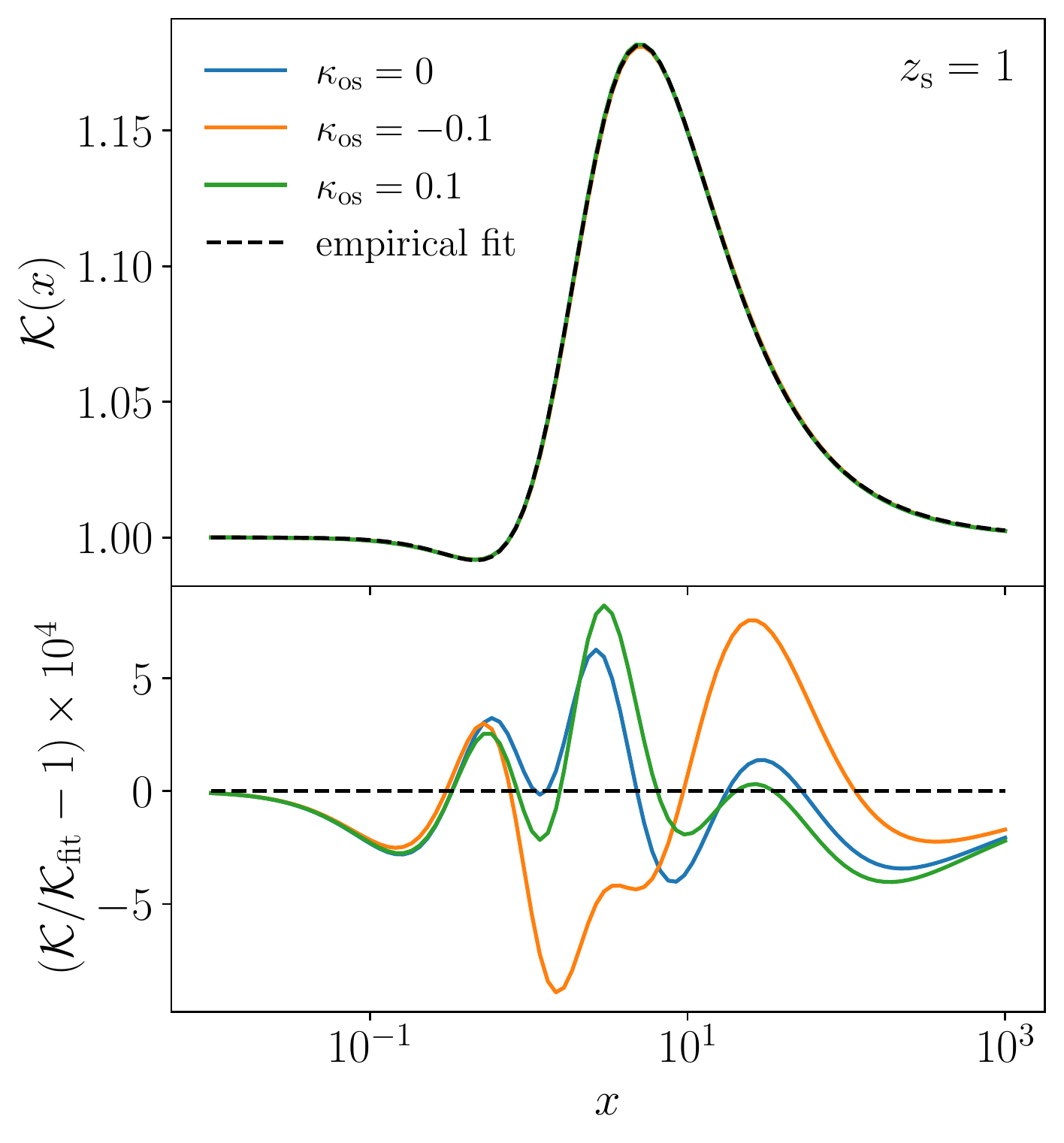}
    \caption{Integral $\mathcal{K}(x)$ defined in \cref{{eq:integral_cal_K_definition}} as a function of $x=(1-\kappa\e{os})^{-7/4}\tau A/A\e{min}$, which encapsulates all the microshear corrections to $p_1(A)$, with the empirical fitting function proposed in \cref{eq:integral_cal_K_fit}. \textit{Left}: Checking the dependence in $\chi\e{s}=\chi(z\e{s})$, for a fixed $\kappa\e{os}=0$. \textit{Right}: Checking the dependence in $\kappa\e{os}$ for $z\e{os}=1$. The bottom panels show the relative accuracy of the empirical fit. We can see that $\mathcal{K}$ is mostly insensitive to both $z\e{s}$ and $\kappa\e{os}$, and that the fitting function is an excellent approximation.}
    \label{fig:integral_cal_K}
\end{figure}

The function $\mathcal{K}(x)$ defined in \cref{eq:integral_cal_K_definition} fully encapsulates the effect of the microshear. In principle, this function depends on the source redshift $z\e{s}$ via $\chi\e{s}=\chi(z\e{s})$, and on the macrostructure along the line of sight via $\kappa\e{os}, \kappa\e{od}, \kappa\e{ds}$. In practice, however, \cref{fig:integral_cal_K} shows that $\mathcal{K}(x)$ is quite insensitive to those parameters. Such an empirical independence of $\mathcal{K}(x)$ in its external parameters encourages us to look for a simple and universal fitting function for it. We find that
\begin{equation}
\mathcal{K}(x) = 1 - 0.254 \, x^{2.33} \pa{1 - 1.30\,x } \pac{1 + \frac{5}{4}\pa{\frac{x}{1.83}}^{5/4}}^{-3.43}
\label{eq:integral_cal_K_fit}
\end{equation}
provides an excellent fit, with an accuracy of a few parts in $10^4$ (see \cref{fig:integral_cal_K}).

The main conclusion of this subsection is that, to an excellent level of precision, the effect of the microshear on $p_1(A)$ mostly depends on the optical depth~$\tau$. It reduces by about \SI{1}{\percent} the probability of amplifications $A\sim 1/\tau$, and enhances larger ones ($A\sim 10/\tau$) by about \SI{15}{\percent}. Since we are considering low values for the optical depths, we can already anticipate that the net impact of shear on reasonable amplifications will be negligible.

\subsection{From one lens to many: the strongest-perturbed-lens prescription}
\label{sec:strongest}

Now that we dispose of an accurate expression for the amplification PDF~$p_1(A)$ of a single perturbed lens within a mesoscopic cone (\cref{fig:mesosopic_cone}), we can generalise it to a large number $N\gg 1$ of such lenses. For that purpose, we shall adapt the strongest-lens prescription of \cref{subsec:strongest_lens_approach}, which consists in assuming that the total amplification~$A$ produced by the $N$ perturbed lenses in the cone, is well-approximated by the amplification due to the strongest of them. Importantly, that is \emph{not} to say that we are entirely neglecting the effect of the other lenses, because it is already encoded in the convergence and microshear corrections. As such, the strongest-perturbed-lens approach must be understood as a statistical prescription that is physically consistent with the set of approximations that we have considered so far.

Let us be more specific. The probability that the strongest individual amplification is smaller than $A$, is equal to the probability that all $N$ lenses individually produce an amplification smaller than $A$. Hence, the probability~$P(A)$ that the strongest amplification is larger than~$A$ reads
\begin{equation}
P(A) = 1 - \pac{ 1 - \int_A^\infty \dd A' \; p_1(A')}^N \ .
\end{equation}
The strongest-lens approximation consists in assuming that the above is a good model for the CDF of the total amplification.

Examining the expression~\eqref{eq:p_1_final} of $p_1(A)$, we notice that it is proportional to $1/N$,
\begin{equation}
p_1(A) \propto 
\frac{\ev{\vartheta\e{E}^2}}{\Theta^2}
= \frac{1}{N} \, \Sigma \pi \ev[2]{\vartheta\e{E}^2}
\approx \frac{1}{N} \, \frac{\tau}{1-\kappa\e{os}} \ ,
\end{equation}
where we recognised the projected angular density of lenses within the mesoscopic cone, $\Sigma=N/(\pi\Theta^2)$, and the microlensing optical depth~$\tau=\Sigma\pi\ev{\theta\e{E}^2}$. We also considered $\ev{\vartheta^2\e{E}}\approx \ev{\theta\e{E}^2}/(1-\kappa\e{os})$, as suggested by \cref{eq:mean_lensed_einstein_radius} where $\kappa\e{od}, \kappa\e{ds}$ only produce minor corrections.\footnote{This implies that the weakly lensed optical depth is approximated as $\Sigma\pi\ev[2]{\vartheta\e{E}^2} \approx \tau/(1-\kappa\e{os})$. In the presence of a positive convergence, i.e. an overdense line of sight, the effective optical depth is thus larger than the one expected without accounting for the convergence corrections.} Thus, in the large-$N$ limit, we have
\begin{equation}
P(A) \approx 1 - \exp\pac{- \int_A^\infty \dd A' \; N p_1(A')} .
\end{equation}
Substituting the explicit expression of $p_1(A)$, and changing the integration variable to $X\define A/A\e{min}$, we finally obtain the main result of this article,
\begin{empheq}[box=\fbox]{equation}
P(A; z\e{s}, \alpha, \bar{\kappa}\e{os})
= 1 - \exp\paac{ -\frac{2\tau}{1-\kappa\e{os}}
                \int_{A/A\e{min}}^{\infty}
                \frac{\dd X}{\pa{X^2-1}^{3/2}} \;
                \mathcal{K}\pac{\frac{\tau X}{(1-\kappa\e{os})^{7/4}}}
                } ,
\label{eq:CDF_final}
\end{empheq}
with the parameters
$\tau = \alpha(\Delta\e{os}+\bar{\kappa}\e{os})$,
$\kappa\e{os} = (1-\alpha)\bar{\kappa}\e{os}-\alpha\Delta\e{os}$
and
$A\e{min} = (1-\kappa\e{os})^{-2}$,
which explicitly depend on the fraction~$\alpha$ of compact objects, the homogeneous convergence deficit~$\Delta\e{os}(z\e{s})$ given in \cref{eq:Delta_os}, and the average weak-lensing convergence~$\bar{\kappa}\e{os}$ that would be observed if all the matter were diffuse.
Note that \cref{eq:CDF_final} is independent of the size~$\Theta$ of the mesoscopic cone that we started with. In the case where the external convergence and shear are neglected, i.e. $\kappa\e{os}=0, A\e{min}=1, \mathcal{K}=1$, we recover the simple result of \cref{eq:CDF_A_low_optical_depth}.

\subsection{Marginalising over the line-of-sight convergence}
\label{subsec:marginalising_over_LOS_convergence}

\Cref{eq:CDF_final} gives the amplification CDF within a mesoscopic area of the sky where $\bar{\kappa}\e{os}$, and hence $\tau, \kappa\e{os}$ can be considered fixed. The full CDF is obtained by marginalising over all mesoscopic lines of sight, that is
\begin{equation}
P(A; z\e{s}, \alpha)
= \int \dd\bar{\kappa}\e{os} \;
    p(\bar{\kappa}\e{os}; z\e{s}) \,
    P(A; z\e{s}, \alpha, \bar{\kappa}\e{os}) \ .
\end{equation}
Just like in \cref{subsec:distribution_optical_depth}, we use the results from simulations and standard cosmology to estimate $p(\bar{\kappa}\e{os}; z\e{s})$, as explained in \cref{app:fit_simulations_convergence}.

\begin{figure}
\centering
\includegraphics[width=0.49\columnwidth]{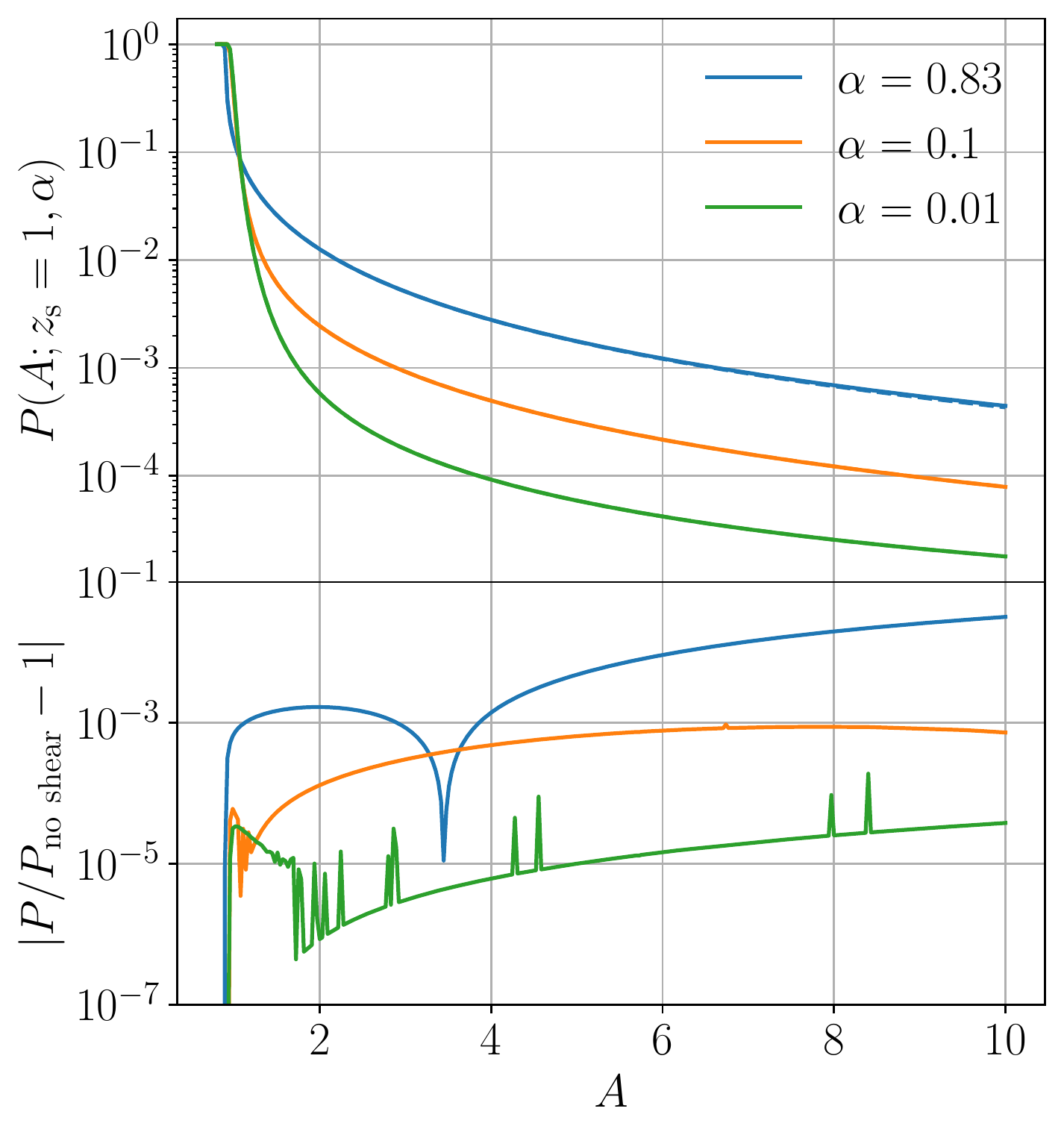}
\hfill
\includegraphics[width=0.49\columnwidth]{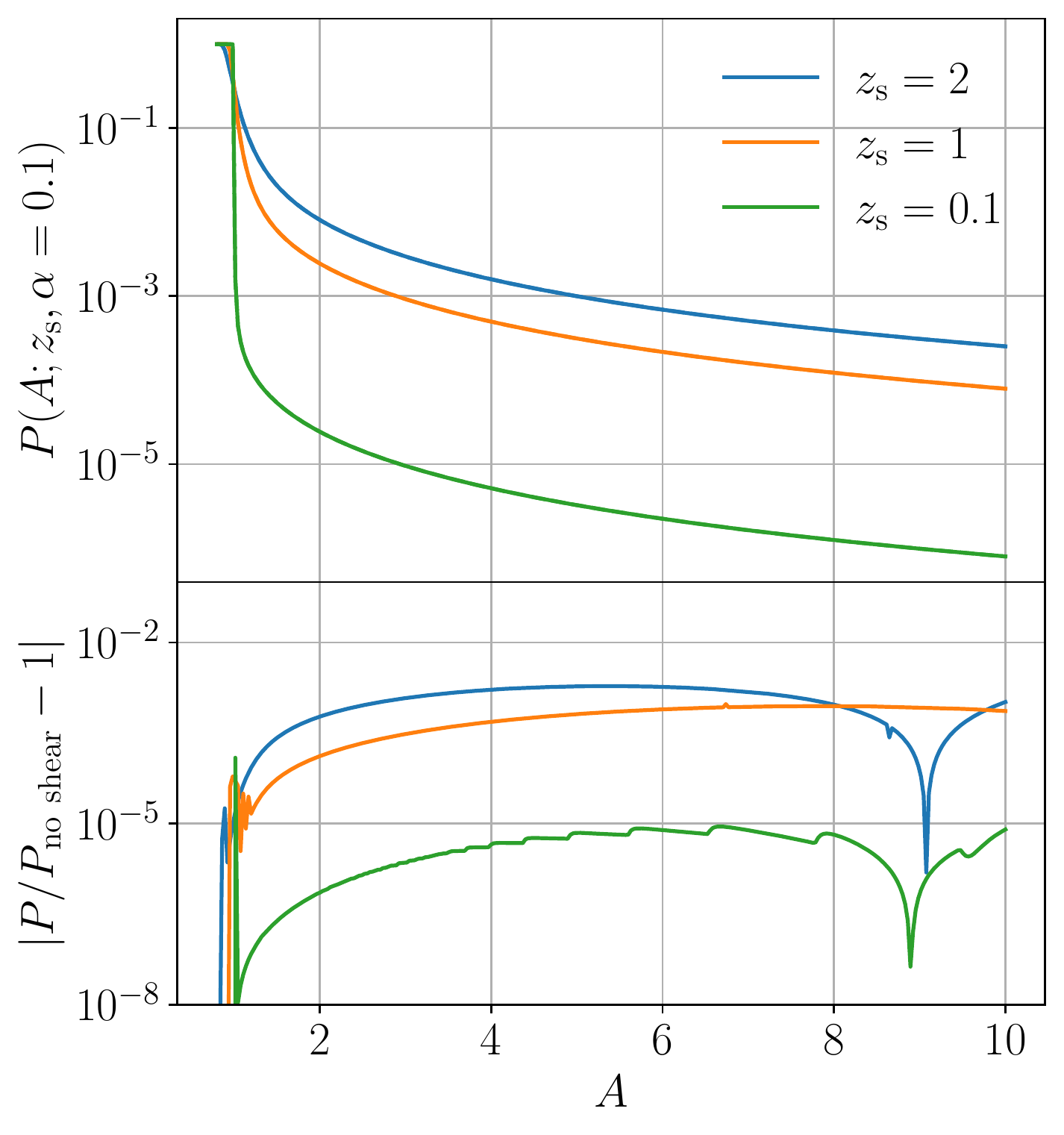}
\caption{Final CDF of the amplification, $P(A; z\e{s}, \alpha)\define \Prob(>A|z\e{s}; \alpha)$, once marginalised over the line-of-sight convergence~$\bar{\kappa}\e{os}$. \textit{Left}: dependence in the fraction~$\alpha$ of compact objects for $z\e{s}=1$. \textit{Right}: dependence in the source redshift~$z\e{s}$, for $\alpha=0.1$. Upper panels indicate both the exact result (solid lines) and the case where the effect of shear is neglected therein, i.e., for $\mathcal{K}=1$ (dashed lines). The solid and dashed lines are superimposed -- their relative difference is depicted in the bottom panels; the rapidly oscillating features are non-physical artefacts due to numerical integration.}
\label{fig:CDF_final}
\end{figure}

The final amplification CDF is depicted in \cref{fig:CDF_final}, for different values of the fraction~$\alpha$ of compact objects and of the source redshift~$z\e{s}$. As expected, the probability of high amplifications increases with both $\alpha$ and $z\e{s}$, because the optical depth~$\tau$ increases with both parameters. For a source at $z\e{s}=1$, the probability that it is amplified by a factor larger than two is \SI{1.6}{\percent} if all the DM (\SI{83}{\percent} of the total matter) in the Universe is made of compact objects. This probability falls to \SI{0.23}{\percent} if \SI{10}{\percent} of matter is compact, and to \SI{0.056}{\percent} if only \SI{1}{\percent} of the matter is compact.

The upper panels of \cref{fig:CDF_final} show both the exact $P(A; z\e{s}, \alpha)$ and the case where microshear is neglected, which corresponds to setting $\mathcal{K}=1$ in \cref{eq:CDF_final},
\begin{align}
P\e{no\ shear}(A; z\e{s}, \alpha, \bar{\kappa}\e{os})
&\define
1 - \exp\pac{ - \frac{2\tau}{1-\kappa\e{os}} \int_{A/A\e{min}}^{\infty}
                \frac{\dd X}{\pa{X^2 - 1}^{3/2}}
                }
\\
&=
1 - \exp\pac{ -\frac{2\tau}{1-\kappa\e{os}}
                \pa{ \frac{A}{\sqrt{A^2 - A\e{min}^2}} - 1 }
                } .
\label{eq:CDF_no_shear}
\end{align}
The associated curves are essentially indistinguishable by eye; their relative difference, $|P/P\e{no\ shear} - 1|$, is shown in the bottom panels of \cref{fig:CDF_final} and is \changed{sub-percent for $A<6$}. Of course, due to the behaviour of the function $\mathcal{K}(x)$, larger amplifications ($A \sim 1/\tau$) are expected to be affected more significantly by the microshear. But in practice such high amplifications are so rare that they have no observational relevance. Hence, the main take-home message of this subsection is that \emph{the effect of shear is negligible in the statistics of extragalactic microlensing}. This implies that, for practical purposes, one may safely use the simple no-shear expression~\eqref{eq:CDF_no_shear} for $P(A; z\e{s}, \alpha; \bar{\kappa}\e{os})$. Such a conclusion could hardly have been guessed from the beginning. The external shear is known to be a crucial parameter in the modelling of strong lenses (e.g.~\cite{1997ApJ...482..604K}), and \cref{fig:point_lens_shear} shows that it generally has a significant impact on the amplification cross section. But since the effect shows up around amplifications~$A\sim 1/\tau$, and that the amplification PDF is already low for $A\gtrsim 1/\tau$, the net integrated effect on $P(A)$ ends up being negligible for interesting values of $A$.\footnote{Another argument is that, for realistic sources of light, large amplifications such that $A\sim 1/\tau$ are very hard to access due to the finite size of the sources (see \cref{sec:extended_sources}). However, with GWs sources, magnification factors of many hundreds are possible and therefore the net effect could not be negligible.} 

\subsection{Comparison with Zumalac\'{a}rregui \& Seljak}

In ref.~\cite{Zumalacarregui:2017qqd} (hereafter ZS17), Zumalac\'{a}rregui \& Seljak have set constraints on the fraction of extragalactic compact objects that would produce a microlensing signal in the supernova data. For that purpose, they used a phenomenological model for the amplification statistics. It is worth comparing the predictions of that model to our approach in order to evaluate what one may call \emph{theoretical systematics} on any analysis of supernova microlensing.

ZS17's model, based on earlier developments by Seljak \& Holz~\cite{Seljak:1999tm} and Metcalf \& Silk~\cite{1999ApJ...519L...1M, 2007PhRvL..98g1302M}, is expressed in terms of a shifted magnification~$\mu$, such that $1+\mu$ represents the magnification of an image with respect to its empty-beam counterpart, i.e., if that image were seen through an empty universe. It is related to our amplification~$A$ as
\begin{equation}
1 + \mu = (1 + \Delta\e{os})^2 \, A \ ,
\end{equation}
where $\Delta\e{os}$ is the same as defined in \cref{eq:Delta_os}. With such conventions, $\mu=0$ corresponds to $A=(1+\Delta\e{os})^{-2}$, which is indeed the empty-beam case. The distribution of $\mu$ is then designed by assuming that $\mu$ can be written as the sum $\mu=\mu\e{s}+\mu\e{c}$ of a weak-lensing contribution from the smooth matter, $\mu\e{s}$, and a microlensing contribution from compact objects, $\mu\e{c}$.

The smooth part is written as $\mu\e{s}=(1-\alpha)\bar{\mu}$, where $\bar{\mu}$ would be the magnification in the absence of compact objects. Note that this is quite similar to our approach described in \cref{subsubsec:physical_origin_convergence_shear}, except that we have worked with convergences rather than shifted magnifications. In ZS17, the statistics of $\bar{\mu}$ are obtained using the \href{http://www.turbogl.org}{\texttt{TurboGL}} code~\cite{Kainulainen:2010at, Kainulainen:2009dw}.

The statistics of microlensing part, $\mu\e{c}$, are based on an empirical model originally used by Rauch to fit ray-shooting simulations in ref.~\cite{1991ApJ...383..466R},
\begin{equation}
p\e{R}(\mu\e{c};\bar{\mu}\e{c})
=
N \pac{ \frac{1 - \ex{-\mu\e{c}/\Delta\mu}}
{(1 + \mu\e{c})^2 - 1} }^{3/2} ,
\end{equation}
where $N$ and $\Delta\mu$ are two functions of $\bar{\mu}\e{c}$ that are chosen so as to ensure that $p\e{R}$ is normalised to 1 and with expectation value $\ev{\mu\e{c}} = \bar{\mu}\e{c}$. This value is set to be $\bar{\mu}\e{c}=\alpha\bar{\mu}$ by the magnification theorem~\cite{2008MNRAS.386..230W}, and plays a role comparable to the optical depth $\tau$ in our approach.

\begin{figure}
\centering
\begin{minipage}{0.6\columnwidth}
\includegraphics[width=1.4\columnwidth]{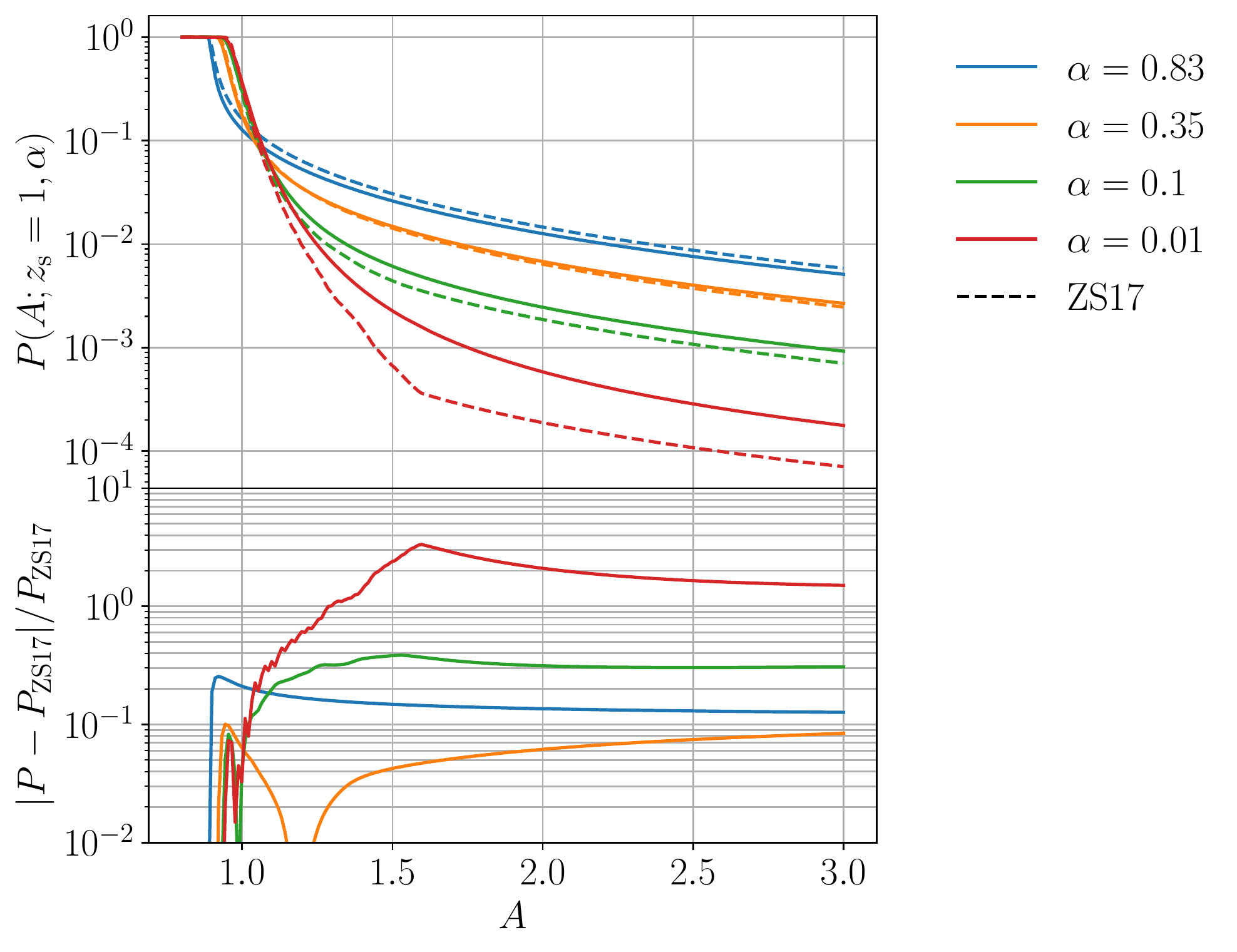}
\end{minipage}
\hfill
\begin{minipage}{0.35\columnwidth}
\vspace*{1cm}
\caption{Comparison of our model for the amplification CDF (solid lines) and the phenomenological model used in ZS17~\cite{Zumalacarregui:2017qqd} (dashed lines), for point-like sources at $z_{\rm s}=1$ and different values of the fraction of compact objects~$\alpha$.}
\label{fig:comparison_ZS17}
\end{minipage}
\end{figure}

In such conditions, ZS17's model for the PDF of the magnification~$\mu=\mu\e{c}+\mu\e{s}$ reads
\begin{equation}
p\e{ZS17}(\mu; z\e{s})
= \int_0^\infty \dd\bar{\mu} \;
    p\e{TurboGL}(\bar{\mu}; z\e{s}) \,
    p\e{R}[\mu - (1-\alpha)\bar\mu; \alpha\bar{\mu}] \ ,
\end{equation}
which undoubtedly has the advantage of simplicity. \Cref{fig:comparison_ZS17} shows a comparison between the predictions of our model with those of ZS17's model, in the case of point-like sources for simplicity. Compared to our approach, ZS17's model tends to overestimate by more than \SI{10}{\percent} the large-amplification events for high values of $\alpha$; for low values of $\alpha$, on the contrary, it tends to underestimate them by more than \SI{100}{\percent}. Coincidentally, both models nearly agree (up to a few percent) for $\alpha=0.35$, which turns out to be the maximum fraction of compact objects allowed at \SI{95}{\percent} confidence level in ZS17. Since $P\e{ZS17}(A)>P(A)$ for smaller values of $\alpha$, this suggest that conducting an analysis similar to ZS17's with our model for amplification statistics would yield slightly weaker constraints on $\alpha$. Such an analysis is beyond the scope of this article, but the present results show that theoretical systematics can generally reach $\SI{100}{\percent}$ for extragalactic microlensing.

\section{Extended sources}
\label{sec:extended_sources}

So far we have considered point-like sources, but the finite size of real light sources is known to have significant effects on the amplification distribution. As a rule of thumb, if a source has an unlensed angular size~$\sigma$, then it smoothes out the amplification map obtained in the point-source case on the angular scale $\sigma$ -- see e.g. fig.~13 of ref.~\cite{Fleury:2019xzr} for illustration. This implies that the effect of small structures, i.e. lenses with small Einstein radii, is suppressed.\footnote{That is why, for example, the constraints on the abundance of PBHs set by SN microlensing in ref.~\cite{Zumalacarregui:2017qqd} only apply to masses larger than $10^{-2}M_\odot$.} In this subsection, we show how to add finite-source corrections to the amplification distributions derived in the previous sections.

\subsection{Extended-source corrections on an isolated point lens}

We have seen in \cref{subsubsec:equivalent_lens} that the problem of a point lens with tidal corrections can be conveniently phrased as an equivalent point lens with a single effective shear correction. \changed{Besides, the analysis leading to \cref{fig:CDF_final} shows that the effect of shear is statistically negligible in the point-source case. Since the extended-source case is deduced from the point-source case by a smoothing of its amplification map, if the effect of shear is small in the latter, it must also be small in the former.} Hence, in all the remainder of this section we shall neglect the shear, so the equivalent lens is a mere unperturbed point lens,
\begin{equation}
\tilde{\vect{\beta}} = \tilde{\vect{\theta}} - \frac{\tilde{\theta}\e{E}^2}{\tilde{\vect{\theta}}} \ ,
\end{equation}
where the twiddled quantities are expressed in terms of the original ones in \cref{eq:beta_tilde,eq:theta_tilde,eq:theta_E_tilde}, all the shears being set to zero.

\begin{figure}[t]
\includegraphics[width=0.49\columnwidth]{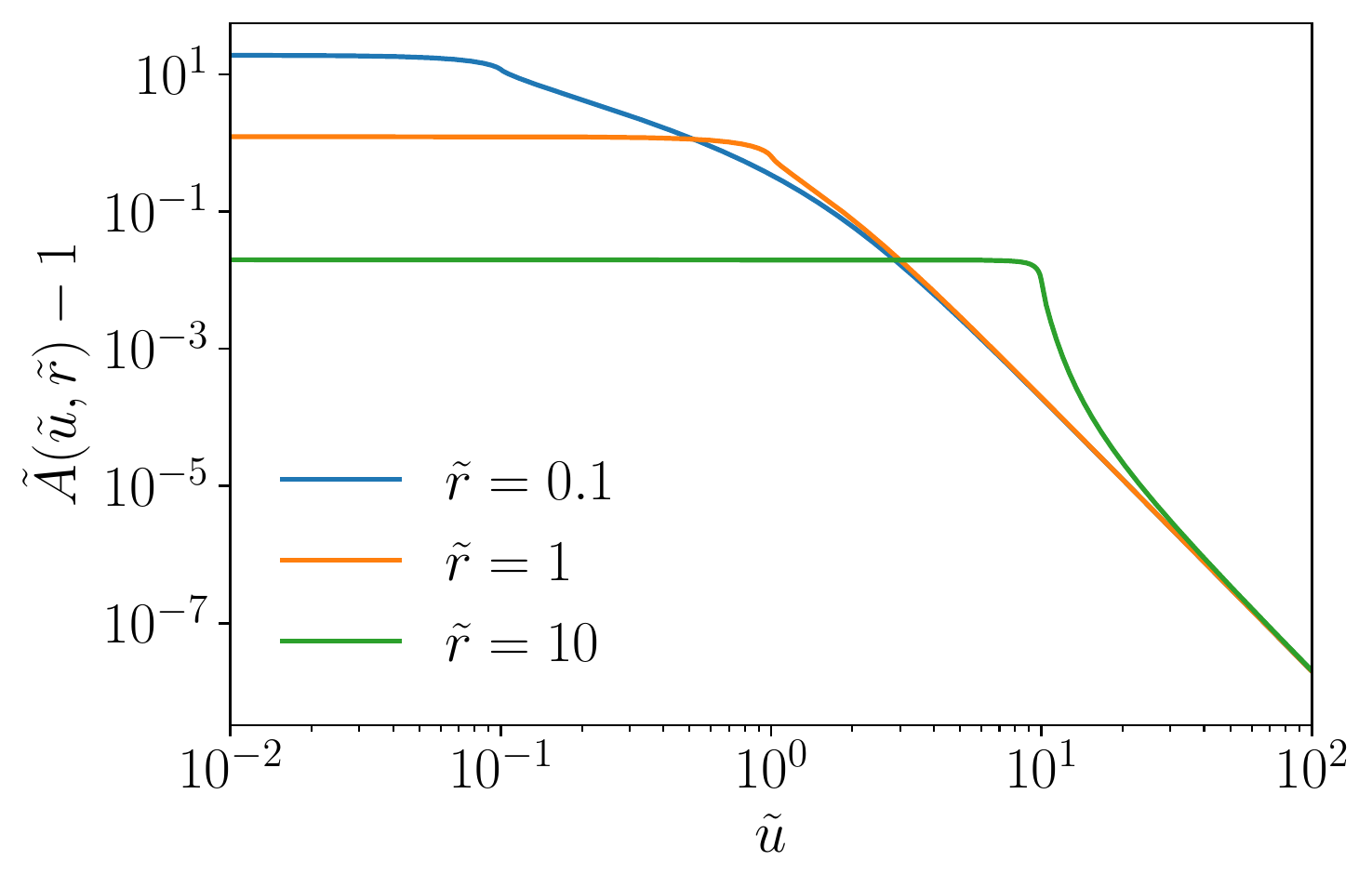}
\hfill
\includegraphics[width=0.49\columnwidth]{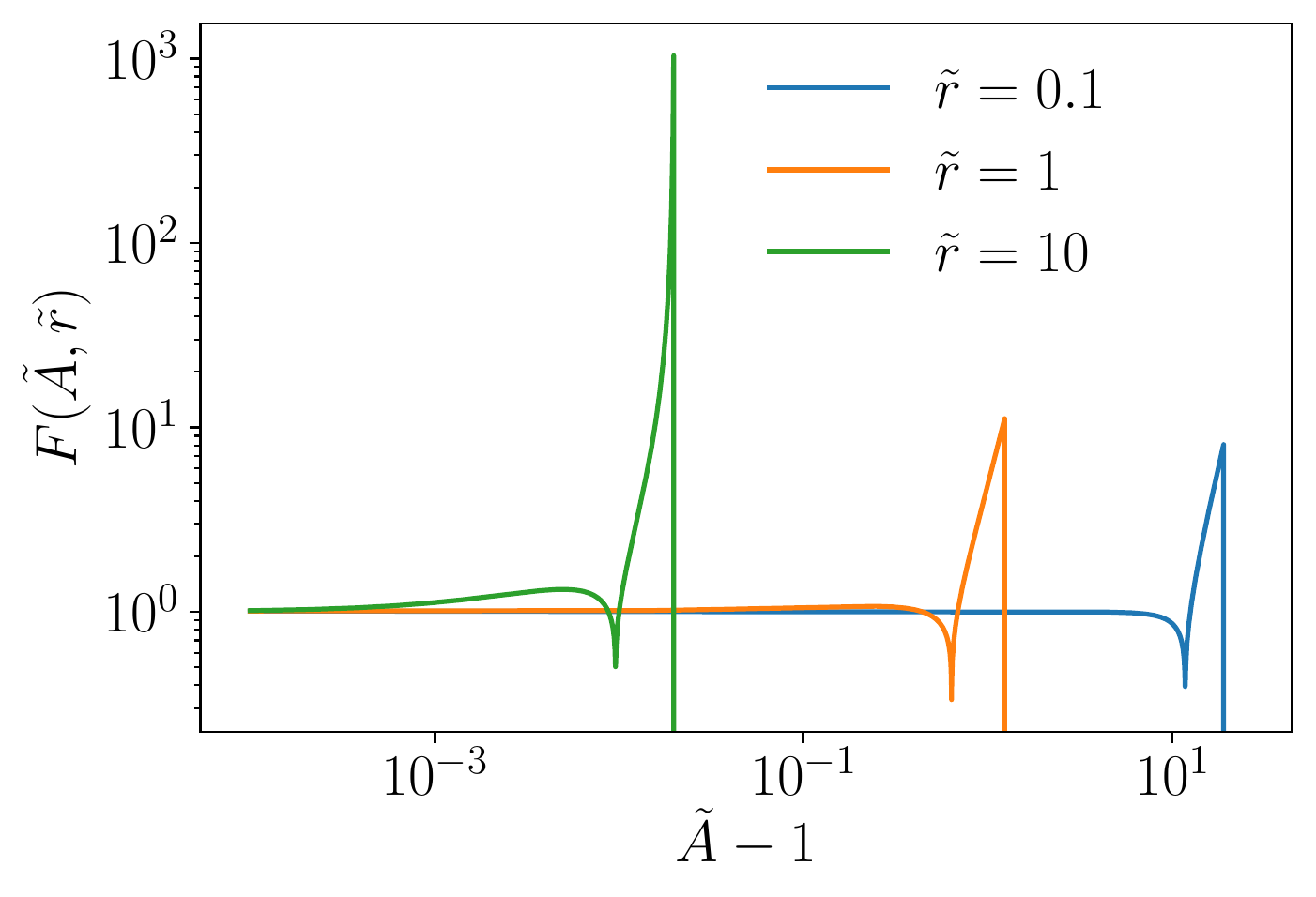}
\caption{Finite-source corrections for the amplification of a homogeneous disk source with angular radius $\tilde{\sigma}=\tilde{r}\tilde{\theta}\e{E}$ by a point lens with Einstein radius $\tilde{\theta}\e{E}$. 
\textit{Left}: Amplification profile $\tilde{A}$ as a function of the reduced impact parameter~$\tilde{u}=\tilde{\beta}/\tilde{\theta}\e{E}$. The larger the source, the smoother the profile and the smaller the maximum amplification.
\textit{Right}: Correction factor to the differential amplification cross section, $F\define\tilde{\Omega}_{\tilde{\sigma}}/\tilde{\Omega}_0$, for the same values of $\tilde{r}$.}
\label{fig:extended_sources_one_lens}
\end{figure}

Let us now consider a source shaped as a disk with angular radius $\sigma$, and whose surface brightness is homogeneous within the disk. Since we are neglecting the shear, the source shape is still a disk in the twiddled world, with radius $\tilde{\sigma}=(1-\kappa\e{od})(1-\kappa\e{ds})^{-1}(1-\kappa\e{os})^{-1}\sigma$. The amplification profile of such a homogeneous disk source by a point lens was derived in ref.~\cite{1994ApJ...430..505W}. If $\tilde{\beta}$ denotes, in the twiddled world, the angle between the centre of the source and the main lens, we define the reduced impact parameter as $\tilde{u}\define\tilde{\beta}/\tilde{\theta}\e{E}$ and the reduced source's radius $\tilde{r}\define\tilde{\sigma}/\tilde{\theta}\e{E}$; the amplification profile then reads

\begin{multline}
\label{eq:A_u_extended}
\tilde{A}(\tilde{u}, \tilde{r}) 
=
\frac{\tilde{u} + \tilde{r}}{2\pi\tilde{r}^2}
\sqrt{4 + (\tilde{u} - \tilde{r})^2} \, \mathrm{E}(m)
-
\frac{\tilde{u}-\tilde{r}}{2\pi\tilde{r}^2}
\frac{8+(\tilde{u}^2-\tilde{r}^2)}{\sqrt{4+(\tilde{u}-\tilde{r})^2}} \, \mathrm{F}(m)
\\
+
\frac{2(\tilde{u}-\tilde{r})^2}{\pi\tilde{r}^2(\tilde{u}+\tilde{r})}\frac{1+\tilde{r}^2}{\sqrt{4+(\tilde{u}-\tilde{r})^2}} \, \Pi(n,m) \ ,
\end{multline}
where
\begin{equation}
n \equiv \frac{4 \tilde{u} \tilde{r}}{(\tilde{u}+\tilde{r})^2} \ ,
\qquad
m \equiv \frac{4n}{4 + (\tilde{u} - \tilde{r})^2}  \ ,
\end{equation}
and the functions $\mathrm{F}$, $\mathrm{E}$ and $\Pi$ are the complete elliptic integrals of the first, second and third type, respectively, in Wolfram's convention for elliptic integrals\footnote{\href{https://reference.wolfram.com/language/guide/EllipticIntegrals.html}{\tt https://reference.wolfram.com/language/guide/EllipticIntegrals.html}}. The maximum amplification is obtained when $\tilde{u} = 0$ and reads
\begin{equation}
\label{eq:maxA}
\tilde{A}_{\mathrm{max}}(\tilde{r})
\equiv
\tilde{A}(0, \tilde{r} ) = \sqrt{1 + \frac{4}{\tilde{r}^2}} \ ,
\end{equation}
which goes to infinity as the source becomes very small ($\tilde{r}\to 0$). More generally, the entire amplification profile of the point-source case~\eqref{eq:A_u} is recovered in that limit. The left panel of \cref{fig:extended_sources_one_lens} illustrates the amplification profile for several values of $\tilde{r}$.

In the next subsubsections, we proceed with the calculation of the amplification probability in the presence of finite-source corrections. This calculation will closely follow the point-source case: amplification cross-section; strongest-lens approximation; and marginalisation over the mesoscopic cone.

\subsection{Amplification cross section of an isolated lens}

With an extended source, the amplification profile changes from \cref{eq:A_u} to \cref{eq:A_u_extended}, so the differential cross section amplification must change as well. In particular, since $\tilde{A}\leq \tilde{A}\e{max}(\tilde{r})$, we must have $\tilde{\Omega}_{\tilde{\sigma}}(\tilde{A}>\tilde{A}\e{max})=0$. Thanks to the axial symmetry of the amplification profile, just like the point-lens case, we have
\begin{equation}
\tilde{\Omega}_{\tilde{\sigma}}(\tilde{A})
= 2\pi \tilde{\beta}(\tilde{A}, \tilde{r}) \abs{\pd{\tilde{\beta}}{\tilde{A}}}
= \pi\tilde{\theta}\e{E}^2 \abs{\frac{\partial\tilde{u}^2}{\partial\tilde{A}}} ,
\end{equation}
except that now $\tilde{u}(\tilde{A}, \tilde{r})$ is the inverse of \cref{eq:A_u_extended} at fixed $\tilde{r}$, which cannot be done analytically.

For later convenience, we introduce the finite-source factor~$F$ as the ratio between the finite-source and point-source amplification cross sections,
\begin{equation}
F(\tilde{A}, \tilde{r})
\define
\frac{\tilde{\Omega}_{\tilde{\sigma}}(\tilde{A})}{\tilde{\Omega}_0(\tilde{A})}
=
\frac{\partial\tilde{\beta}^2}{\partial\tilde{A}} \,
\frac{\dd\tilde{A}}{\dd\tilde{\beta}_0^2} \ .
\end{equation}
This way, finite-source corrections are fully encapsulated in a single function, just like shear corrections were encapsulated in the function~$k$ in \cref{subsec:amplification_cross_section}. A significant difference, however, is that the function~$F$ has two variables while $k$ had only one, which makes the analysis technically harder. The right panel of \cref{fig:extended_sources_one_lens} shows three examples of $\tilde{A}\mapsto F(\tilde{A}, \tilde{r})$, for $\tilde{r}=0.1, 1, 10$. As expected, for low amplifications $F\approx 1$, which translates the fact that far from the lens, the finiteness of the source has essentially no effect. For larger amplifications, the cross section is enhanced for $\tilde{A}\lesssim\tilde{A}\e{max}(\tilde{r})$ and then suddenly drops to zero beyond.

The amplification cross section in the original problem (non-twiddled world) is obtained from $\tilde{\Omega}(\tilde{A})$ similarly to \cref{subsubsec:back_original_problem}. The calculation uses that, in terms of the original quantities, $\tilde{r}=\sqrt{A\e{min}}\,\sigma/\vartheta\e{E}$. The final result is
\begin{empheq}[box=]{equation}
\label{eq:cross_section_final_extended}
\Omega_\sigma(A)
=
2\pi\vartheta\e{E}^2 \,
F\pa{\frac{A}{A\e{min}}, \sqrt{A\e{min}} \, \frac{\sigma}{\vartheta\e{E}}}
\frac{A\e{min}^2}{(A^2 - A\e{min}^2)^{3/2}}
\ .
\end{empheq}

\subsection{Amplification probabilities with extended sources}

Just like in the point-lens case, the amplification PDF for one lens in a mesoscopic cone with half angle $\Theta$ reads
\begin{equation}
\label{eq:p_1_definition_extended}
p_1(A; \sigma)
=
\frac{1}{\pi\Theta^2}
\int \dd m \, \dd\chi \, \; p(m, \chi) \,
\Omega_\sigma(A; m, \chi) \ .
\end{equation}
Substituting \cref{eq:cross_section_final_extended} and $p(m, \chi)=(3\chi^2/\chi\e{s}^3) \, p(m)$, we may again gather all the finite-source corrections within a single function as
\begin{empheq}[box=]{equation}
\label{eq:p_1_final_extended}
p_1(A; \sigma)
=
\frac{2\ev{\vartheta\e{E}^2}}{\Theta^2} \,
\frac{A\e{min}^2}{(A^2-A\e{min}^2)^{3/2}} \,
\mathcal{F}\pa{\frac{A}{A\e{min}}, \sigma, \kappa\e{os}} \ ,
\end{empheq}
with $\mathcal{F}\define\ev{\vartheta\e{E}^2 F}/\ev{\vartheta\e{E}^2}$, that is, explicitly,
\begin{equation}
\mathcal{F}(X, \sigma, \kappa\e{os})
=
\frac{
    \int_0^{\chi\e{s}} \frac{\dd\chi}{a(\chi)} \;
    \frac{1-\kappa\e{ds}}{1-\kappa\e{od}} \,
    \chi(\chi\e{s}-\chi)
    \int \dd m \; \frac{m}{\ev{m}} \,
    F\pac{X,
        \sigma\sqrt{\frac{1-\kappa\e{od}}
                        {(1-\kappa\e{os})(1-\kappa\e{ds})}
                    \frac{a(\chi)\chi\chi\e{s}}
                        {4Gm(\chi\e{s}-\chi)}
                    }
        }
    }
    {
    \int_0^{\chi\e{s}} \frac{\dd\chi}{a(\chi)} \;
    \frac{1-\kappa\e{ds}}{1-\kappa\e{od}} \,
    \chi(\chi\e{s}-\chi)
    } \ ,
\end{equation}
where $X=A/A\e{min}$. \Cref{eq:p_1_final_extended} is formally quite similar to \cref{eq:p_1_final}, which was the case of point-like sources with external shear. However, here the correction factor~$\mathcal{F}(X, \sigma, \kappa\e{os})$ has three variables, instead of one for the function~$\mathcal{K}(x)$ of \cref{eq:p_1_final}

\begin{figure}[t]
\centering
\includegraphics[width=0.49\columnwidth]{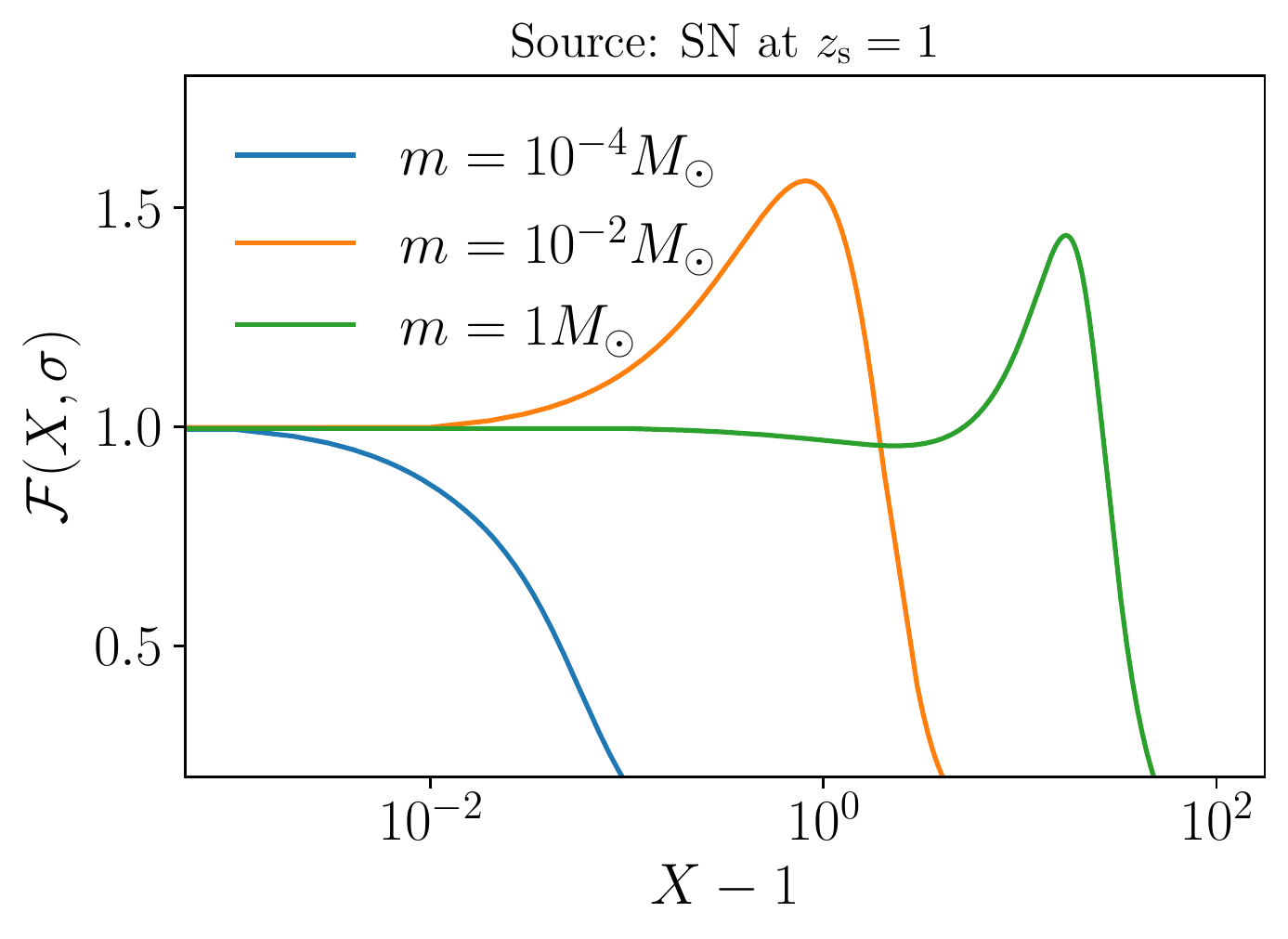}
\hfill
\includegraphics[width=0.49\columnwidth]{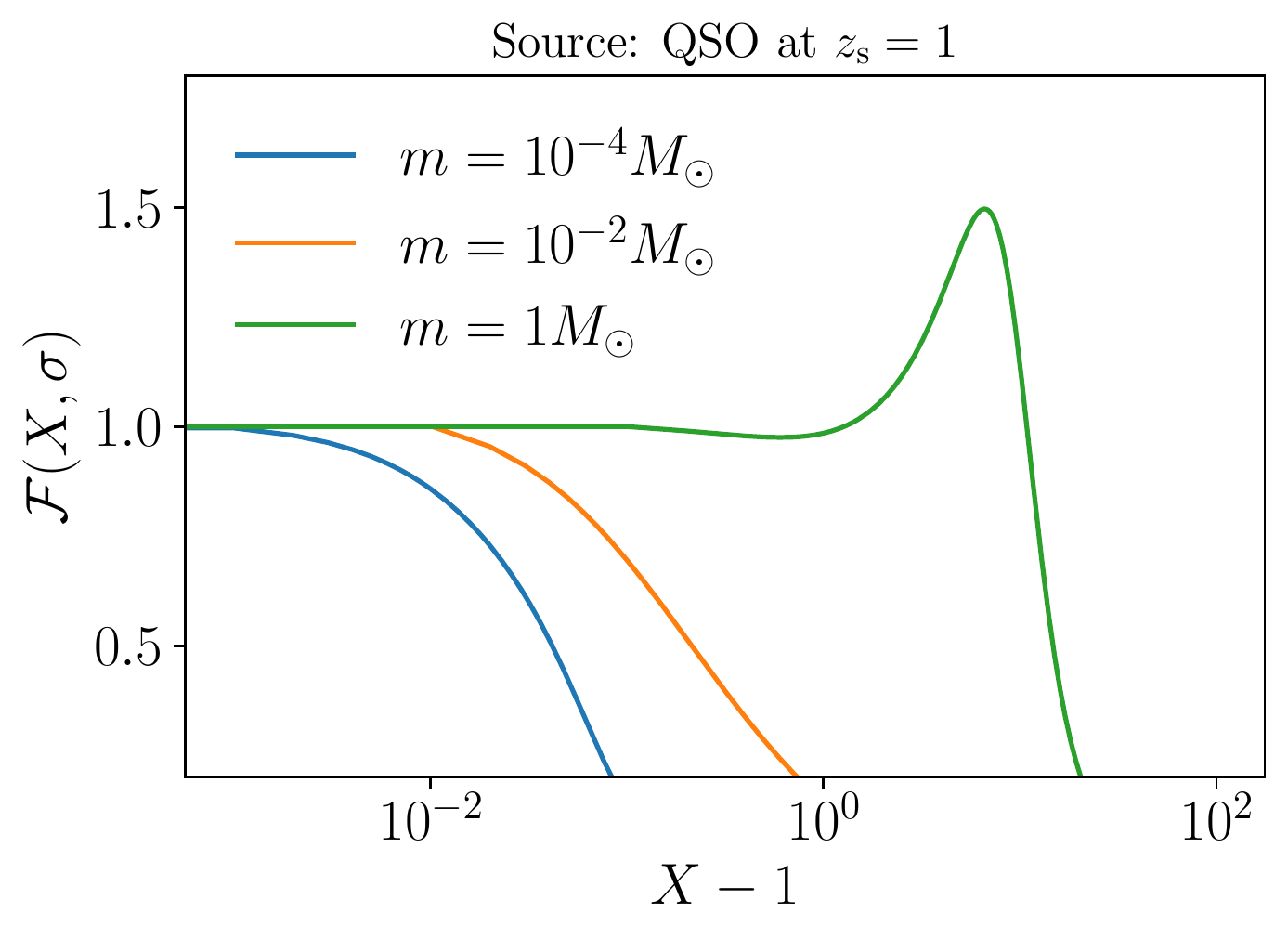}
\caption{Examples of the extended-source correction factor~$\mathcal{F}(X, \sigma, \kappa\e{os}=0)$ involved in \cref{eq:p_1_final_extended}, for two different kind of sources at $z\e{s}=1$, in the case where all the compact objects have the same mass $m$. \textit{Left}: The sources are SNe, with physical radius \SI{300}{\astronomicalunit}, i.e. $\sigma\e{SN}=\mathcal{O}(10^{-13})$ rad at $z\e{s}=1$. \textit{Right}: The sources are QSOs, with physical radius 4 light days, i.e. $\sigma\e{QSO}=\mathcal{O}(10^{-9})$ rad at $z\e{s}=1$.}
\label{fig:integral_cal_F}
\end{figure}

\Cref{fig:integral_cal_F} shows examples of the $\mathcal{F}$ function for two types of sources (SNe and QSOs)\footnote{The typical size of type-Ia SNe can be inferred from the typical expansion velocity of \SI{20000}{\kilo\meter\per\second} of the luminous envelope about a month after explosion, which gives around 2~light-days, or \SI{300}{\astronomicalunit}. The size of the inner region of QSO, which suffers the effect of microlensing, is also about 4 to 8~light-days~\cite{Esteban-Gutierrez:2022pql,Hawkins:2022vqo}.} at $z\e{s}=1$, in the case where all the compact objects have the same mass $m$; we have set all the convergences to zero for simplicity. As expected, $\mathcal{F}$ being a smoothed version of $F$, it preserves some of its features; in particular, the amplification probability is suppressed beyond a critical value of $X=A/A\e{min}$. For a given source size~$\sigma$, the lower the deflectors' mass~$m$, the larger the values of $r$ and hence the smaller the critical amplification; this is apparent in both panels of \cref{fig:integral_cal_F}, where the curves are displaced to the left as $m$ decreases. For relatively large values of the deflectors' mass, $\mathcal{F}(X, \sigma)$ is almost self-similar. As expected, the lens mass required to allow large amplifications to happen is much larger for QSOs than for SNe, because the latter is much closer to a point source than the former.

Finally, the total amplification CDF, produced by an infinite population of lenses in the mesoscopic cone, is derived from \cref{eq:p_1_final_extended} following the exact same method as in \cref{sec:strongest}, i.e. in the framework of the strongest-perturbed lens approximation. The result is
\begin{empheq}[box=\fbox]{equation}
P(A; \sigma, z\e{s}, \alpha, \bar{\kappa}\e{os})
= 1 - \exp\paac{
                \frac{-2\tau}{1-\kappa\e{os}} \int_{A/A\e{min}}^{\infty}
                \frac{ \; \dd X}
                {(X^2-1)^{3/2}} \,
                \mathcal{F}(X, \sigma, \kappa\e{os})
                } ,
\label{eq:CDF_final_extended}
\end{empheq}
with
$\tau = \alpha(\Delta\e{os}+\bar{\kappa}\e{os})$,
$\kappa\e{os} = (1-\alpha)\bar{\kappa}\e{os}-\alpha\Delta\e{os}$
and
$A\e{min} = (1-\kappa\e{os})^{-2}$. Averaging over the mesoscopic cone is obtained by marginalising over $\bar{\kappa}\e{os}$, as discussed in \cref{subsec:marginalising_over_LOS_convergence}.

\begin{figure}
\centering
\includegraphics[width=0.49\columnwidth]{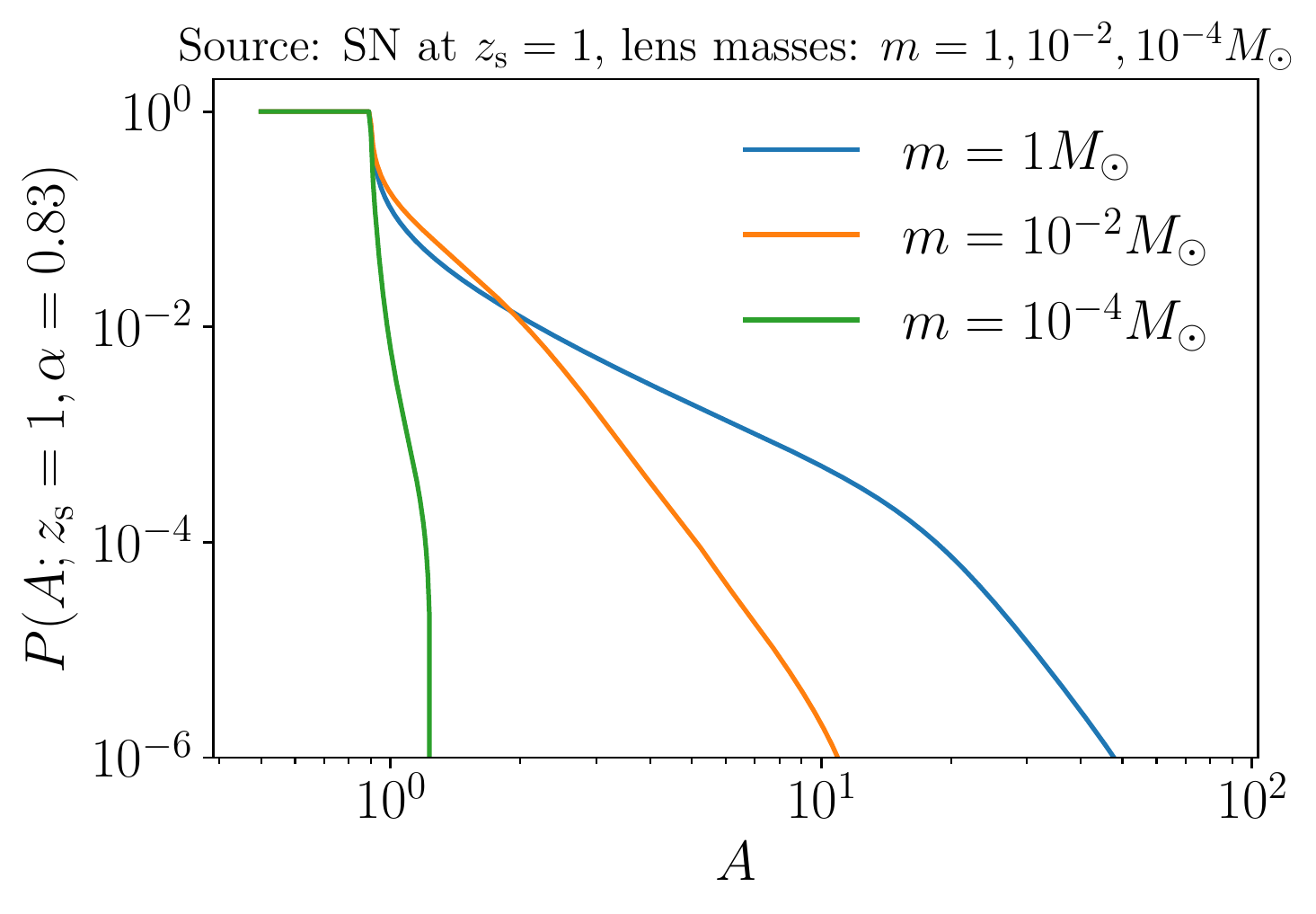}
\hfill
\includegraphics[width=0.49\columnwidth]{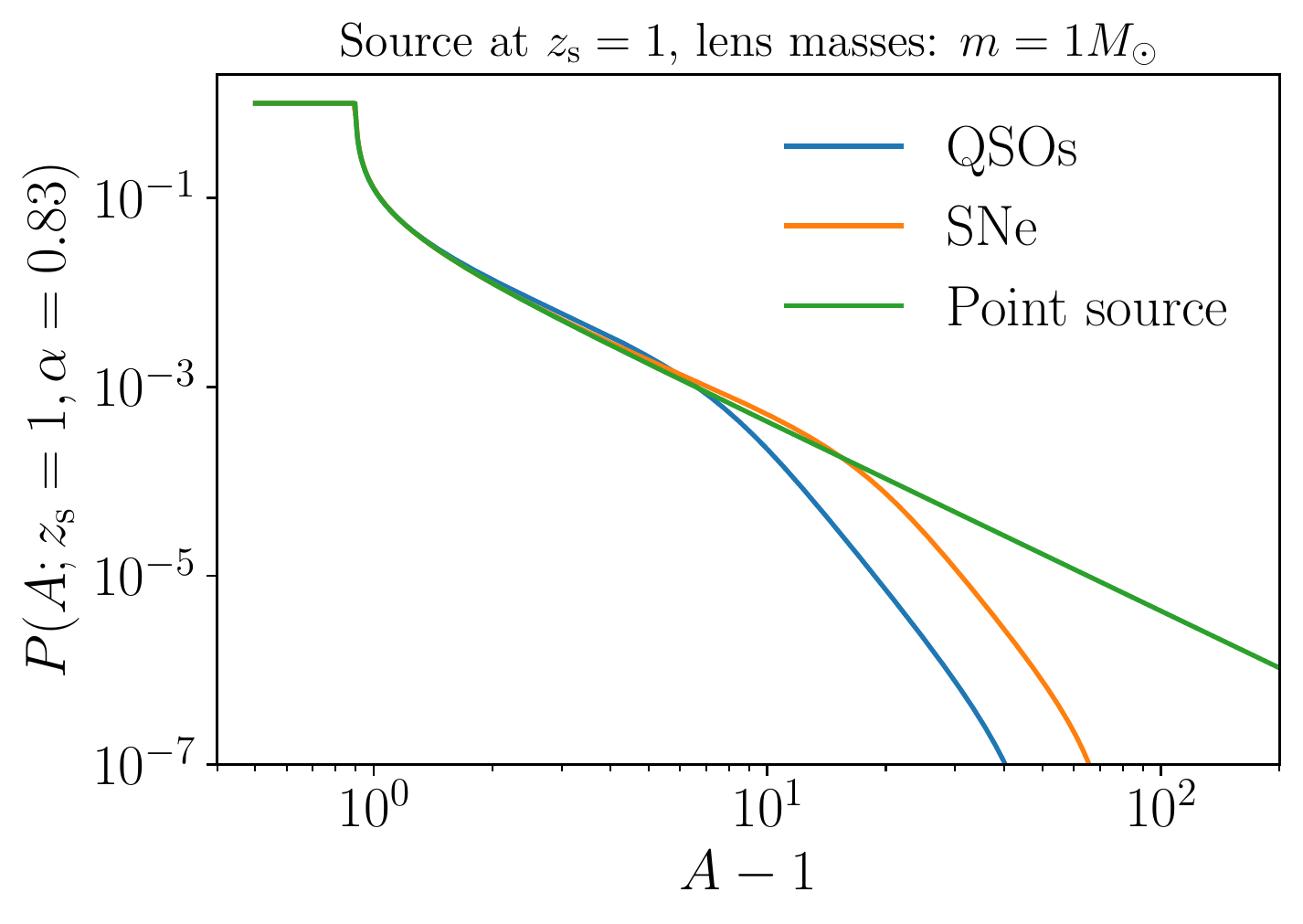}\\
\includegraphics[width=0.49\columnwidth]{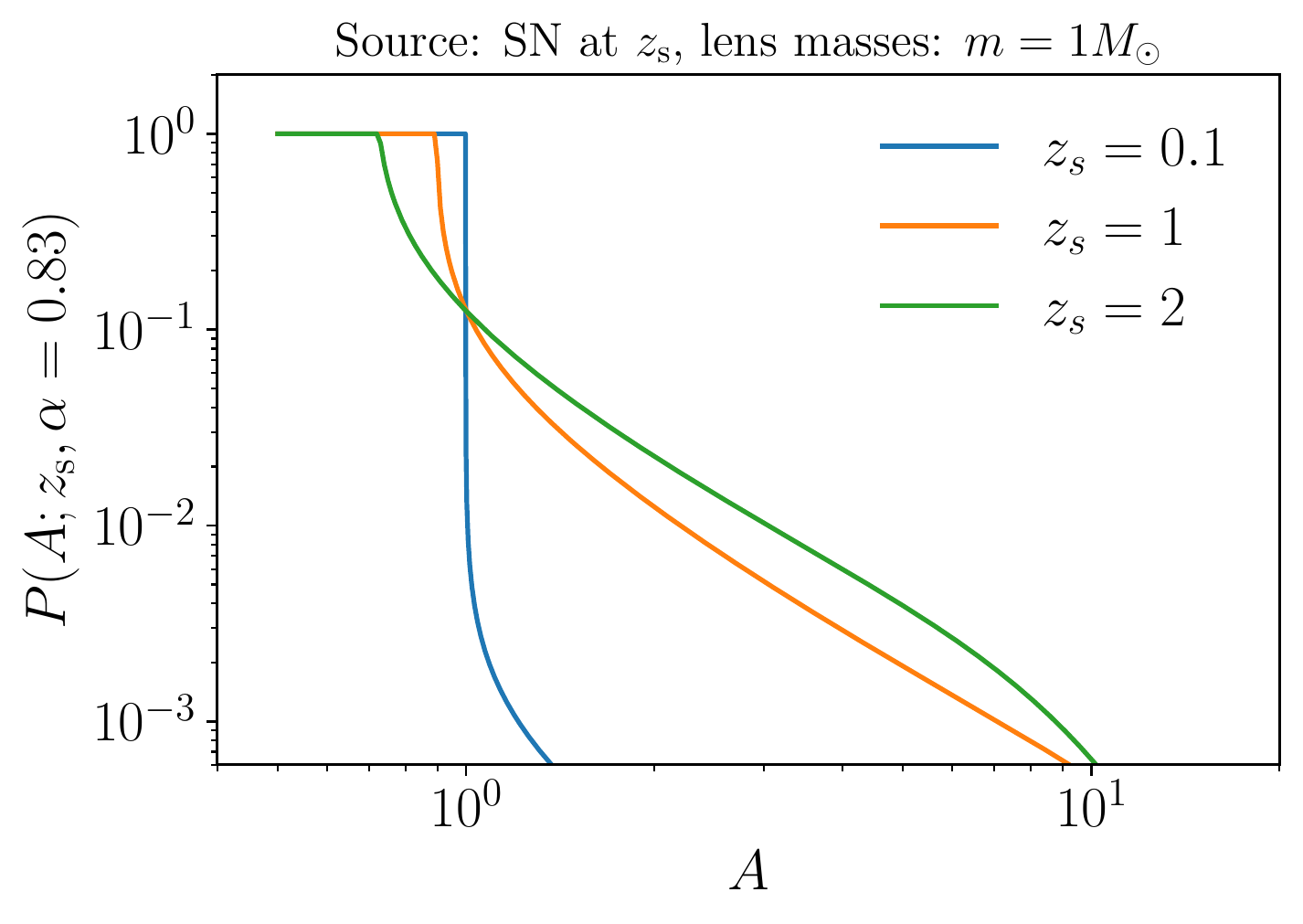}
\hfill
\includegraphics[width=0.49\columnwidth]{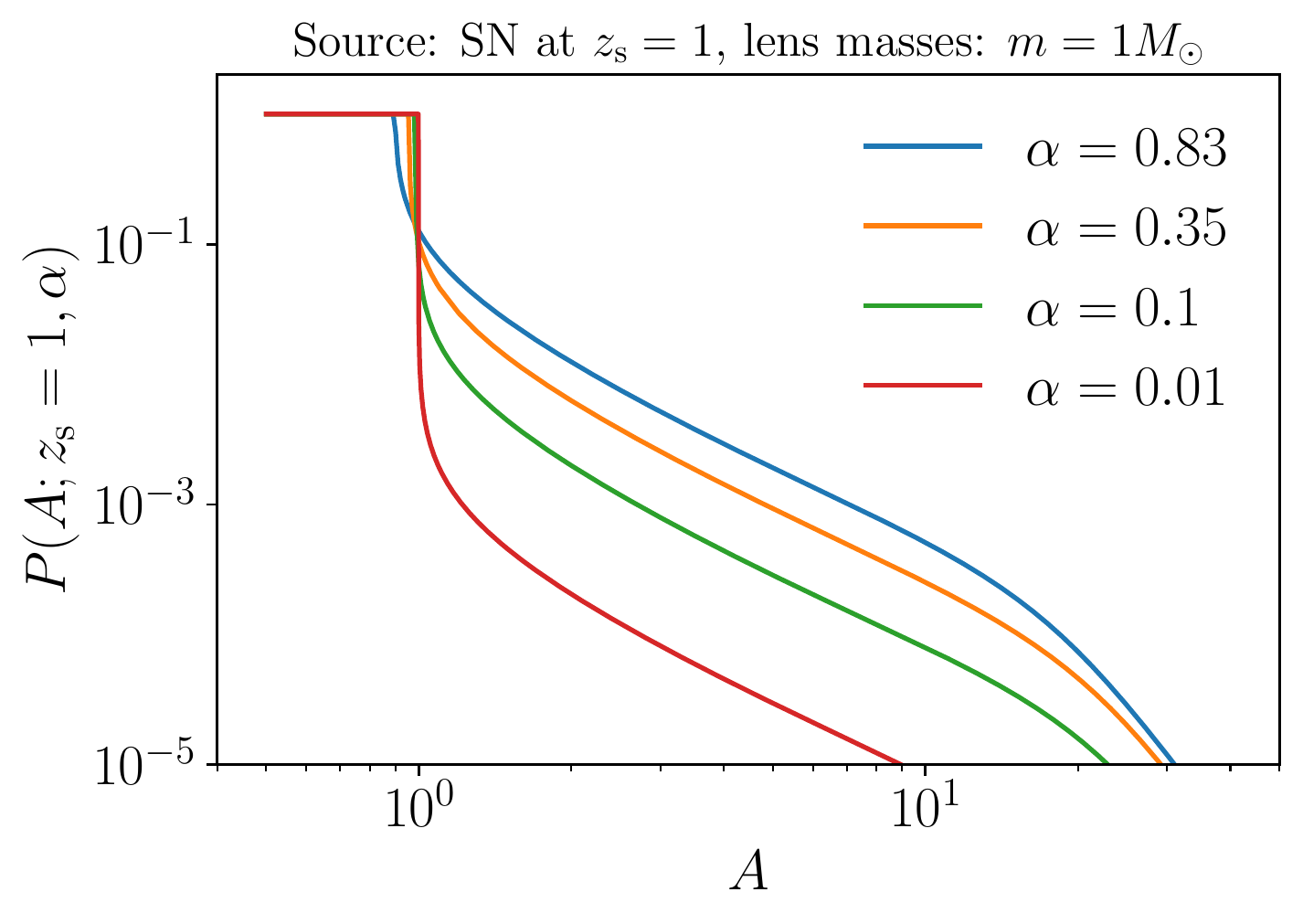}
\caption{Examples of the amplification CDF $P(A; \sigma, z\e{s}, \alpha, \bar{\kappa}\e{os})$ with extended sources, as given by \cref{eq:CDF_final_extended}, for a line of sight with $\bar{\kappa}\e{os}=0$. \textit{Top left:} Varying the mass~$m$ of the compact objects. \textit{Top right:} Changing the source type, namely point sources, SNe [physical radius \SI{300}{\astronomicalunit} i.e. $\sigma\e{SN}=\mathcal{O}(10^{-13})$ at $z\e{s}=1$], and QSOs [physical radius 4 light days i.e. $\sigma\e{QSO}=\mathcal{O}(10^{-9})$ at $z\e{s}=1$]. \textit{Bottom left}: Varying the source redshift~$z\e{s}$. \textit{Bottom right}: Varying the fraction~$\alpha$ of compact objects.}
\label{fig:CDF_extended}
\end{figure}

Examples of $P(A; z\e{s}, \alpha, \bar{\kappa}\e{os}, \sigma)$ for various values of its parameters are depicted in \cref{fig:CDF_extended}; we fixed $\bar{\kappa}\e{os}=0$ for simplicity. Compared to the point-source case, the amplification probability is slightly enhanced near some critical value that depends on the lens masses, source redshift and size, after what it gets quickly suppressed. As expected from our analysis of the function~$\mathcal{F}$, finite-source effects are stronger as the source size increases and the mass of the compact objects decreases.

\section{Conclusion}
\label{sec:conclusion}

Extragalactic microlensing is a potentially powerful probe of the nature of dark matter (DM). In particular, a handful of past studies have used supernova microlensing to set constraints on the fraction of intergalactic DM that could be made of compact objects. Those analyses relied on a simple phenomenological modelling of the microlensing amplification statistics, which did not account for the coupling between the main deflector responsible for the amplification and its environment -- the other lenses and the large-scale cosmic structures. Such an approximation is expected to be valid in the limit of very low optical depths, and for lines of sight that are representative of the mean homogeneous and isotropic model.

In this work, we started assessing the validity of the very-low-optical-depth assumption, and found that for observationally interesting values of the microlensing amplification, relevant optical depths are low to mild ($\tau\lesssim 0.1$). This first result, together with the known fact that environmental effects are generally non-negligible in strong lensing, suggests that environmental and line-of-sight corrections may be significant in extragalactic microlensing. Hence, they must be taken into account in order to accurately predict the probability of microlensing amplification by a cosmic population of compact objects.

We have derived, from first principles, an expression for the amplification probability that we expect to be valid up to mild optical depths. Our approach, which may be referred to as the ``strongest perturbed lens model'', consistently accounts for: (i) the external convergences due to overdensities or underdensities in the smooth matter distribution along the line of sight; and (ii) the external shears produced by the large-scale structure and the lenses near the line of sight. This result and its derivation constitute the main focus of the article. The derivation was performed in the case of point-like sources of light, but we also explicitly derived the extended-source corrections for completeness. In numerical illustrations, the statistical distributions of the line-of-sight convergences and shears were extracted from ray tracing in $N$-body simulations, for which we found interesting fitting functions.

From this new model of microlensing amplification probabilities, two conclusions turn out to be particularly noteworthy. First, in observationally relevant situations, the effect of external shear (both due to the large-scale structure and to compact objects near the line of sight) is statistically negligible -- corrections are at most on the order of a part in a thousand. Second, however, the predictions of our model are still quantitatively discrepant from the literature, with relative differences larger than \SI{100}{\percent} in some cases. Such differences might be explained from our non-linear treatment of the external convergences and our careful embedding of microlenses within the cosmic large-scale structure. This result emphasises the crucial importance of an elaborate theoretical modelling of amplification statistics in order to extract accurate constraints on the fraction of compact objects in the Universe.

The next step of this work naturally consists in applying its result to, e.g., SN data similarly to what was done in refs.~\cite{Seljak:1999tm, 1999ApJ...519L...1M, 2007PhRvL..98g1302M, Zumalacarregui:2017qqd, Garcia-Bellido:2017imq}. This will require an efficient numerical implementation of our model for the amplification probability, which can be technically challenging for extended-source corrections. Application to data requires to properly deal with their outliers, in order to distinguish between lensed and intrinsically anomalous SNe.

\section*{Acknowledgements}

We warmly thank Michel-Andrès Breton for his help with the ray-tracing data for convergence and shear. VB received the support of a fellowship from ``la Caixa'' Foundation (ID 100010434). The fellowship code is LCF/BQ/DI19/11730063.  PF received the support of a fellowship from ``la Caixa'' Foundation (ID 100010434). The fellowship code is LCF/BQ/PI19/11690018. The authors  acknowledge support from the Research  Project  PGC2018-094773-B-C32  and  the Centro de Excelencia Severo Ochoa Program SEV-2016-0597.

\appendix

\section{Weak-lensing statistics with RayGalGroupSims}
\label{app:fit_simulation}

This appendix is dedicated to our analysis of the statistics of weak-lensing convergence and shear from a numerical simulation. Specifically, we have used results from a dark-matter-only simulation performed with the $N$-body code RAMSES~\cite{teyssier2002cosmological, guillet2011simple}. The simulation has been performed with the best-fit parameters of WMAP-7~\cite{komatsu2011seven}, a comoving length of $2625h^{-1}~\text{Mpc}$ and a particle mass of $1.88 \times 10^{10}h^{-1} M_\odot$. Fully relativistic ray tracing has been performed through this simulation~\cite{Breton:2018wzk} using the \textsc{Magrathea} library~\cite{reverdy2014propagation, Breton}. Healpix maps with various lensing quantities, such as convergence, shear and magnification, are publicly available.\footnote{\href{https://cosmo.obspm.fr/public-datasets/raygalgroupsims-relativistic-halo-catalogs}{\tt https://cosmo.obspm.fr/public-datasets/raygalgroupsims-relativistic-halo-catalogs}} We focus here on the PDF of convergence~$\bar{\kappa}\e{os}$ and (macro)shear $\bar{\gamma}\e{os}$.

\subsection{Convergence}
\label{app:fit_simulations_convergence}

\begin{figure}
\centering
\includegraphics[width=0.49\columnwidth]{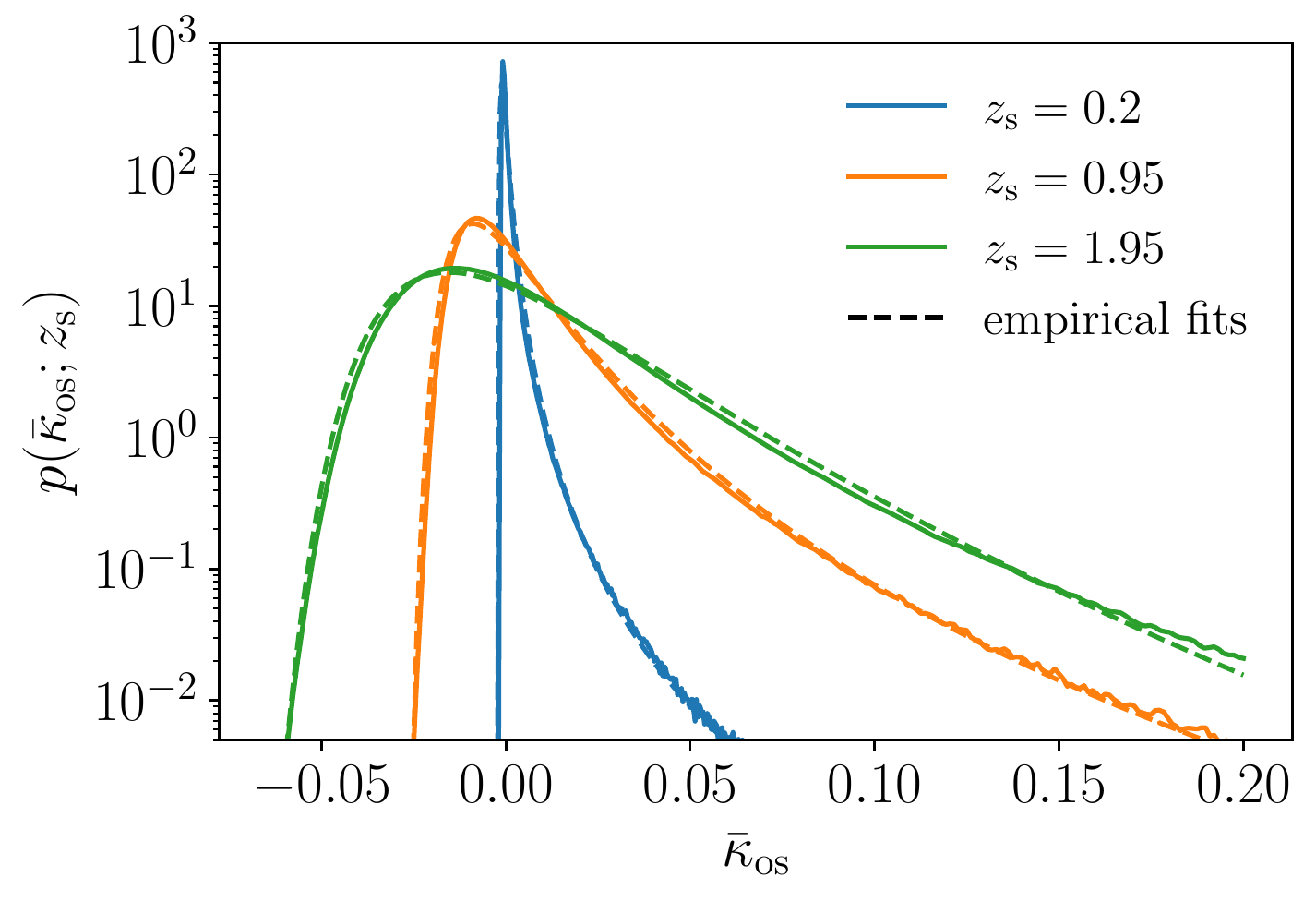}
\hfill
\includegraphics[width=0.49\columnwidth]{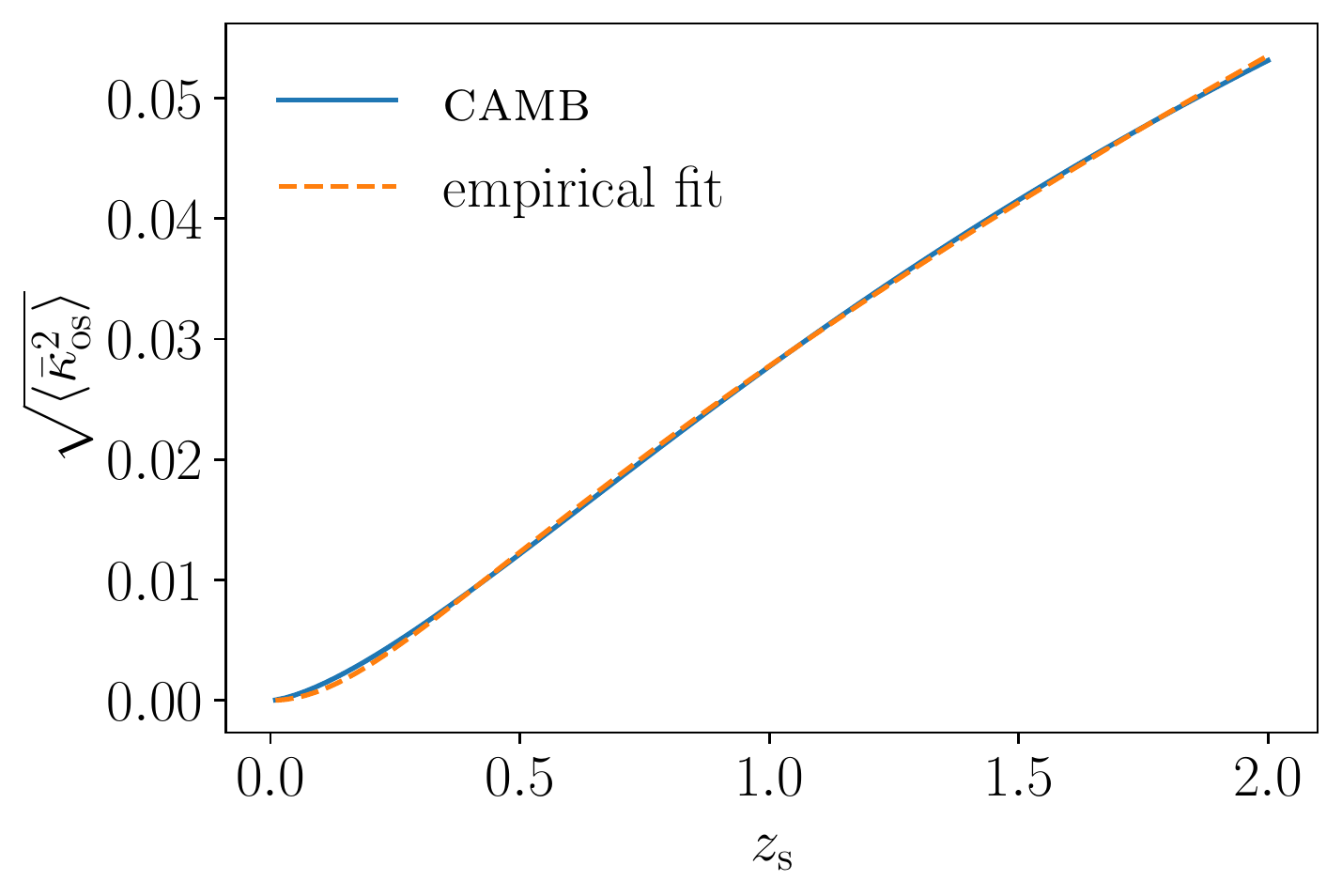}
\caption{Statistics of the weak-lensing convergence~$\bar{\kappa}\e{os}$. \textit{Left}: comparison between the PDF of the convergence~$\bar{\kappa}\e{os}$ obtained from ray tracing in an $N$-body simulation (solid lines) and the ansatz of \cref{eq:PDF_kappa_ansatz} (dashed lines), with the constraint~\eqref{eq:constraint_convergence}, $\nu(z\e{s})=2.3(1+z\e{s})$ and $\Delta\kappa$ left as a free parameter. \textit{Right}: standard deviation of the convergence computed from \textsc{camb} with a \emph{Planck}-2018 cosmology, compared with the empirical fit of \cref{eq:sigma_kappa_fit}.}
\label{fig:stats_kappa}
\end{figure}

We analysed the PDF of the weak-lensing convergence~$\bar{\kappa}\e{os}$ obtained by ray tracing. In the redshift range $z\e{s}<2$, we found that the following ansatz provides a good fit to the data,
\begin{equation}
\label{eq:PDF_kappa_ansatz}
p(\bar{\kappa}\e{os}; z\e{s})
= \ddf{}{\bar{\kappa}\e{os}} \exp
    \paac{ - \pac{
                \frac{\Delta\kappa(z\e{s})}
                    {\bar{\kappa}\e{os}-\kappa_0(z\e{s})}
                }^{\nu(z\e{s})}
            + \pac{
                \frac{\Delta\kappa(z\e{s})}
                    {1-\kappa_0(z\e{s})}
                }^{\nu(z\e{s})}
        } ,
\end{equation}
where the parameters $\nu$, $\kappa_0$ and $\Delta\kappa$ depend on the source redshift~$z\e{s}$. Note that \cref{eq:PDF_kappa_ansatz} is normalised to 1 for $\bar{\kappa}\e{os}\in[\kappa_0,1]$ by definition; $\kappa_0(z\e{s})<0$ thus denotes the minimum convergence for sources at a redshift $z\e{s}$. Imposing that the convergence averages to $0$ imposes the following constraint between the model parameters,
\begin{equation}
\label{eq:constraint_convergence}
\kappa_0 = - \Gamma\pac{\frac{\nu-1}{\nu}} \Delta\kappa \ ,
\end{equation}
where $\Gamma$ denotes the usual Gamma function. Together with the above, we find that
\begin{equation}
\nu(z\e{s}) = 2.3 \, (1+z\e{s})
\end{equation}
fits well the data as $\Delta\kappa$ is left as a free parameter. The accuracy of this empirical fit is illustrated in the left panel of \cref{fig:stats_kappa}.

As could be guessed from \cref{eq:PDF_kappa_ansatz}, $\Delta\kappa$ is related to the variance of the convergence. Specifically, we have
\begin{equation}
\label{eq:variance_convergence_Delta_kappa}
\ev[2]{\bar{\kappa}\e{os}^2}
=
\pac{\Gamma\pa{\frac{\nu-2}{\nu}} - \Gamma^2\pa{\frac{\nu-1}{\nu}}} \Delta\kappa^2 \ .
\end{equation}
The variance of the convergence significantly depends on the cosmology. In the weak-lensing regime, at linear order and in Limber's approximation, it is known to read
\begin{equation}
\label{eq:variance_convergence_power_spectrum}
\ev[2]{\bar{\kappa}\e{os}^2}
=
\int_0^\infty \frac{\ell\dd\ell}{2\pi} \; P_\kappa(\ell, z\e{s}) \ ,
\end{equation}
where $P_\kappa$ denotes the convergence angular power spectrum, which is directly related to the matter power spectrum~\cite{Bartelmann:1999yn}. Since the simulation data at our disposal used slightly outdated cosmological parameters, we thus expect the resulting $\ev[2]{\bar{\kappa}\e{os}^2}$ to be outdated as well. In order to circumvent this issue, we estimated $\ev[2]{\bar{\kappa}\e{os}^2}$ from \cref{eq:variance_convergence_power_spectrum} using \textsc{camb}.\footnote{\href{https://camb.info/}{\tt https://camb.info/}} For a \textit{Planck}-2018 cosmology~\cite{Planck:2018vyg}, we find that the standard deviation of the convergence is well fit by
\begin{equation}
\label{eq:sigma_kappa_fit}
\sqrt{\ev[2]{\bar{\kappa}\e{os}^2}}
= 0.0218 \pac{ \pa{1 + 12.6\,z^2}^{0.315} - 1 },
\end{equation}
as illustrated in the right panel of \cref{fig:stats_kappa}. In practice, we substitute this expression into \cref{eq:variance_convergence_Delta_kappa} to determine $\Delta\kappa(z\e{s})$ for application in this article.

\subsection{Macroshear}
\label{app:fit_simulation_shear}

\begin{figure}
    \centering
    \includegraphics[width=0.49\columnwidth]{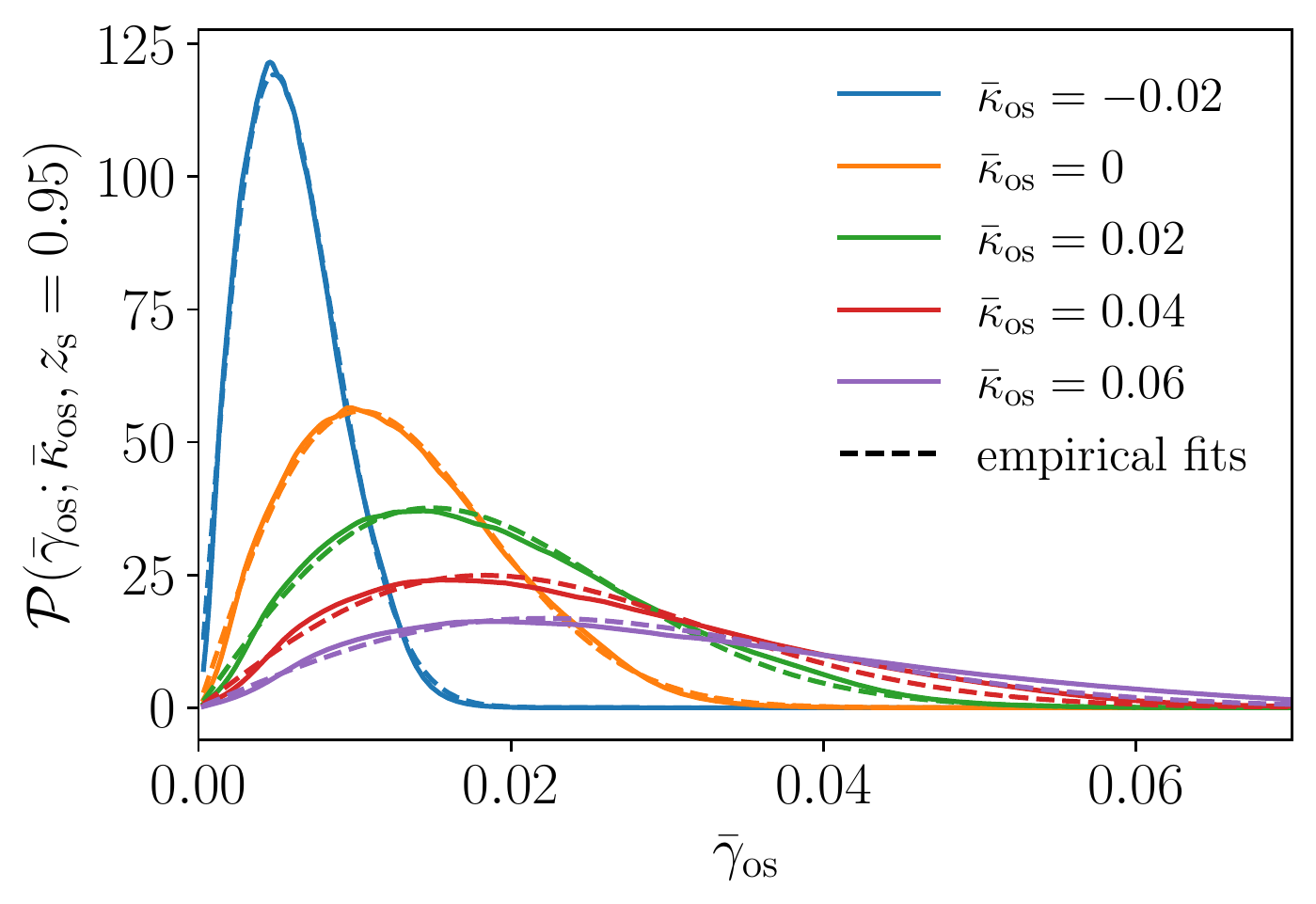}
    \hfill
    \includegraphics[width=0.49\columnwidth]{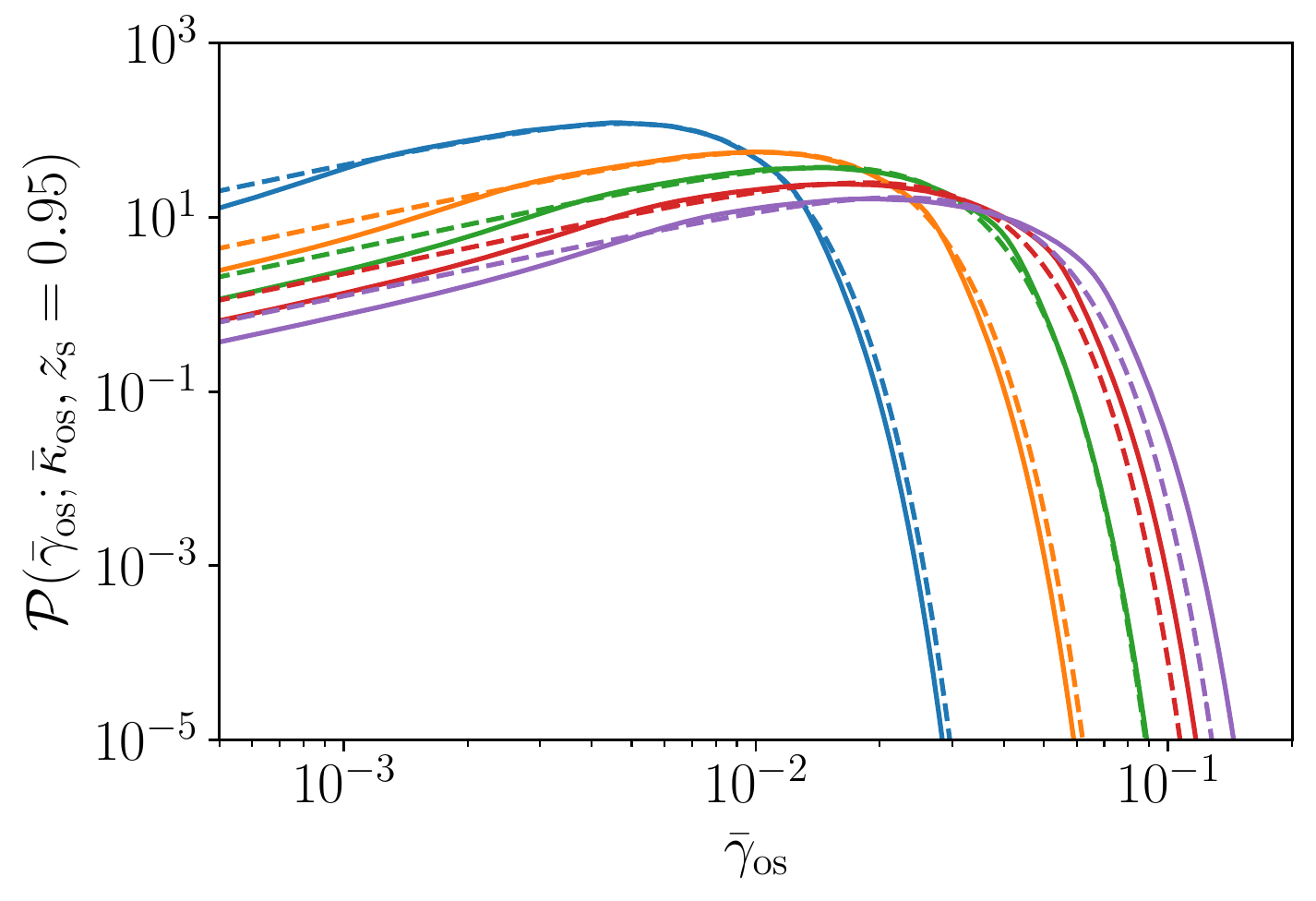}
    \caption{Conditional PDF of the magnitude of weak-lensing shear~$|\bar{\gamma}\e{os}|$ at fixed values of the convergence~$\bar{\kappa}\e{os}$, for sources at $z\e{s}=0.95$. The figures compare results from simulations (solid lines) with the ansatz of \cref{eq:PDF_macroshear_fit} (dashed lines). \emph{Left}: linear scale. \emph{Right}: logarithmic scale.}
    \label{fig:stats_shear}
\end{figure}

In the range of redshift relevant for the present discussion, we find that the conditional PDF for the shear at a fixed convergence, $p(\bar{\gamma}\e{os};\bar{\kappa}\e{os}, z\e{s})$, is surprisingly well fit by a two-dimensional Gaussian distribution,
\begin{equation}
p(\bar{\gamma}\e{os};\bar{\kappa}\e{os}, z\e{s}) \, \dd^2\bar{\gamma}\e{os}
= \frac{1}{2\pi \sigma^2(\bar{\kappa}\e{os}, z\e{s})}
    \exp\pac{ - \frac{|\bar{\gamma}\e{os}|^2}{2\sigma^2(\bar{\kappa}\e{os}, z\e{s})} }
    \dd^2\bar{\gamma}\e{os} \ .
\end{equation}
Since the Universe is statistically isotropic, there is no preferred orientation for the complex shear, and hence the conditional PDF of its magnitude takes the form
\begin{align}
\mathcal{P}(|\bar{\gamma}\e{os}|;\bar{\kappa}\e{os}, z\e{s})
&= 2\pi |\bar{\gamma}\e{os}| \, p(\bar{\gamma}\e{os};\bar{\kappa}\e{os}, z\e{s})
\\
&= \frac{|\bar{\gamma}\e{os}|}{\sigma^2(\bar{\kappa}\e{os}, z\e{s})}
    \exp\pac{ - \frac{|\bar{\gamma}\e{os}|^2}{2\sigma^2(\bar{\kappa}\e{os}, z\e{s})} }
\\
\label{eq:PDF_macroshear_fit}
&= \frac{\dd}{\dd |\bar{\gamma}\e{os}|}
    \exp\pac{ - \frac{|\bar{\gamma}\e{os}|^2}{2\sigma^2(\bar{\kappa}\e{os}, z\e{s})} } .
\end{align}
\Cref{fig:stats_shear} shows a comparison between the numerical data and the ansatz~\eqref{eq:PDF_macroshear_fit} for $z\e{s}=0.95$; at that redshift we find the empirical expression $\sigma(\bar{\kappa}\e{os}, z\e{s}=0.95)=0.01 + 0.26\,\bar{\kappa}\e{os}$.

\section{Derivation of the microshear distribution}
\label{app:microshear_distribution}

This appendix is dedicated to the derivation of the distribution of the effective reduced microshear that was given in \cref{eq:distribution_microshear}.

\subsection{PDF of a sum of complex shears}

Consider $N$ randomly distributed lenses $\ell$, each one producing a complex shear $s_\ell\in\mathbb{C}$. The total shear is the sum of all those contributions,
\begin{equation}
s = \sum_{\ell=1}^{N} s_\ell \ .
\end{equation}
The PDF of $s$ is, therefore, the convolution product of the $N$ PDFs of the individual shears. Assuming -- without loss of generality -- that the $N$ lenses are indistinguishable and have identical statistical properties, we have
\begin{equation}
\label{eq:convolution_shear_PDF}
p_N(s)
= \underbrace{(p_1 \star \ldots \star p_1)}_{\text{$N$ times}}(s)
\define p_1^{\star N}(s) \ ,
\end{equation}
where a $\star$ denotes a convolution product and $p_1$ is the PDF of the shear for $1$ lens, accounting for the randomness of its position, mass, etc. The convolution product is better handled in Fourier space. We define here the Fourier transform in a way that acknowledges the spin-$2$ character of the complex shear,
\begin{equation}
\tilde{p}(q) \define \int \dd^2s \; \ex{-2\ii \Re(q^*s)} \, p(s) \ ,
\qquad
p(s) = \int \frac{\dd^2 q}{(2\pi)^2} \; \ex{2\ii\Re(q^*s)} \, \tilde{p}(q) \ ,
\end{equation}
with $q\in\mathbb{C}$ the Fourier variable dual to $s$; ${}^*$ denotes complex conjugation and the differential elements are $\dd^2 q=\dd q_1 \dd q_2$, $\dd^2s=\dd s_1 \dd s_2$. We shall also use polar components for both $s=S\ex{2\ii\ph}$ and $q=Q\ex{2\ii\psi}$, with $\ph, \psi\in[0,\pi)$, in which case $\dd^2 s= 2S\dd S\dd\ph$ and $\dd^2 q = 2Q\dd Q\dd\psi$. In Fourier space with the above convention, \cref{eq:convolution_shear_PDF} becomes
\begin{equation}
\tilde{p}_N(q) = \tilde{p}_1^N(q) \ .
\end{equation}

Consider now the case where the lenses are all axisymmetric -- this is valid for the application that will eventually interest us, namely point lenses. In that case the PDF of each lens only depends on $S\define |s|$,
\begin{equation}
p_1(s) \, \dd^2 s = \mathcal{P}_1(S) \dd S \, \frac{\dd\ph}{\pi} \,
\qquad \text{with} \quad
\mathcal{P}_1(S) \define 2\pi S \, p_1(S)
\end{equation}
the PDF of the magnitude of the shear for single lens. The polar angle can be integrated out in the expression of the Fourier transform, which then only depends on $Q=|q|$,
\begin{align}
\tilde{p}_1(Q)
=
\int_0^{\pi} \frac{\dd\ph}{\pi}
\int_0^\infty \dd S \;
\mathcal{P}_1(S) \, \ex{-\ii Q S \cos2(\ph-\psi)}
=
\int_0^\infty \dd S \;
\mathcal{P}_1(S) \, J_0(QS)  \ ,
\end{align}
where $J_0$ is the zeroth order Bessel function. From the above, we notice that $\tilde{p}_1(q)$ may be also be interpreted as the expectation value of that Bessel function for a single lens,
\begin{equation}
\label{eq:p_tilde_1_Bessel}
\tilde{p}_1(Q) = \ev{J_0(QS)}_1 \ ,
\end{equation}
where $\ev{\ldots}_1$ denotes the average over statistics of a single lens.

In the same manner, since $\tilde{p}_N(Q)$ does not depend on the polar angle $\psi$ of $q$, we may integrate this angle out in its inverse Fourier transform,
\begin{align}
p_N(S)
=
\int_0^\infty \frac{Q \, \dd Q}{2\pi}
\int_0^\pi \frac{\dd\psi}{\pi} \;
\ex{\ii Q S \cos 2(\psi-\ph)} \, \tilde{p}_N(Q)
=
\frac{1}{2\pi} \int_0^\infty \dd Q \; Q \, J_0(QS) \, \tilde{p}_N(Q) \ .
\end{align}
The PDF of the sole magnitude~$S$ of the sum all all $N$ complex shears is, therefore,
\begin{equation}
\label{eq:PDF_magnitude_shear_formal}
\mathcal{P}_N(S)
= 2\pi S \, p_N(S)
= \int_0^\infty \dd Q \; QS \, J_0(QS) \, \tilde{p}_N(Q) \ .
\end{equation}

\subsection{Large-\texorpdfstring{$N$}{N} limit}

Now consider the setup depicted in \cref{fig:mesosopic_cone}: the $N\gg 1$ lenses are distributed within a mesoscopic cone with half angle $\Theta$. Now as $\Theta$ is much larger than the typical Einstein radius of the lenses, it quite clear that $\mathcal{P}_1(S)$ must approach $\delta(S)$.\footnote{A more technical, though heuristic, argument goes as follows. Just like the amplification PDF, the shear PDF of a single lens may be expressed as the ratio of a shear cross section with the solid angle of the cone, so that $\mathcal{P}_1(S)\propto (\pi\Theta^2)^{-1}$. Since $\Theta$ is much larger than the typical angular scale characterising a single lens, we expect $\mathcal{P}_1(S)\to 0$ for $S\neq 0$. But since $\mathcal{P}_1(S)$ is a PDF it must be normalised to $1$. The only way out consists in having $\mathcal{P}_1(S\to 0)$ very large, in agreement with the intuition that it is very likely that a single lens lost in a huge domain produces a tiny shear.} Since $J_0(0)=1$, we conclude that $\tilde{p}_1(Q)=\ev{J_0(QS)}_1\approx 1$. This suggests the following manipulation
\begin{align}
\label{eq:p_N_shear_Bessel}
\tilde{p}_N(Q)
= \ev{J_0(QS)}_1^N
= \ev{1 + [J_0(QS) - 1]}_1^N
\approx \exp\pac{ N\ev{J_0(QS)-1}_1 }
\end{align}
in the large-$N$ limit.

\subsection{Application to the effective reduced microshear due to point lenses}

The quantity of interest here is the effective reduced microshear, due to the compact objects located near the line of sight of the dominant lens at $\chi$,
\begin{equation}
s \define
\frac{s\e{os}}{1-\kappa\e{os}}
- \frac{s\e{od}}{1-\kappa\e{od}}
- \frac{s\e{ds}}{1-\kappa\e{ds}}
= \sum_{\ell=1}^N s_\ell \ ,
\qquad
s_\ell \define
\frac{4G m_\ell}{(\cplx{\beta}_\ell^*)^2} \, W(\chi_\ell) ,
\end{equation}
with
\begin{equation}
W(\chi_\ell)
\define (1+z_\ell)
\times
\begin{cases}
\displaystyle
\frac{\chi\e{s}-\chi_\ell}{(1-\kappa\e{os})\chi_\ell\chi\e{s}}
- \frac{\chi\e{d}-\chi_\ell}{(1-\kappa\e{od})\chi_\ell\chi\e{d}}
& \chi_\ell \leq \chi\e{d}
\\[5mm]
\displaystyle
\frac{\chi\e{s}-\chi_\ell}{(1-\kappa\e{os})\chi_\ell\chi\e{s}}
- \frac{(\chi_\ell-\chi\e{d})(\chi\e{s}-\chi_\ell)}
    {(1-\kappa\e{ds})\chi_\ell^2(\chi\e{s}-\chi\e{d})}
& \chi_\ell \geq \chi\e{d}
\end{cases}
\end{equation}
where $m_\ell, z_\ell, \chi_\ell$ denote the mass, redshift, comoving position of lens~$\ell$, and $\cplx{\beta}_\ell$ its complex unlensed angular position with respect to the line of sight; $\chi\e{d}, \chi\e{s}$ are the comoving positions of the main deflector and source.

We assume for simplicity that the lenses are uniformly distributed in comoving space, with masses independent of the positions, so that within the mesoscopic cone of \cref{fig:mesosopic_cone} we have
\begin{equation}
p(\chi_\ell, \beta_\ell, m_\ell) \, \dd\chi_\ell \dd\beta_\ell \dd m_\ell
= \frac{3\chi_\ell^2\dd\chi_\ell}{\chi\e{s}^3} \,
\frac{2\beta_\ell\dd\beta_\ell}{\Theta^2} \,
p(m_\ell) \dd m_\ell \ .
\end{equation}
In the remainder of this appendix we shall drop the subscript $\ell$ to alleviate notation.\footnote{This implies that \emph{in this appendix only} $\chi\define\chi_\ell$ refers to the comoving position of a secondary deflector; in the main text we have instead $\chi\define\chi\e{d}$.}

In such conditions, the Fourier transform~$\tilde{p}_1(Q)$ of the one-lens shear, interpreted as the expectation value of $J_0(SQ)$ following \cref{eq:p_tilde_1_Bessel} reads
\begin{equation}
\tilde{p}_1(Q)
= \ev{J_0(SQ)}_1
= \int \frac{\dd(\chi^3)}{\chi\e{s}^3} \,
\frac{\dd(\beta^2)}{\Theta^2} \,
p(m) \dd m \;
J_0\pac{\frac{4 G m}{\beta^2} \, W(\chi) Q} - 1.
\end{equation}
We may then perform the change of variable $\beta^2\mapsto x=4Gm W Q/\beta^2$ to get
\begin{equation}
\ev{J_0(SQ)}_1 - 1
=
\frac{4 G Q}{\Theta^2}
\int \dd m \; m \, p(m)
\int_0^{\chi\e{s}} \frac{\dd(\chi^3)}{\chi\e{s}^3} \; W(\chi)
\int_{4GmW Q/\Theta^2}^\infty \frac{\dd x}{x^2} \; [J_0(x)-1] .
\end{equation}
In the limit where $\Theta$ is very large, the lower limit in the integral over $x$ can be set to zero, in which case the integral is known,
\begin{equation}
\int_0^\infty \frac{\dd x}{x^2} \; [J_0(x)-1]
= -1 \ ,
\end{equation}
so that
\begin{equation}
\ev{J_0(SQ)}_1 - 1
\approx
-\frac{4 G \ev{m} Q}{\Theta^2}
\int_0^{\chi\e{s}} \frac{\dd(\chi^3)}{\chi\e{s}^3} \; W(\chi) \ .
\end{equation}

The last steps of the calculation consist in (i) substituting the above in \cref{eq:p_N_shear_Bessel} and (ii) computing the inverse Fourier transform to get $p_N(S)$. Step (i) yields
\begin{equation}
\label{eq:p_tilde_N_shear_exponential}
\tilde{p}_N(Q)
= \ex{ - f \tau Q } ,
\end{equation}
with the optical depth $\tau = N\ev{\theta\e{E}^2}/\Theta^2$ corrected by the factor
\begin{align}
f
&\define
\frac{
    \int_0^{\chi\e{s}} \frac{\dd(\chi^3)}{\chi\e{s}^3} \; W(\chi)
    }
    {
    \int_0^{\chi\e{s}} \frac{\dd(\chi^3)}{\chi\e{s}^3} \;
                        \frac{\chi\e{s}-\chi}{a(\chi)\chi\chi\e{s}}
    }
\\
&=
\frac{
    \int_0^{\chi\e{d}} \frac{\dd\chi}{a(\chi)}
        \pac{
            \frac{\chi(\chi\e{s}-\chi)}{(1-\kappa\e{os})\chi\e{s}}
            - \frac{\chi(\chi\e{d}-\chi)}{(1-\kappa\e{od})\chi\e{d}}
            }
    +
    \int_{\chi\e{d}}^{\chi\e{s}} \frac{\dd\chi}{a(\chi)}
        \pac{
            \frac{\chi(\chi\e{s}-\chi)}{(1-\kappa\e{os})\chi\e{s}}
            - \frac{(\chi-\chi\e{d})(\chi\e{s}-\chi)}{(1-\kappa\e{ds})(\chi\e{s}-\chi\e{d})} }
    }
    {
    \int_0^{\chi\e{s}} \frac{\dd\chi}{a(\chi)} \;
        \frac{\chi(\chi\e{s}-\chi)}{\chi\e{s}}
    } .
\end{align}
Note that, in the large-$N$ limit, $\tilde{p}_N(Q)$ is independent on $N$. The last step (ii) is performed by substituting \cref{eq:p_tilde_N_shear_exponential} into \cref{eq:PDF_magnitude_shear_formal}, which finally yields
\begin{equation}
\mathcal{P}_N(S)
= \int_0^\infty \dd Q \; QS \, J_0(QS) \ex{-f\tau Q}
= \frac{f\tau S}{\pac{(f\tau)^2 + S^2}^{3/2}} \ .
\end{equation}

\bibliographystyle{JHEP.bst}
\bibliography{bibliography.bib}

\end{document}